\documentclass{aa}  
\usepackage{chemist}
\usepackage[separate-uncertainty=true]{siunitx}
\usepackage{natbib}
\usepackage{multirow}
\usepackage{subfig}
\usepackage{float}
\usepackage{txfonts}
\usepackage{graphicx}
\usepackage{hyperref}
\hypersetup{
    colorlinks=true,
    linkcolor=black,
    citecolor=black,
    filecolor=black,
    urlcolor=black,
}
\bibpunct{(}{)}{;}{a}{}{,}
\newcommand{\chem}[1]{${\mathrm{#1}}$} 
\newcommand*{\pmt}{$\pm$}  
\newcommand*\diff{\mathop{}\!\mathrm{d}} 
\newcommand*{\trans}[2]{$#1 - #2$}

\begin{document} 

\title{The water line emission and ortho-to-para ratio in the Orion Bar photon-dominated region}

\author{T. Putaud\inst{1}, X. Michaut\inst{1}, F. Le Petit\inst{1}, E. Roueff\inst{1},\and D. C. Lis\inst{2,1}}

\institute{Sorbonne Université, Observatoire de Paris, Université PSL, CNRS, LERMA, F-75014, Paris, France
\and Jet Propulsion Laboratory, California Institute of Technology, 4800 Oak Grove Drive, Pasadena, CA 91109, USA }

\date{$\copyright 2019$. All rights reserved. Received date / Accepted date} 

\abstract
{The ortho-to-para ratio (OPR) of water in the interstellar medium (ISM) is often assumed to be related to the formation temperature of water molecules, making it a potentially interesting tracer of the thermal history of interstellar gas.}
{A very low OPR of \numrange[range-phrase=--]{0.1}{0.5} was previously reported in the Orion Bar photon-dominated region (PDR), based on observations of two optically thin \chem{H_2^{18}O} lines which were analyzed by using a single-slab large velocity gradient (LVG) model. The corresponding spin temperature does not coincide with the kinetic temperature of the molecular gas in this UV-illuminated region. This was interpreted as an indication of water molecules being formed on cold icy grains which were subsequently released by UV photodesorption.}
{A more complete set of water observations in the Orion Bar, including seven \chem{H_2^{16}O} lines and one \chem{H_2^{18}O} line, carried out using \textit{Herschel}/HIFI instrument, was reanalyzed using the Meudon PDR code to derive gas-phase water abundance and the OPR. The model takes into account the steep density and temperature gradients present in the region.}
{The model line intensities are in good agreement with the observations assuming that water molecules formed with an OPR corresponding to thermal equilibrium conditions at the local kinetic temperature of the gas and when solely considering gas-phase chemistry and water gas-grain exchanges through adsorption and desorption. Gas-phase water is predicted to arise from a region deep into the cloud, corresponding to a visual extinction of $A_{\mathrm{V}} \sim 9$, with a \chem{H_2^{16}O} fractional abundance of $\sim$ \num{2e-7} and column density of \SI{1.4(8)e15}{cm^{-2}} for a total cloud depth of $A_{\mathrm{V}}=15$. A line-of-sight average ortho-to-para ratio of \num{2.8(2)} is derived.}
{The observational data are consistent with a nuclear spin isomer repartition corresponding to the thermal equilibrium at a temperature of \SI{36(2)}{K}, much higher than the spin temperature previously reported for this region and close to the gas kinetic temperature in the water-emitting gas.}

\keywords{ISM: molecules -- ISM: individual objects: Orion Bar -- (ISM:) photon-dominated region (PDR) -- ISM: lines and bands -- Submillimeter: ISM}

\titlerunning{The water line emission and OPR in the Orion Bar PDR}
\authorrunning{T. Putaud et al.}
\maketitle

\section{Introduction}

\begin{table*}
\centering
\caption{Observations used for the analysis of water line emission in the Orion Bar. \label{tab:ObsUsed}}
\begin{tabular}{ccclccccclll}
\hline \hline
Molecule                           & S\tablefootmark{a}   & Transition                                 & \multicolumn{1}{c}{$\nu$\tablefootmark{b}} & $E_{up}$\tablefootmark{b} &$g_{up}$\tablefootmark{c}& HPBW\tablefootmark{d}& $\eta_{mb}$\tablefootmark{d}& HIFI & \multicolumn{1}{c}{Observing} &\multicolumn{1}{c}{$t_{int}$\tablefootmark{e}}& \multicolumn{1}{c}{Observation} \\
                                   &                      &                                            & \multicolumn{1}{c}{(\si{GHz})}             & (\si{K})                  &                         & (\si{''})            &                             & band & \multicolumn{1}{c}{mode}      &\multicolumn{1}{c}{(sec)}                     & \multicolumn{1}{c}{Id.}         \\ \hline
\multirow{17}{*}{\chem{H_2^{16}O}} & \multirow{3}{*}{o}   & \multirow{3}{*}{\trans{1_{10}}{1_{01}}} & \multirow{3}{*}{\num{556.936}}             & \multirow{3}{*}{60.96}    & \multirow{3}{*}{9}      & \multirow{3}{*}{37}  & \multirow{3}{*}{0.62}       & 1B   & Load Chop                     &103.3                                         &1342215923\tablefootmark{f}      \\ 
                                   &                      &                                            &                                            &                           &                         &                      &                             & 1A   & Load Chop                     &126.3                                         &1342218527\tablefootmark{g}      \\
								   &                      &                                            &                                            &                           &                         &                      &                             & 1B   & On-The-Fly                    &4.0                                           &1342215922                       \\\cline{2-12}
                                   & \multirow{2}{*}{p}   & \multirow{2}{*}{\trans{1_{11}}{0_{00}}} & \multirow{2}{*}{\num{1113.343}}            & \multirow{2}{*}{53.43}    & \multirow{2}{*}{3}      & \multirow{2}{*}{19}  & \multirow{2}{*}{0.63}       & 4B   & Load Chop                     &105.8                                         &1342217720                       \\
								   &                      &                                            &                                            &                           &                         &                      &                             & 4B   & On-The-Fly                    &2.9                                           &1342215970                       \\\cline{2-12}
                                   & \multirow{3}{*}{o}   & \multirow{3}{*}{\trans{2_{12}}{1_{01}}} & \multirow{3}{*}{\num{1669.905}}            & \multirow{3}{*}{114.38}   & \multirow{3}{*}{15}     & \multirow{3}{*}{12}  & \multirow{3}{*}{0.58}       & 6B   & Fast DBS                      &1274                                          &1342229840\tablefootmark{h}      \\
								   &                      &                                            &                                            &                           &                         &                      &                             & 6B   & Fast DBS                      &134.6                                         &1342218426\tablefootmark{i}      \\            
								   &                      &                                            &                                            &                           &                         &                      &                             & 6B   & On-The-Fly                    &4.0                                           &1342251054                       \\\cline{2-12}
                                   & \multirow{2}{*}{p}   & \multirow{2}{*}{\trans{2_{02}}{1_{11}}} & \multirow{2}{*}{\num{987.927}}             & \multirow{2}{*}{100.85}   & \multirow{2}{*}{5}      & \multirow{2}{*}{21}  & \multirow{2}{*}{0.64}       & 4A   & Load Chop                     &105.8                                         &1342218628                       \\
								   &                      &                                            &                                            &                           &                         &                      &                             & 4A   & On-The-Fly                    &2.9                                           &1342218217                       \\\cline{2-12}
                                   & \multirow{5}{*}{p}   & \multirow{5}{*}{\trans{2_{11}}{2_{02}}} & \multirow{5}{*}{\num{752.033}}             & \multirow{5}{*}{136.94}   & \multirow{5}{*}{5}      & \multirow{5}{*}{28}  & \multirow{5}{*}{0.65}       & 2B   & Load Chop                     &94.0                                          &1342216378                       \\
								   &                      &                                            &                                            &                           &                         &                      &                             & 2B   & On-The-Fly                    &3.9                                           &1342190848                       \\
								   &                      &                                            &                                            &                           &                         &                      &                             & 2B   & On-The-Fly                    &4.9                                           &1342190849                       \\
								   &                      &                                            &                                            &                           &                         &                      &                             & 2B   & On-The-Fly                    &2.9                                           &1342203221                       \\
								   &                      &                                            &                                            &                           &                         &                      &                             & 2B   & On-The-Fly                    &3.9                                           &1342203222                       \\\cline{2-12}
                                   &o                     & \trans{2_{21}}{2_{12}}                  & \num{1661.008}                             & \num{194.09}              & 15                      & 13                   & 0.58                        & 6B   & Fast DBS                      &1274                                          &1342229840                       \\ \cline{2-12}
                                   &o                     & \trans{3_{12}}{3_{03}}                  & \num{1097.365}                             & \num{249.44}              & 21                      & 19                   & 0.63                        & 4B   & Load Chop                     &105.8                                         &1342217720                       \\\hline\hline
\multirow{3}{*}{\chem{H_2^{18}O}}  & \multirow{2}{*}{o}   & \multirow{2}{*}{\trans{1_{10}}{1_{01}}} & \multirow{2}{*}{\num{547.676}}             & \multirow{2}{*}{60.46}    & \multirow{2}{*}{9}      & \multirow{2}{*}{38}  & \multirow{2}{*}{0.62}       & 1A   & Freq. Switch                  &5278                                          &1342205273                       \\ 
								   &                      &                                            &                                            &                           &                         &                      &                             & 1A   & On-The-Fly                    &3.96                                          &1342216355                       \\\cline{2-12}
                                   &p                     & \trans{1_{11}}{0_{00}}                  & \num{1101.697}                             & \num{52.87}               & 3                       & 19                   & 0.63                        & 4B   & Load Chop                     &105.8                                         &1342217720\tablefootmark{j}      \\                       
\hline
\end{tabular}
\tablefoot{
\tablefoottext{a}{Nuclear spin state, o stands for ortho and p for para.}
\tablefoottext{b}{Frequency of the transition (up-down) and rotational energy of the upper level in relation to the para $0_{00}$ level from the HITRAN2016 database~\citep{Hitran2016}}
\tablefoottext{c}{Statistical weight of the upper level.}
\tablefoottext{d}{Half-power beam width and main-beam efficiency of the \textit{Herschel} telescope, averaged between H and V polarizations, at the water line frequency~\citep{Mueller2012} .}
\tablefoottext{e}{Integration time from FITS header.}
\tablefoottext{f}{Hereafter, this line is labeled \chem{H_2^{16}O} \trans{1_{10}}{1_{01}}-a.}
\tablefoottext{g}{Hereafter, this line is labeled \chem{H_2^{16}O} \trans{1_{10}}{1_{01}}-b.}
\tablefoottext{h}{Hereafter, this line is labeled \chem{H_2^{16}O} \trans{2_{12}}{1_{01}}-a.}
\tablefoottext{i}{Hereafter, this line is labeled \chem{H_2^{16}O} \trans{2_{12}}{1_{01}}-b.}
\tablefoottext{j}{\chem{H_2^{18}O} \trans{1_{11}}{0_{00}} para line is not unambiguously detected.}
}

\end{table*}

Photon-dominated regions (PDRs) are surface layers of molecular clouds irradiated by a strong UV photon flux, such as those present in star-forming regions. Studies of the physical and chemical properties of molecules in such regions are of great interest for understanding the feedback of young stars on their surrounding medium.

The Orion Molecular Cloud 1 (OMC-1) contains two embedded star-forming regions: Orion BN/KL and Orion South, as well as a group of young massive stars known as the Trapezium Cluster. The Trapezium, and in particular the brightest O6 type star $\Theta^1$ Ori C, has irradiated the surrounding molecular cloud creating an \chem{H}{\tiny II} region, which is bordered by the Orion Bar on its southeastern side and by the Orion Ridge on its western side~\citep{RodriguezFrancoAA1998}. Close to Earth (\SI{414}{pc};~\citealt{MentenAA2007}) and with a nearly edge-on geometry~\citep{HogerheijdeAA1995,JansenAA1995} explained by the blister model (see for example \citealt{WenAPJ1995} and references therein), the Orion Bar is an excellent place to test PDR models~\citep{TielensApJ1985,SternbergApJ1989,LePetitApJSS2006,HollenbachApJ2009,AndreeLabschAA2017}. 

The Orion Bar has been observed extensively in multiple gas tracers~\citep{FuenteAA1996,LarssonAA2003,LeuriniAA2006,vanderTakAA2012,vanderTakAA2013,CuadradoAA2015,CuadradoAA2017,NagyAA2017,JoblinAA2018}. Previous studies have led to a UV flux of $G_0=\numrange[range-phrase=-]{1}{4e4}$ in Habing units~\citep{TielensApJ1985,MarconiAA1998} impinging upon a molecular cloud with a mean \chem{H_2} density of about \SI{5e4}{cm^{-3}}~\citep{WyrowskiAA1997} and a kinetic temperature of \SI{85}{K}~\citep{HogerheijdeAA1995}. A model of the Orion Bar with a clumpy medium, based on high-density clumps ($n_{\mathrm{H}}=\numrange[range-phrase=-]{e6}{e7}~\si{cm^{-3}}$) embedded in an interclump medium ($n_{\mathrm{H}}=\numrange[range-phrase=-]{5e4}{e5}~\si{cm^{-3}}$), was proposed to explain the following: the excitation of atomic lines, some excited \chem{CO} lines, and warm \chem{H_2} observations~\citep{ParmarApJl1991,TauberApJ1994,vanderWerfAA1996}. The presence of high-density clumps was also suggested by observations of \chem{H_2CO}, \chem{HCO^{+}}, and \chem{HCN}~\citep{HogerheijdeAA1995,YoungOwlApJ2000,LisApJl2003}. However, the clumpy model is challenged by the very high resolution observations provided by the Atacama Large Millimeter/submillimeter Array (ALMA) radiotelescope~\citep{GoicoecheaNature2016,GoicoecheaAA2017}, which show that, to the first order, the structure of the PDR is a compressed layer at high-pressure where a warm chemistry takes place, leading to the presence of molecules such as \chem{SH^{+}} and high-J excited \chem{CO}~\citep{JoblinAA2018}. This structure is also observed in other PDRs, such as Trumpler 14 in the Carina nebula~\citep{WuAA2018}. High-density structures exist inside the PDR, as seen in \chem{OH}~\citep{GoicoecheaAA2011,ParikkaAA2017} and \chem{H^{13}CN} maps~\citep{LisApJl2003}, as well as in ALMA maps~\citep{GoicoecheaNature2016}. The high-pressure layer can be interpreted by the photo-evaporation effect as explained by~\cite{BronAA2018} with the PDR hydrodynamical code Hydra.

An important tracer for studying the physical and chemical evolution of the interstellar medium (ISM) is \chem{H_2O}. This molecule (like other hydrides) has two identical protons and exists in two nuclear configurations, called nuclear spin isomers, where the hydrogen nuclear spins are either parallel (ortho) or antiparallel (para). At the thermal equilibrium, the ortho-to-para ratio (OPR) of gas-phase water molecules is determined by the kinetic temperature of the gas. Some non-thermal equilibrium OPRs have been reported in different regions~\citep{HogerheijdeScience2011,LisJPCA2013,ChoiAA2014}. In particular, \cite{ChoiAA2014} derived a very low OPR of \numrange[range-phrase=--]{0.1}{0.5} in the Orion Bar, based on observations of two \chem{H_2^{18}O} emission lines. With such a low OPR, the so-called spin temperature does not exceed \SI{10}{K}, which is much lower than the gas kinetic temperature in this region.

To explain the discrepancy between kinetic and spin temperatures, it was suggested that the spin temperature might reflect the temperature of the grains at the time the molecules formed. Indeed, a PDR model which considers oxygen grain chemistry has shown that the photodesorption of water molecules formed on grains could be an efficient pathway for gas-phase water production~\citep{HollenbachApJ2009}. After desorption from grains, the non-thermal equilibrium OPR could have been preserved due to the very low pressure of the ISM~\citep{CaccianiPRA2012}. However, a recent laboratory study of water molecules photodesorbed from cold surfaces using a UV laser produced a gas-phase water with an OPR in agreement with the high thermal equilibrium value of three~\citep{HamaScience2016}.

Further studies of the impact of grains on the abundance and quantum state of gas-phase molecules have yet to be carried out. However, the reanalysis of the observational data, using more sophisticated models, has sometimes modified the reported OPR of hydride molecules in the ISM. Better calibrated observational data and improved computational models have indeed shown that the low OPR value measured in TW Hydrae protoplanetary disk~\citep{HogerheijdeScience2011} is very model-dependent~\citep{SalinasAA2016}. Furthermore, the observations are consistent, within the observational uncertainties, with the high-temperature equilibrium value.

We present a new analysis of the \chem{H_2^{16}O} and \chem{H_2^{18}O} lines observed toward the Orion Bar with \textit{Herschel}/HIFI, which are reduced using the latest pipeline and with the latest calibration~\citep{Mueller2012}. Careful corrections are applied to the spectra to account for the spatial offsets and beam coupling effects. The Meudon PDR code~\citep{LePetitApJSS2006} is used to constrain the physical conditions inside the Orion Bar and derive the gas-phase water OPR. The impact of grain processes, such as water adsorption and thermal or photo-induced desorption, is also investigated.

\section{Observations}

Observations of the Orion Bar discussed here were carried out at the \chem{CO^{+}} peak position~\citep{StoerzerAA1995} ($\alpha_{2000}=05^{\si{h}}35^{\si{m}}20.61^{\si{s}}$, $\delta_{2000}=-05\si{\degree}\ang{;25;}\ang{;;14.0}$) as part of the HEXOS\footnote{\textit{Herschel} observations of the EXtra-Ordinary Sources}~\citep{BerginAA2010} guaranteed time key program (GTKP). The Heterodyne Instrument for the Far-Infrared~(HIFI, \citealt{deGraauwAA2010}) aboard the \textit{Herschel} Space Observatory~\citep{PilbrattAA2010}, operating with the wide-band spectrometer (WBS) as a backend, provided spectra at \SI{1.1}{MHz} resolution for H and V orthogonal polarizations. The data were reduced with the \textit{Herschel} Interactive Processing Environement~(HIPE, \citealt{Ott2010}) version 14.1.0. Within the whole HIFI's frequency range (\numrange[range-phrase=--]{480}{1250}~\si{GHz} and \numrange[range-phrase=--]{1410}{1910}~\si{GHz}), seven rotational emission lines of \chem{H_2^{16}O} and one of \chem{H_2^{18}O} were detected from March 2011 to October 2011, either in load chop, fast dual beam switch (DBS), or frequency switch observing mode. The upper limit of the ground state para \chem{H_2^{18}O} line was also included in the analysis. Observations used in this article are listed in Table~\ref{tab:ObsUsed}. Most of these lines were already presented in an extensive spectral survey carried out by \cite{NagyAA2017}.

In addition to the spectral scan observations, nine on-the-fly (OTF) HIFI maps using the WBS as a backend were also analyzed (see Table~\ref{tab:ObsUsed}). Within the HEXOS GTKP~\citep{BerginAA2010}, \cite{HabartProp2011} proposal, and calibration programs\footnote{calibration\_pvhifi\_37 and calibration\_pvhifi\_85}, observations were obtained from February 2010 to September 2012. The total field of view is roughly $\ang{;;50}\times\ang{;;100}$ (except for \chem{H_2^{16}O} \trans{1_{10}}{1_{01}} and \trans{2_{12}}{1_{01}} where only a \ang{;;100} perpendicular strip through the Bar was performed) with a position angle perpendicular to the Bar. The maps were regridded using the GILDAS\footnote{https://www.iram.fr/IRAMFR/GILDAS/} software leading to a spatial resolution typically \SI{5}{\%} lower than the \textit{Herschel} telescope half-power beam width (HPBW). All water spectra and water maps are shown in the Appendix~\ref{app:LineUsed}.

Further calibration corrections must be applied to the data. First, the antenna temperature on the $T_A^{*}$ scale, given by the pipeline, should be corrected by the ratio of the forward efficiency ($\eta_l=0.96$, \citealt{RoelfsemaAA2012}) and the main-beam efficiency ($\eta_{mb}$) to produce spectra on the main-beam brightness temperature scale $T_{mb}$~\citep{Mueller2012}.
\begin{equation}
\label{eq:MainBeamScale}
T_{mb}=\frac{\eta_l}{\eta_{mb}}T_A^{*}
\end{equation}  
This correction only considers the main detection lobe of the telescope. The latest values of $\eta_{mb}$ and the HPBW, derived from Mars observations and complete optical model of the \textit{Herschel} telescope~\citep{Mueller2012,ShipmanAA2017}, were used in the analysis. These numbers differ up to \SI{20}{\percent} from the previous estimates assuming a simplified Gaussian beam shape~\citep{RoelfsemaAA2012}. The exact values used for each H and V observations are listed in Table~\ref{tab:IntensityCal} in the Appendix~\ref{app:LineUsed}.

Furthermore, owing to the variation in the HPBW with frequency, the beam coupling correction has to be carefully considered, as discussed below. To quantify this effect, three high angular resolution \chem{CO} maps were convolved to the beam widths corresponding to the various water transitions. The beam coupling factor $\Omega$ is defined as the ratio between the convolved intensity at the targeted coordinates and the maximum intensity toward the Bar at high resolution. The \chem{CO} \trans{1}{0} map is part of the CARMA-NRO Orion Survey~\citep{KongAPJS2018} and was obtained by combining the CARMA\footnote{Combined Array for the Research in Millimeter Astronomy} interferometric data with NRO45\footnote{Nobeyama Radio Observatory \SI{45}{m} telescope} single-dish observations leading to a resolution of \ang{;;9} over a \ang{2;;}$\times$\ang{2;;} field of view. The \chem{^{12}CO} \trans{6}{5} and \chem{^{13}CO} \trans{6}{5} maps were obtained using the Caltech Submillimeter Observatory (CSO)~\citep{LisApJl2003} at \ang{;;11} resolution and cover respectively a \ang{;3;}$\times$\ang{;4;} and \ang{;3;}$\times$\ang{;2.5;} regions.

\section{Results}

\begin{figure}
\centering
\includegraphics[scale=1]{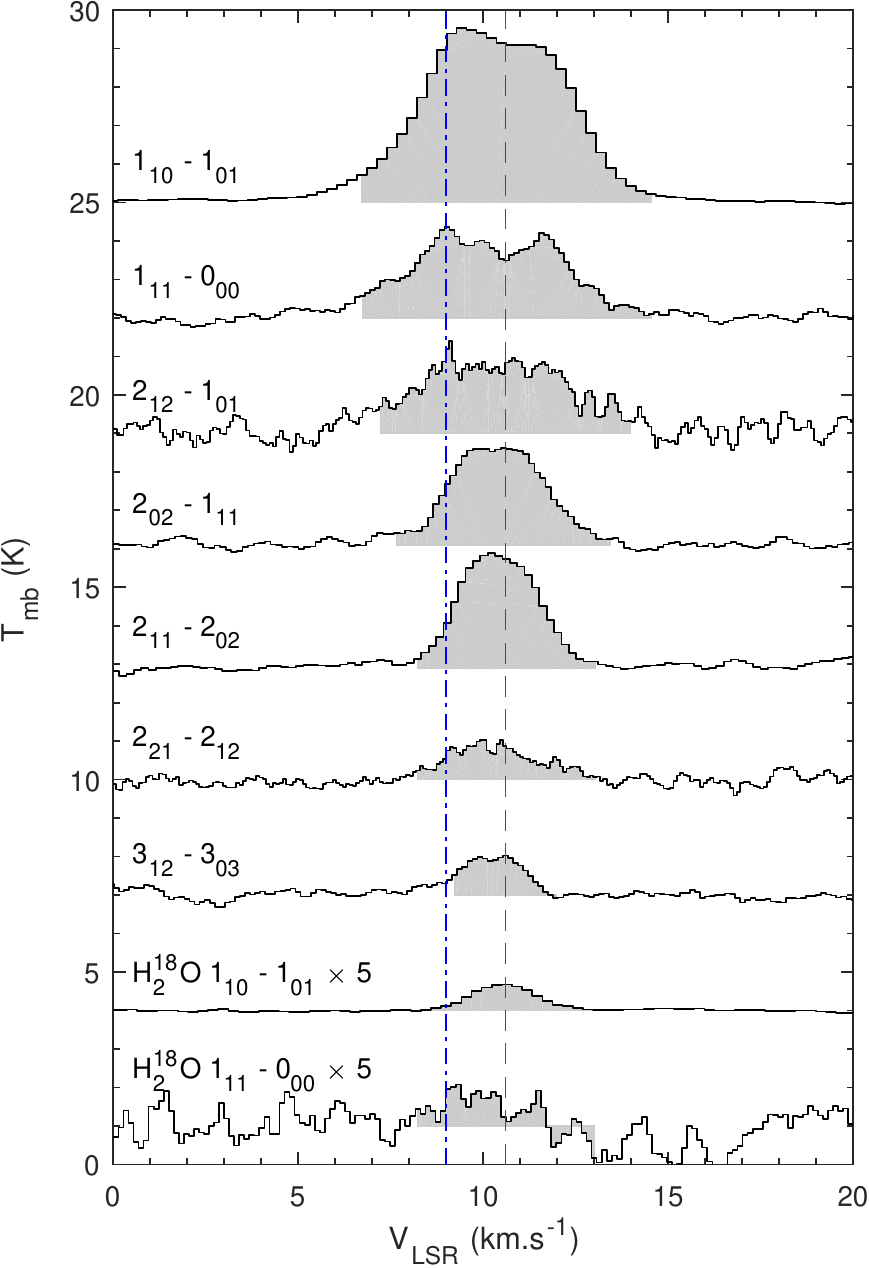}
\caption{Averaged spectra of \chem{H_2^{16}O} and \chem{H_2^{18}O} lines toward the Orion Bar on the main-beam brightness temperature scale, not corrected for the beam coupling factor. The gray areas represent the velocity integration range. The blue dash-dotted line and red dashed line are the expected velocity components from Orion Ridge (\SI{9.0}{km.s^{-1}}) and Orion Bar (\SI{10.6}{km.s^{-1}}). \chem{H_2^{18}O} lines were multiplied by a factor of 5. \label{fig:DataIntegration}}
\end{figure}

Integrated line intensities of the water spectra shown in Fig.~\ref{fig:DataIntegration} are given in Table~\ref{tab:IntensitySet}. On average, the integration was performed from a local standard of rest velocity ($V_{\mathrm{LSR}}$) of \SI{7.7}{km.s^{-1}} to \SI{13.5}{km.s^{-1}}. Precise integration ranges for each line are listed in Table~\ref{tab:IntensityCal}. Linear baseline subtraction was carefully applied except for frequency switch observation mode where a fourth order polynomial baseline was used because of residual standing waves. The uncertainties of the integrated line intensities, in units of \si{K.km.s^{-1}}, are computed as
\begin{equation}
\Delta I =\sqrt{\frac{N_{int}}{N}\sum_i^N T_{mb,i}^2}\Delta V
\end{equation}
where the summation runs over all the channels $N$ outside the line integration velocity range, $N_{int}$ is the number of channels inside the integration velocity range and $\Delta V$ is the width of a velocity channel.

The average intensities of the H and V polarizations agree with the values previously reported~\citep{NagyAA2017,Choi2015}, but are a factor of two higher than the intensities derived from SPIRE observations averaged along the Bar~\citep{HabartAA2010}. The ortho \chem{H_2^{18}O} line intensity is also very close to the one derived by \cite{ChoiAA2014}. However the para \chem{H_2^{18}O} line (a 4.7$\sigma$ detection in \citealt{ChoiAA2014}) is not clearly detected in our data set given the noise level in the spectrum.

\subsection{Line profiles}

Figure~\ref{fig:DataIntegration} shows the line profiles of the eight detected lines, and the spectrum in the para \chem{H_2^{18}O} \trans{1_{11}}{0_{00}} line region. Some of the lines have a profile inconsistent with a single Gaussian, in particular the \chem{H_2^{16}O} ground state lines (\trans{1_{11}}{0_{00}}, \trans{1_{10}}{1_{01}}, and \trans{2_{12}}{1_{01}}) exhibit two velocity components around \SI{9.5}{km.s^{-1}} and \SI{11.5}{km.s^{-1}}. Previous studies of tracers arising close to the \chem{CO^{+}} peak reported velocity components from the Orion Bar at \SI{10.6}{km.s^{-1}}, as well as features from the Orion Ridge at \SI{9}{km.s^{-1}}~\citep{vanderTakAA2013,GoicoecheaAPJ2015,NagyAA2017,CuadradoAA2017}. Moreover lines between $V_{LSR} = \SI{8}{km.s^{-1}}$ and $V_{LSR} = \SI{9}{km.s^{-1}}$ were attributed to OMC-1 as suggested by IRAM \SI{30}{m} telescope observations of small hydrocarbons~\citep{CuadradoAA2015}.
Strips across the Bar have shown that \chem{CH^{+}} peak velocity shifts from \SI{9}{km.s^{-1}} in front of the Bar to \SI{11}{km.s^{-1}} behind~\citep{ParikkaAA2017}.

In the case of water emission, the two-component line profiles are more likely explained by self-absorption in the low-energy lines~\citep{Choi2015}, which are expected to be optically thicker than the excited lines. Such an explanation is supported by the fact that excited lines exhibit a single component at an intermediate velocity, close to the expected velocity of the Bar~\citep{NagyAA2017}. The self-absorption dip also matches this velocity. Furthermore, a full width at half maximum (FWHM) of $\sim$~\SI{4.5}{km.s^{-1}} is measured for the lines showing a self-absorption dip in good agreement with optically thick lines in the Orion Bar~\citep{HogerheijdeAA1995,NagyAA2017} and $\sim$~\SI{2.5}{km.s^{-1}} for the other ones which is close to the typical line width observed in this region~\citep{ChoiAA2014,CuadradoAA2015,NagyAA2017}. The FWHM line widths are reported in Table~\ref{tab:IntensityCal}.

Nevertheless, \chem{CO} lines~\citep{JoblinAA2018} show a weak asymmetry at low velocities that could represent signal coming from the Orion Ridge or OMC-1. Assuming a peak velocity of \SI{10.6}{km.s^{-1}} and considering a symmetric line profile, velocity integration range is derived for each line from the half width measured at high velocity (see Table~\ref{tab:IntensityCal}). Even if a possible contamination of the line by surrounding features could still lead to an overestimate of the intensity, the resulting intensities are our best-estimates, because the optically thick character of the lines makes a fitting method unreliable. 

\subsection{Beam dilution}

\begin{figure}
\centering
\includegraphics[scale=0.45]{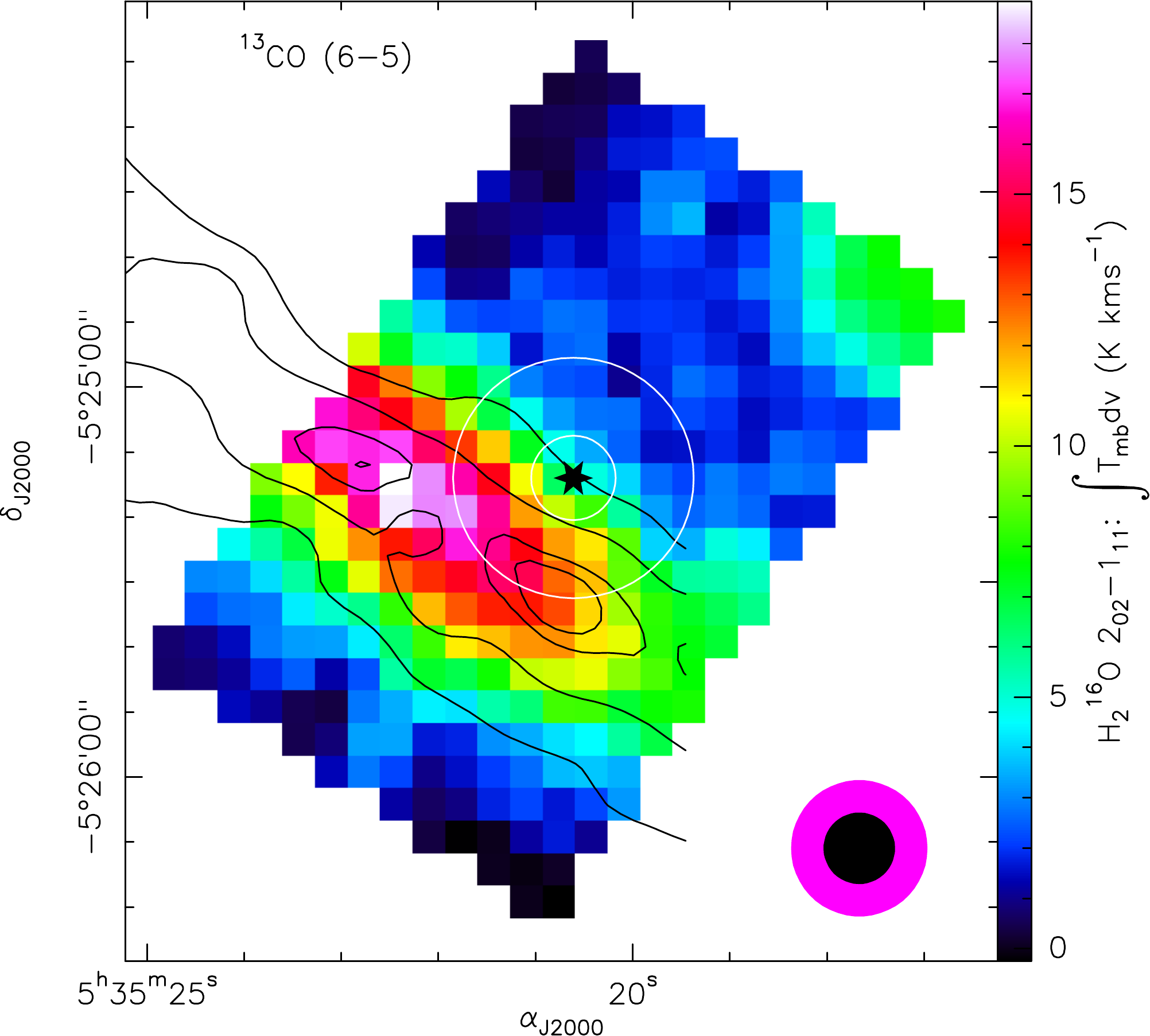}
\caption{\chem{H_2^{16}O} \trans{2_{02}}{1_{11}} emission map at \SI{988}{GHz} integrated from \SI{7.7}{km.s^{-1}} to \SI{13.5}{km.s^{-1}}. Contours represent the emission of \chem{^{13}CO} \trans{6}{5} at \si{95}{\%}, \si{90}{\%}, \si{75}{\%} and \si{50}{\%} of the maximum intensity, integrated from \SI{2}{km.s^{-1}} to \SI{17}{km.s^{-1}}, observed with the CSO at a resolution of \ang{;;11}. Black star marks the \chem{CO^{+}} peak position, white circles show the HIFI HPBW at the extreme frequencies, black and magenta disks show the CSO and \SI{988}{GHz} HIFI beam width (see Table~\ref{tab:ObsUsed}). FUV field from the Trapezium cluster comes from the upper right corner.\label{fig:COWaterMap}}
\end{figure}

\begin{table*}
\centering
\caption{Observed intensities of water lines toward the Orion Bar and their correction for the beam coupling factor. \label{tab:IntensitySet}}
\begin{tabular}{cccc|cccc|c}
\hline \hline
Molecule                          & Transition                & $I_{Obs}$\tablefootmark{a} & $I_{Corr}$         & \multicolumn{4}{c|}{$\Omega$\tablefootmark{b}}                                                                                                                    & $1/\Omega$\tablefootmark{c}\\
                                  &                           & (\si{K.km.s^{-1}})         & (\si{K.km.s^{-1}}) & \chem{H_2O} & \chem{^{12}CO} \trans{1}{0}\tablefootmark{d} & \chem{^{12}CO} \trans{6}{5}\tablefootmark{e} & \chem{^{13}CO} \trans{6}{5}\tablefootmark{e}               \\ \hline
\multirow{7}{*}{\chem{H_2^{16}O}} & \trans{1_{10}}{1_{01}}    &21.13\pmt0.95               & 27.48\pmt1.19      &0.83\pmt0.07 & 0.80\pmt0.02                                    & 0.85\pmt0.02                                    & 0.66\pmt0.04                                    &1.32\pmt0.17 \\ 
                                  & \trans{1_{11}}{0_{00}}    &10.28\pmt0.75               & 12.57\pmt0.89      &0.61\pmt0.04 & 0.86\pmt0.03                                    & 0.85\pmt0.03                                    & 0.73\pmt0.08                                    &1.24\pmt0.13  \\
                                  & \trans{2_{12}}{1_{01}}    &8.91\pmt0.60                & 12.46\pmt0.83      &0.78\pmt0.04 & 0.86\pmt0.03                                    & 0.79\pmt0.01                                    & 0.72\pmt0.04                                    &1.28\pmt0.11  \\
                                  & \trans{2_{02}}{1_{11}}    &8.06\pmt0.59                & 9.41\pmt0.66       &0.3\pmt0.2   & 0.85\pmt0.04                                    & 0.86\pmt0.03                                    & 0.73\pmt0.07                                    &1.27\pmt0.13 \\
                                  & \trans{2_{11}}{2_{02}}    &7.82\pmt0.60                & 9.95\pmt0.75       &0.77\pmt0.01 & 0.82\pmt0.03                                    & 0.87\pmt0.03                                    & 0.69\pmt0.05                                    &1.24\pmt0.16  \\
                                  & \trans{2_{21}}{2_{12}}    &2.62\pmt0.31                & 3.13\pmt0.34       &             & 0.87\pmt0.03                                    & 0.79\pmt0.01                                    & 0.74\pmt0.04                                    &1.26\pmt0.10  \\ 
                                  & \trans{3_{12}}{3_{03}}    &1.85\pmt0.18                & 2.24\pmt0.20       &             & 0.86\pmt0.03                                    & 0.85\pmt0.03                                    & 0.73\pmt0.08                                    &1.24\pmt0.13  \\\hline
\multirow{2}{*}{\chem{H_2^{18}O}} & \trans{1_{10}}{1_{01}}    &0.29\pmt0.02                & 0.36\pmt0.02       &0.6 \pmt0.4  & 0.79\pmt0.03                                    & 0.85\pmt0.02                                    & 0.65\pmt0.05                                    &1.33\pmt0.18  \\ 
                                  & \trans{1_{11}}{0_{00}}    &$<$0.34\pmt0.15             & $<$0.42\pmt0.16    &             & 0.86\pmt0.03                                    & 0.85\pmt0.03                                    & 0.73\pmt0.08                                    &1.24\pmt0.13  \\ 
\hline
\end{tabular}
\tablefoot{
\tablefoottext{a}{Integration range is around \SI{7,7}{km.s^{-1}} and \SI{13.5}{km.s^{-1}}. Precise integration range for each line is given in Table~\ref{tab:IntensityCal}.}
\tablefoottext{b}{H and V averaged beam coupling factors from each tracer.}
\tablefoottext{c}{Mean global corrective coefficients from \chem{CO} maps.} 
\tablefoottext{d}{\cite{KongAPJS2018}.}
\tablefoottext{e}{\cite{LisApJl2003}. Values for \chem{^{13}CO} are obtained with a \ang{;;4} shift toward the ionization front.}
}
\end{table*}

The observed water lines cover a wide range of frequencies, leading to a significant variation in the \textit{Herschel} telescope beam width. Extreme beam sizes at the spectral scan coordinates are shown as white circles in the \chem{H_2^{16}O} \trans{2_{02}}{1_{11}} map in Fig.~\ref{fig:COWaterMap}. In this map, corresponding to an intermediate frequency, the maximum HIFI beam size is larger than the spatial width of the Bar. Thus the coupling of the beam to the water-emitting region has to be considered when comparing the observed line intensities at different frequencies with model predictions. In addition, Figure~\ref{fig:COWaterMap} clearly shows that water emission peaks deeper in the molecular cloud with respect to the \chem{CO^{+}} peak position. Thus the observations were not centered on the water emission peak in the Orion Bar and only a fraction of the emission is included in the beam. 

The offset effect can be easily taken into account using water maps by comparing the intensity at the targeted coordinates with the mean intensity along the Bar. This gives a rough estimate of the beam coupling factor of about 0.8. Nevertheless, applying this method does not consider the fact that the maximum intensity is already impacted by the size of the beam, in particular for low frequency transitions for which the HPBW of the \textit{Herschel} telescope is the largest.

To retrieve the maximal intensity of each line in the Bar, the effect of the pointing offset and the beam width variations is investigated from high resolution maps. The intensity along the Bar at the full resolution is derived and compared to the intensity at the observed coordinates after convolution at the HPBW corresponding to the line frequency. Three \chem{CO} maps were used to derive the beam coupling factor. These maps have a higher resolution than the most resolved water maps (\ang{;;9} for \chem{^{12}CO} \trans{1}{0}, \citealt{KongAPJS2018} and \ang{;;11} for \chem{^{12}CO} \trans{6}{5} and \chem{^{13}CO} \trans{6}{5}, \citealt{LisApJl2003}) and are large enough to be convolved without edge effect. Moreover, they span different excitation conditions and different spatial extensions. Indeed, \chem{^{12}CO} emission is slightly wider than the water one, whereas \chem{^{13}CO} peaks slightly deeper into the cloud (\ang{;;4}, see \chem{H_2^{16}O} \trans{1_{11}}{0_{00}} and \trans{2_{12}}{1_{01}} maps in Fig.~\ref{fig:WaterOffset}). So averaging the beam coupling factors derived from each \chem{CO} maps should give a good estimate of its value and its uncertainty (Table.~\ref{tab:IntensitySet}). We note that the \ang{;;4} offset between \chem{^{13}CO} and water emission corresponds to a typical pointing error of the CSO telescope at this high frequency. Since only the spatial extension of the emission is considered in the estimation of the beam coupling factor, the \chem{^{13}CO} map was shifted toward the ionization front by \ang{;;4} to match the water emission, in order not to add artificial bias due to the position.

The beam coupling factors adopted for each observed transitions are given in Table~\ref{tab:IntensityCal} together with our best-estimates for the intensity of each lines. The beam coupling factors are close to 0.8 whereas the values retrieved for high-J \chem{CO} lines assuming a \ang{;;2} filament for the Bar range from 0.07 to 0.23~\citep{JoblinAA2018}. Each H and V spectrum is corrected separately for beam coupling to account for a pointing offset between the two polarizations and the resulting spectra are then averaged with root mean square (rms) weighting: 
\begin{equation}
T_{\mathrm{Weight}}=\frac{w_HT_H+w_VT_V}{w_H+w_V}
\end{equation}
where $T_{H/V}$ is the main-beam brightness temperature for H and V polarizations, corrected for the beam coupling factor, and $w_{H/V}$ the weight for each polarizations determined by $1/\sigma^2$, with $\sigma$ the rms noise measured outside the line window. The averaged spectra without beam coupling corrections are shown in Fig.~\ref{fig:DataIntegration} and the line intensities are listed in Table~\ref{tab:IntensitySet}. The final uncertainties for the corrected intensities are computed as the quadratic sum of the spectral rms related to the signal-to-noise ratio, the absolute intensity calibration error~\citep{TeyssierHB2017}, the main-beam coefficient uncertainty~\citep{Mueller2012}, and the beam coupling factor uncertainty.

\section{Analysis}

To model the water vapor emission in the Orion Bar, we use the Meudon PDR code\footnote{https://ism.obspm.fr}~\citep{LePetitApJSS2006} to fit the observed line intensities. The Meudon PDR code simulates a stationary plane-parallel slab of gas and dust and computes the radiative transfer along a line-of-siglht. At each position within the cloud, thermal and chemical balances are computed, as well as non-local thermodynamic equilibrium (non-LTE) level populations. Level populations and resulting line intensities are computed considering collisional and radiative processes, as well as chemical formation and destruction in various rotational levels. For radiative processes, the code takes into account non-local pumping by the continuum (background and dust emission) and line emission, as explained in~\cite{GonzalezGarciaAA2008}.

The incoming UV radiation field is based on the interstellar radiation field (ISRF)~\citep{MathisAA1983}, scaled by a multiplicative factor $G_0$. Mathis UV radiation field is about 1.3 factor as high as Habing one~\citep{HabingBAIN1968} in the spectral range of 91.2 to \SI{111}{nm}~\citep{WuAA2018}. 

In the currently released version of the code (1.5.2), the formation of water molecules only includes gas-phase chemistry. The nuclear spin distribution of the water molecules formed follows the kinetic temperature at each point within the cloud with a non-thermal equilibrium rotational excitation. 

\subsection{Gas-phase chemistry model}

Investigation of the input parameters of the Meudon PDR code such as the inclination angle, cloud depth, UV radiation field intensity, and density or thermal pressure (considering either isochoric or isobaric models) is performed by using the Interstellar Medium Database\footnote{http://ismdb.obspm.fr} (ISMDB) to compare the observed line intensities with grids of model predictions (precomputed with the Meudon PDR code 1.5.2 and available online). Intensities given by the ISMDB (in cgs units) are compared with observations using the following conversion:
\begin{equation}
I=\frac{2k}{\lambda^3}\int T_{mb} \diff V
\end{equation}
with $k$ the Boltzmann constant (\si{erg.K^{-1}}) and $\lambda$ the wavelength (\si{cm}). Intensities in cgs units are given in Table~\ref{tab:PDRIntensities}.

\begin{table}
\centering
\caption{Best-fit input parameters of the Meudon PDR code for the corrected intensities. \label{tab:InputPDR}}
\begin{tabular}{lll}
\hline \hline
Parameters                                          & ISMDB                   & Ads/Des                   \\ 
                                                    & model\tablefootmark{a}  & model\tablefootmark{b}    \\\hline
$\theta$ (\si{\degree})                             & 60                      & 60                        \\ 
$A_{\mathrm{V}}^{tot}$                                       & 20                      & 20                        \\ 
$G_0$ (Mathis Unit)                                 & \num{5e4}               & \num{3.1e4}                 \\
$P_{th}$ (\si{K.cm^{-3}})                           & \num{1e8}               & \num{2.8e8}               \\ \hline
Cosmic rays (\si{s^{-1}} per \chem{H_2})            & \num{1e-16}             & \num{5e-16}               \\
Rv                                                  & 3.1                     & 5.62                      \\
$\mathrm{N}${\chem{_H}}/E(B-V) (\si{cm^{-2}})       & \num{5.8e21}            & \num{1.05e22}             \\
Mass grains / Mass gas                              & 0.01                    & 0.01                      \\
Grain size distribution                             & $\propto \alpha^{-3.5}$ & $\propto \alpha^{-3.5}$   \\
min grain radius (\si{cm})                          & \num{1e-7}              & \num{3e-7}                \\
max grain radius (\si{cm})                          & \num{3e-5}              & \num{3e-5}                \\
\chem{H_2O} binding energy (K)                      &                         & 5600                      \\
\chem{H_2O} photodesorption yield\tablefootmark{c}  &                         & \num{2e-3}                \\
\chem{H_2} internal energy used                     & No                      & Yes                       \\
\hline
\end{tabular}
\tablefoot{
\tablefoottext{a}{Pure gas-phase chemistry model with standard ISMDB inputs.}
\tablefoottext{b}{Adsorption and desorption optimized model with \cite{JoblinAA2018} inputs and \chem{H_2} internal energy. No additional scaling factor was applied to correct for a more edge-on geometry.}
\tablefoottext{c}{The water photodesorption yield is in units of molecules per incident photon.}
}
\end{table}

\begin{table}
\centering
\caption{Intensities of water lines toward the Orion Bar corrected for the beam coupling factor, in \si{erg.cm^{-2}.s^{-1}.sr^{-1}}, and preticted intensities from the Meudon PDR code using input parameters in Table~\ref{tab:InputPDR} for the pure gas-phase chemsitry ISMDB model and the adsorption and desorption optimized model with \chem{H_2} internal energy. \label{tab:PDRIntensities}}
\begin{tabular}{ccccc}
\hline \hline
Mol.                              & Trans.                    & Corrected                                     & ISMDB                                     & Ads/Des\\
                                  &                           & intensities\tablefootmark{a}                  &  model                                    & model \\ \hline
\multirow{7}{*}{\chem{H_2^{16}O}} & \trans{1_{10}}{1_{01}} &\num[exponent-product = \cdot,separate-uncertainty=false]{4.9(2) e-6}     &\num[exponent-product = \cdot]{4.9e-6}     &\num[exponent-product = \cdot]{3.5e-6}\\ 
                                  & \trans{1_{11}}{0_{00}} &\num[exponent-product = \cdot,separate-uncertainty=false]{1.8(1) e-5}     &\num[exponent-product = \cdot]{1.8e-5}     &\num[exponent-product = \cdot]{1.5e-5} \\
                                  & \trans{2_{12}}{1_{01}} &\num[exponent-product = \cdot,separate-uncertainty=false]{6.0(4) e-5}     &\num[exponent-product = \cdot]{3.4e-5}     &\num[exponent-product = \cdot]{3.1e-5}\\
                                  & \trans{2_{02}}{1_{11}} &\num[exponent-product = \cdot,separate-uncertainty=false]{9.3(7) e-6}     &\num[exponent-product = \cdot]{1.2e-5}     &\num[exponent-product = \cdot]{1.4e-5}\\
                                  & \trans{2_{11}}{2_{02}} &\num[exponent-product = \cdot,separate-uncertainty=false]{4.3(3) e-6}     &\num[exponent-product = \cdot]{5.9e-6}     &\num[exponent-product = \cdot]{3.2e-6}\\
                                  & \trans{2_{21}}{2_{12}} &\num[exponent-product = \cdot,separate-uncertainty=false]{1.5(2) e-5}     &\num[exponent-product = \cdot]{1.6e-5}     &\num[exponent-product = \cdot]{1.3e-5}\\ 
                                  & \trans{3_{12}}{3_{03}} &\num[exponent-product = \cdot,separate-uncertainty=false]{3.0(3) e-6}     &\num[exponent-product = \cdot]{2.9e-6}     &\num[exponent-product = \cdot]{1.9e-6}\\\hline
\multirow{2}{*}{\chem{H_2^{18}O}} & \trans{1_{10}}{1_{01}} &\num[exponent-product = \cdot,separate-uncertainty=false]{6.1(3) e-8}     &\num[exponent-product = \cdot]{1.4e-7}     &\num[exponent-product = \cdot]{1.0e-7}\\ 
                                  & \trans{1_{11}}{0_{00}} &$<$\num[exponent-product = \cdot,separate-uncertainty=false]{5.8(22) e-7}   &\num[exponent-product = \cdot]{6.6e-7}     &\num[exponent-product = \cdot]{3.6e-7}\\
\hline
\end{tabular}
\tablefoot{
\tablefoottext{a}{Observed intensity uncertainties are given in parentheses.}
}\end{table}

\begin{figure}
\centering
\includegraphics[scale=1]{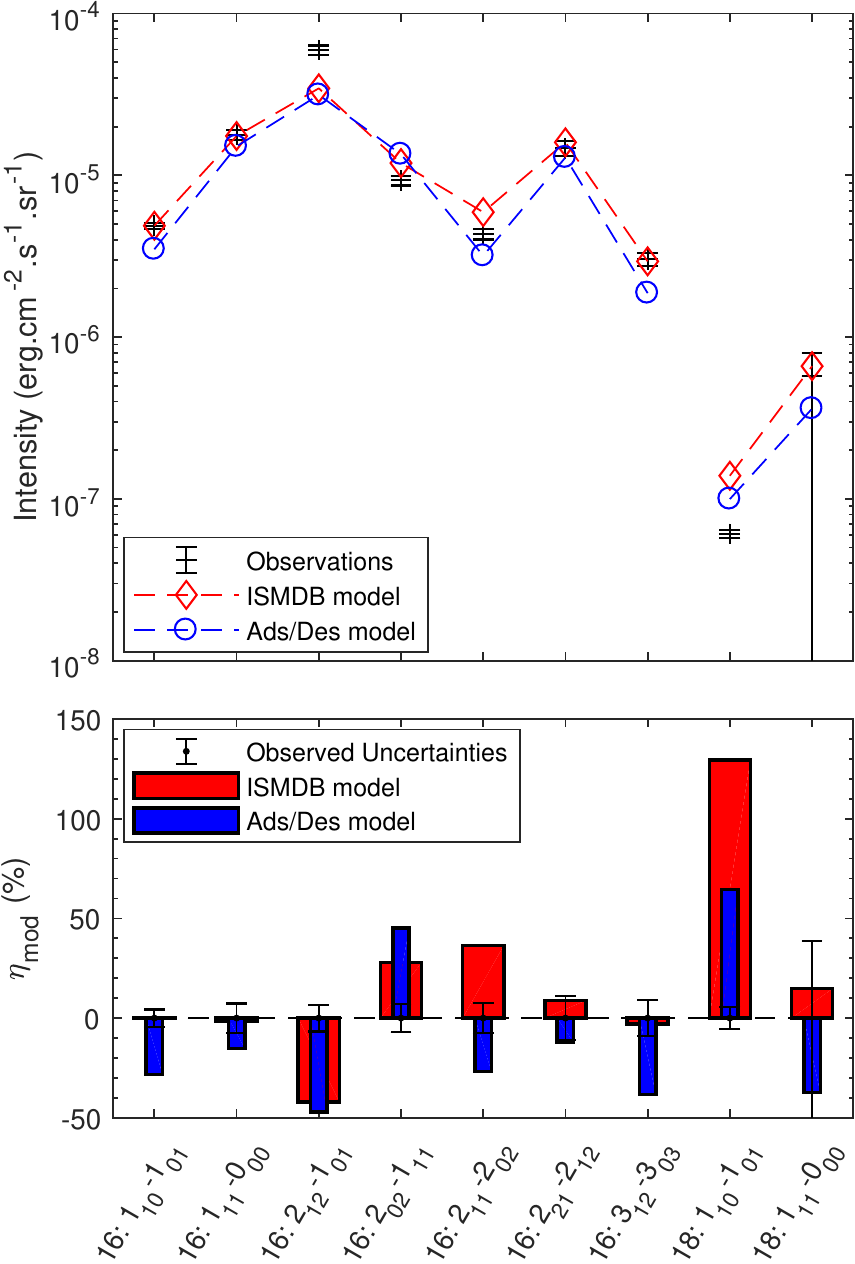}
\caption{Comparison between observed intensities and the ones predicted by the pure gas-phase chemistry ISMDB model and the adsorption and desorption optimized model with \chem{H_2} internal energy (Table~\ref{tab:InputPDR}). Bottom panel gives the relative difference between predictions and observations. \label{fig:PDRIntensities}}
\end{figure}

\begin{figure}
\centering
\includegraphics[scale=1]{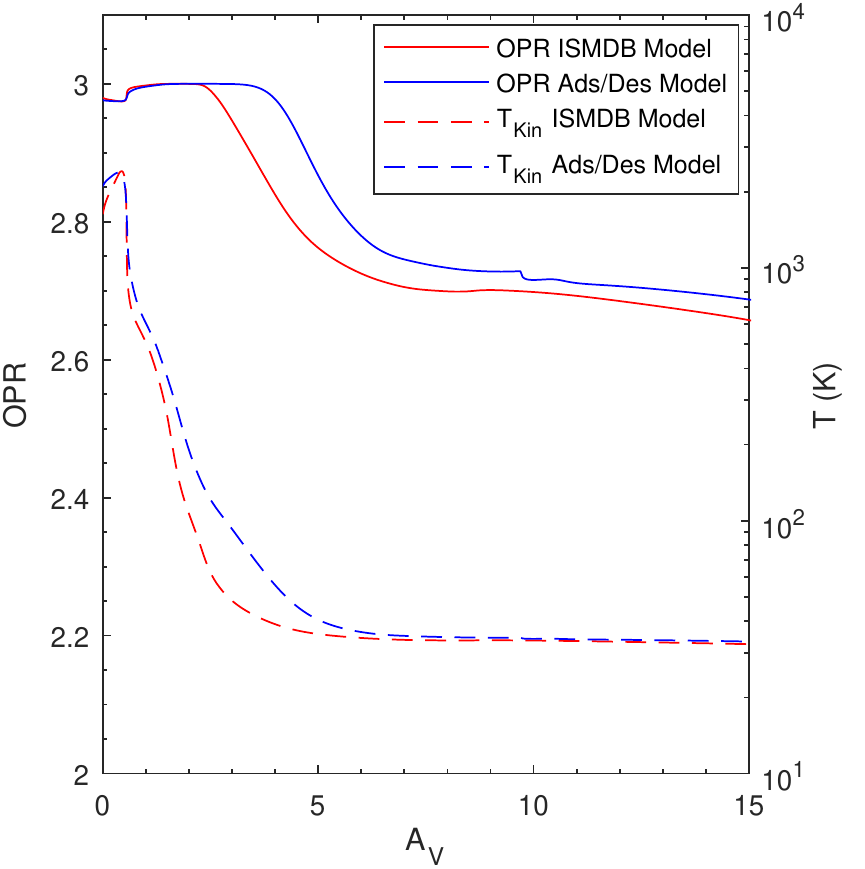}
\caption{Evolution of the the water ortho-to-para ratio (left axis) in agreement with the kinetic temperature (right axis) as a function of the visual extinction inside the molecular cloud for the pure gas-phase chemistry ISMDB model and the adsorption and desorption optimized model with \chem{H_2} internal energy (Table~\ref{tab:InputPDR}). \label{fig:OPRPDR}}
\end{figure}

As already shown using more than twenty lines emitted by nine species, the line emission at the edge of the Orion Bar is better explained by an isobaric model than by a constant density model~\citep{JoblinAA2018}. This is understandable because a steep density and temperature gradient is expected to exist at the edge of a UV-illuminated molecular cloud (see Figs.~\ref{fig:OPRPDR} and \ref{fig:DensityAds}). This agrees with other studies of the Orion Bar~\citep{MarconiAA1998,AllersApJ2005}, as well as in the northwest PDR in NGC 7023~\citep{JoblinAA2018} and Carina PDR~\citep{WuAA2018}. A theoretical explanation, based on the photo-evaporation process, was provided by~\cite{BronAA2018}. Consequently, an isobaric model is adopted.

To reproduce the nearly edge-on geometry of the Bar~\citep{WenAPJ1995,HogerheijdeAA1995,JansenAA1995,WalmsleyAA2000}, the maximum viewing angle of \SI{60}{\degree} accepted by the code is adopted. This angle is defined as the angle between the normal to the ionization front and the line-of-sight, meaning that a viewing angle of \SI{90}{\degree} corresponds to an edge-on PDR. Using an isobaric model, the best-fit to our best-estimate \chem{H_2^{16}O} intensities is obtained for a cloud depth corresponding to a visual extinction of $A_{\mathrm{V}}^{tot}=20$, a UV field of $G_0=\num{3.5e4}-\num{8e4}$ in Mathis units and a thermal pressure $P_{th}=\num{6e7}-\SI{2e8}{K.cm^{-3}}$. The \trans{2_{11}}{2_{02}} and \trans{2_{02}}{1_{11}} line intensities require a UV field \si{40}{\%} lower ($G_0=\num{3e4}$ in Mathis units) to match the observations, whereas the \chem{H_2^{16}O} \trans{2_{12}}{1_{01}} line requires a higher pressure by a factor of five. It is worth noting that the intensity derived for this line is \si{20}{\%} higher than the one reported by \cite{NagyAA2017} measured from a short integration-time observation\footnote{\chem{H_2^{16}O} \trans{2_{12}}{1_{01}}-b, Obs. Id: 1342218426}. \chem{H_2^{18}O} lines are not used in the fitting procedure because the online ISMDB does not include \chem{H_2^{18}O} predictions. Input parameters for the ISMDB model of the Meudon PDR code are listed in Table~\ref{tab:InputPDR}. 

The best-fit parameters obtained for the observed \chem{H_2^{16}O} intensities are close to those derived from high-J \chem{CO} lines~\citep{JoblinAA2018} or small hydrocarbons and complex organic molecules (COM) emission~\citep{CuadradoAA2015,CuadradoAA2017}. Small deviations from the high-J~\chem{CO} line fit could be explained by the fact that water emits $\sim$ \ang{;;5} deeper in the cloud than \chem{CO}~\citep{ParikkaAA2017}. However these deviations may also result from the data reduction method applied. Main-beam efficiencies reported by \cite{RoelfsemaAA2012} were used by \cite{JoblinAA2018} along with a mean value between the antenna temperature and the main-beam brightness temperature to deal with a spatial emission in-between point-source and extended source~\citep{OssenkopfAA2013}. This method could lead to an underestimate of the main-beam brightness temperature by a factor close to \num{1.3}. Moreover specific parameters of the code have been optimized for high-J~\chem{CO} fitting. In particular, the grain size distribution was adapted to model the extinction curve toward $\Theta^1$ Ori C~\citep{FitzpatrickApJS1990,MarconiAA1998,JoblinAA2018}, and a scaling factor was applied to correct the bias for the assumed geometry of the Bar, such as the beam coupling factor and the viewing angle, and get closer to an edge-on model. 

The stand-alone version of the Meudon PDR code (1.5.2) was then used with the best-fit parameters derived from the water line observations to compute the intensities of water transitions and in particular those of \chem{H_2^{18}O}. This is performed by extending the gas-phase \chem{H_2^{16}O} chemistry to \chem{H_2^{18}O}, introducing the possible \chem{^{18}O} fractionation reactions~\citep{LoisonMNRAS2019}. The radiative excitation processes involving \chem{H_2^{18}O} have been implemented from the HITRAN2016\footnote{https://hitran.org/} database~\citep{Hitran2016}. The \chem{H_2^{18}O} de-excitation collisional rates are assumed to be identical to those of \chem{H_2^{16}O}. The upward collisional excitation rates of \chem{H_2^{18}O} are computed by introducing the appopriate energy defect for \chem{H_2^{18}O} and considering the different energy order involving highly-J excited rotational levels. Collisional rates for rotational excitation of \chem{H_2^{16}O} by \chem{He} are taken from \cite{GreenApJS1993} for which analytical temperature dependence was derived by \cite{GonzalezGarciaAA2008}. The collisional rates due to \chem{H_2} are computed from these values by introducing the relative reduced mass factor. The computed intensities are given in Table~\ref{tab:PDRIntensities} and compared with the corrected observed line intensities in Fig.~\ref{fig:PDRIntensities}. 

The observed intensities are quite well reproduced by the Meudon PDR code with these parameters. Lines outside the observed uncertainty range are those mentioned above (\trans{2_{12}}{1_{01}}, \trans{2_{02}}{1_{11}}, and \trans{2_{11}}{2_{02}}) and the \chem{H_2^{18}O} ortho ground state line, which is clearly overestimated by the model. The value obtained for the \chem{H_2^{18}O} \trans{1_{11}}{0_{00}} para line agrees with the upper limit observed.

In the Meudon PDR code, water vapor is produced with nuclear spin populations in agreement with the local kinetic temperature. The computed kinetic temperature varies from \SI{2000}{K} close to the ionization front down to \SI{30}{K} in the deepest part of the cloud (see Fig.~\ref{fig:OPRPDR}). The ortho-to-para ratio of water as a function of the depth into the cloud is shown in Fig.~\ref{fig:OPRPDR}. To obtain the mean OPR of the water molecules along the line-of-sight, the OPR is averaged along the maximum density region of water vapor. Figure~\ref{fig:DensityAds} shows that the density of gas-phase water peaks between $A_{\mathrm{V}}=5$ and $A_{\mathrm{V}}=10$. Thus the gas-phase water-emitting region is characterized by a temperature of \SI{34.0(5)}{K}, implying an OPR of \num{2.7(1)}. Computing the OPR with the ratio of ortho and para column densities of gas-phase water in the cloud up to $A_{\mathrm{V}}=15$ leads to a value in good agreement within the uncertainty range. However this value is clearly higher than the value of \numrange{0.1}{0.5} derived by~\cite{ChoiAA2014} from \chem{H_2^{18}O} observations. 

The OPR uncertainty derived above only considers the predicted OPR variations through the cloud. To take into account the uncertainty due to discrepancies between the observations and the predicted intensities, models at the boundary UV field and thermal pressure regions giving a good fit lead to an OPR uncertainty of 0.1.

\subsection{Adsorption and desorption model}

\begin{figure}
\centering
\includegraphics[scale=1]{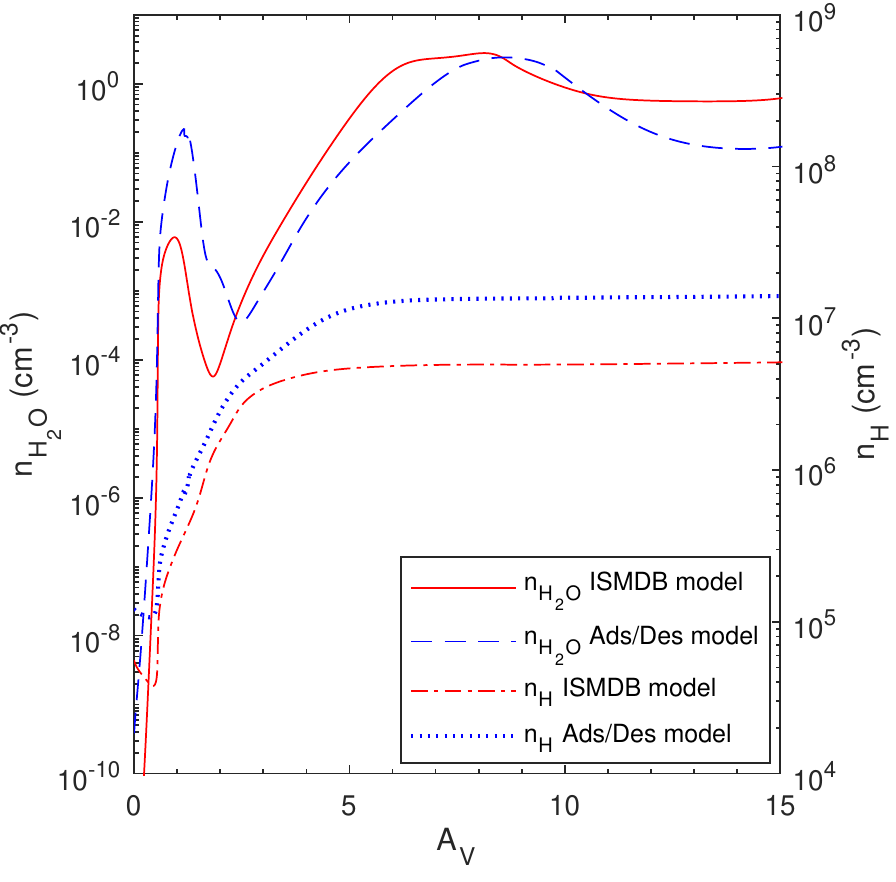}
\caption{Comparison of the spatial density of gas-phase water (left axis) and \chem{H} density (right axis) as a function of the visual extinction inside the molecular cloud for the pure gas-phase chemistry ISMDB model and the adsorption and desorption optimized model with \chem{H_2} internal energy (Table~\ref{tab:InputPDR}). \label{fig:DensityAds}}
\end{figure}

\begin{figure}
\centering
\includegraphics[scale=1]{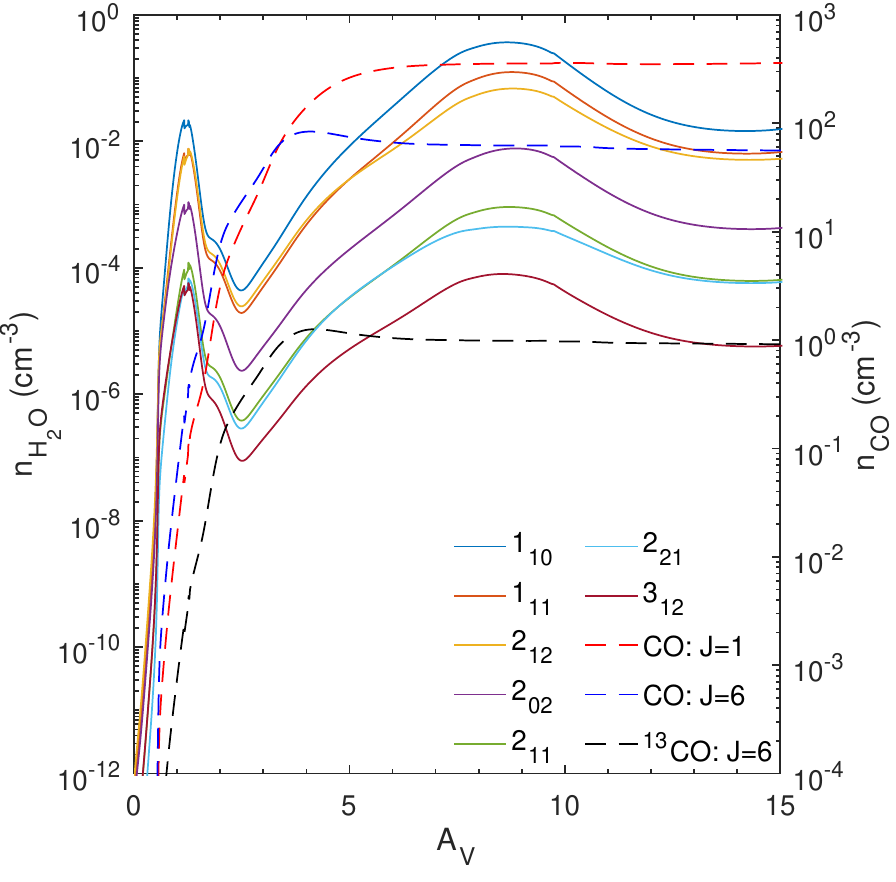}
\caption{Evolution of the gas-phase rotational populations of \chem{H_2^{16}O} (left axis), \chem{CO} and \chem{^{13}CO} (right axis) as a function of the visual extinction inside the molecular cloud for the adsorption and desorption optimized model with \chem{H_2} internal energy (Table~\ref{tab:InputPDR}). \label{fig:OutputPDR}}
\end{figure}

\begin{figure}
\centering
\includegraphics[scale=1]{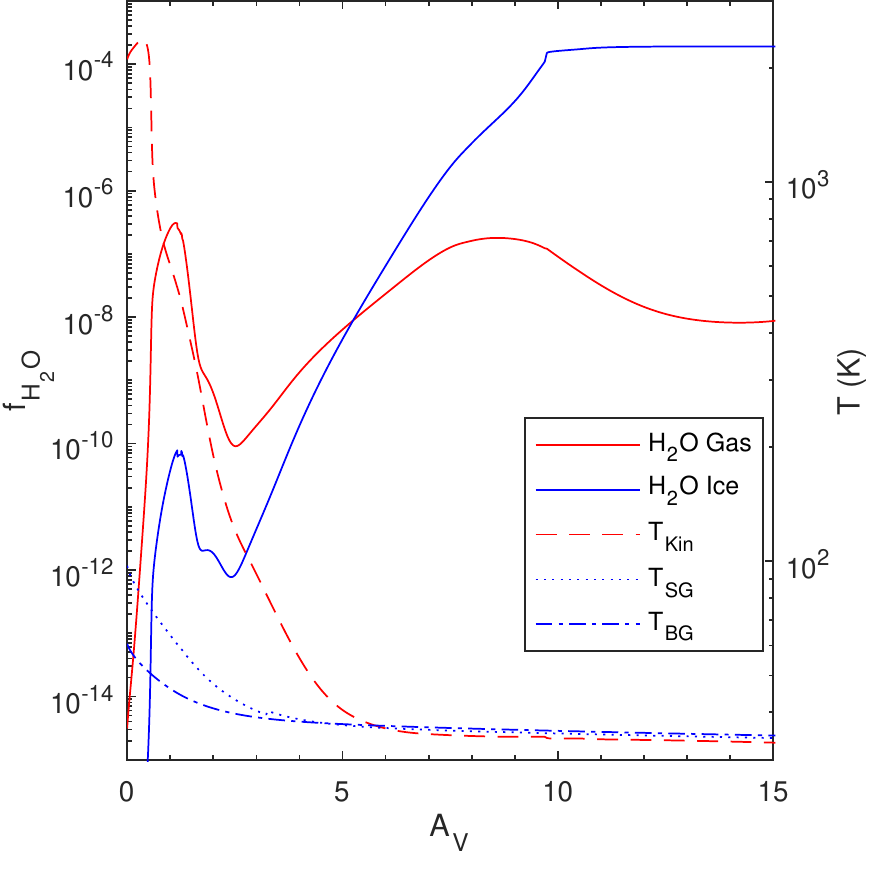}
\caption{Evolution of the fractional abundance of \chem{H_2^{16}O} gas and solid-phase (left axis) as a function of the visual extinction inside the molecular cloud for the adsorption and desorption optimized model with \chem{H_2} internal energy (Table~\ref{tab:InputPDR}). Dashed lines show the gas-phase temperature and the smallest (SG~$\sim$ \SI{3}{nm}) and biggest (BG~$\sim$ \SI{300}{nm}) grain temperature (right axis).} \label{fig:Tdust}
\end{figure}

We stress that gas-phase water population and spin distribution in the currently released version of the Meudon PDR code is only governed by gas-phase chemistry and thermal equilibrium. The presence of grains in the model only affects the thermal balance through photoelectric heating, while attenuating the UV flux~\citep{LePetitApJSS2006} and catalysing the formation of \chem{H_2} via surface chemistry~\citep{LePetitAA2009,LeBourlotAA2012}. The need of surface chemistry was less important in previous studies, for which tracers that appear in a warmer medium (Fig.~\ref{fig:OutputPDR}) were used~\citep{CuadradoAA2015,CuadradoAA2017,JoblinAA2018}. Up to $A_{\mathrm{V}}=3$, \cite{EspluguesAA2016} have shown, with a PDR model considering adsorption and desorption processes and surface chemistry, that the gas-phase abundances of several species, such as \chem{H_2O}, \chem{H_2CO}, and \chem{CH_3OH}, are independant of the dust surface chemistry. Indeed, close to the ionization front, the high kinetic temperature of the gas and grains (see Fig.~\ref{fig:Tdust}) is not suitable for long time residence of these molecules on grains~\citep{HollenbachApJ2009}. However, for water molecules, surface chemistry could have an important role to play. Indeed the oxygen surface chemistry, introduced in the \cite{HollenbachApJ2009} PDR model, appears to dominate the gas-phase water production. Figures~\ref{fig:OPRPDR} and \ref{fig:DensityAds} show that the gas-phase water maximum density region from $A_{\mathrm{V}}=5$ to $10$ corresponds to a medium with a gas temperature below~\SI{40}{K}. In addition, the water vapor abundance remains relatively high for $A_{\mathrm{V}}>10$, where a strong freeze-out of water molecules on grains is expected~\citep{HollenbachApJ2009,EspluguesAA2016}. Thus the line emission of water molecules could be strongly modified by their interactions with cold grains.

The implementation of state-of-the-art surface chemistry processes in the Meudon PDR code is in progress. For this work, we have added adsorption and desorption processes of water. The adsorption is governed by a sticking coefficient proportional to the inverse of the square root of the temperature for a kinetic temperature higher than \SI{10}{K} and equals to 1 otherwise. Two desorption processes were considered. The thermal desorption is dependent on a binding energy of \SI{5600}{K}~\citep{GarrodApJ2009,WakelamMolAstr2017} and the photodesorption is related to a yield of \SI{5e-4}{} molecules per incident UV photon~\citep{ObergApJ2009}.

Using our ISMDB best-fit parameters and adsorption and desorption processes, the overall water density was computed with the Meudon PDR code. Depletion of gas-phase water is clearly observed for $A_{\mathrm{V}} >5$ when the adsorption process is included, in good agreement with~\cite{HollenbachApJ2009}. As expected, the depletion of water molecules decreases the computed intensity of water lines with an accentuated effect for the excited lines. The depletion is sufficiently significant that very high UV flux and thermal pressure are needed to match the observational data using regular parameters in the Meudon PDR code.

Gas-phase water is mainly formed by two processes (\citealt{vanDishoeckChemRev2013} and references therein) represented by each water density maximum in Fig.~\ref{fig:DensityAds}. The first peak around $A_{\mathrm{V}}=1$ is the result of the balance between water formation via \chem{OH} and photodissociation.
\begin{equation}
\nonumber
\begin{array}{l}
\mathrm{OH} + \mathrm{H_2} \rightarrow \mathrm{H_2O} + \mathrm{H}\\
\mathrm{H_2O} + \mathrm{photon} \rightarrow \mathrm{H} + \mathrm{OH}
\end{array}
\end{equation}
Then the gas-phase \chem{OH} abundance decreases and the photodissociation reduces the gas-phase water density. From $A_{\mathrm{V}}=3$ to $A_{\mathrm{V}}=5$, the increase in water vapor abundance is caused by the recombination of \chem{H_3O^{+}} with electrons, still softened by the photodissociation.
\begin{equation}
\nonumber
\mathrm{H_3O^{+}} + \mathrm{e^{-}} \rightarrow \mathrm{H_2O} + \mathrm{H}
\end{equation}
Adsorption and desorption processes compete with \chem{H_3O^{+}} recombination leading to an increase in the water ice abundance up to $A_{\mathrm{V}}=10$ as represented in Fig.~\ref{fig:Tdust}. Eventually, the electronic recombination is weakened and the gas-phase water abundance is reduced.

Without water surface chemistry, grains act as water trap which explains the discrepancy between observed and computed intensities. To reproduce the observed intensities, water should be formed or released from the grains more efficiently. 

Close to the \chem{H}/\chem{H_2} transition, an activation energy of \SI{3240}{K} has to be overcome to initiate the following reaction and increase the water precursor reservoir (\citealt{vanDishoeckChemRev2013} and references therein).
\begin{equation}
\nonumber
\mathrm{O} + \mathrm{H_2} \rightarrow \mathrm{OH} + \mathrm{H}
\end{equation}
In the standard 1.5.2 version of the Meudon PDR code, energy for chemical reactions is only provided by the kinetic energy of the reactants due to thermal motion. However, close to the \chem{H}/\chem{H_2} transition, \chem{H_2} is highly ro-vibrationally excited~\citep{HollenbachARAA1997}. This energy can be used to overcome the activation energy~\citep{AgundezApJ2010}, as was already studied for the \chem{C^{+}+H_2} reaction~\citep{ZanchetApJ2013,HerraezAguilarPCCP2014} and the \chem{S^{+}+H_2} reaction~\citep{ZanchetAJ2013}. To implement this effect in the model, the activation energy is thus taken as the difference between the activation energy and the ro-vibrational energy of \chem{H_2}. Reactions with vibrationally excited \chem{H_2} were already included in the Meudon PDR code to reproduce the \chem{OH} emission in the Orion Bar~\citep{GoicoecheaAA2011}. Considering this effect enhances the intensity predicted for the excited lines up to a factor of 2.5 due to the increase in the excited populations between $A_{\mathrm{V}}=6$ and $A_{\mathrm{V}}=8$.

Deeper into the cloud, where the gas is colder, gas-phase water is formed by the recombination of \chem{H_3O^{+}} with electrons. This reaction is initiated by cosmic ray ionization producing \chem{H_3^{+}} and leading to \chem{H_3O^{+}} through ion-neutral reactions~\citep{GerinAA2010}. Moreover, in the cold region, where water is depleted by adsorption, desorption is mainly achieved by photodesorption triggered by the UV secondary photons induced by cosmic rays~\citep{PrasadApJ1983}. Following these two processes, the cosmic ray ionization rate should have a main impact on line intensities coming from the deepest part of the cloud. From our \chem{H} column density $(N_\mathrm{H})$ estimate in the water region, a range of cosmic ray ionization rate can be estimated. For $A_{\mathrm{V}}>5$, $N_\mathrm{H}$ is higher than \SI{e22}{cm^{-2}} and the cosmic ray ionization rate could be increased by a factor of ten in relation to the value used by~\cite{JoblinAA2018}, up to $\zeta_{\mathrm{H_2}}=\SI{5e-16}{s^{-1}}$ per \chem{H_2}~\citep{PadovaniAA2018}. This leads to an increase in the line intensities by a factor of the order of three, except for \chem{H_2^{16}O} \trans{2_{11}}{2_{02}} for which a factor of seven is obtained. Such an enhancement is mainly produced by the increase in the \chem{H_3O^{+}} reservoir.

Finally, the water photodesorption yield used in the PDR Code could be tuned up. Recent study of \cite{CruzDiazMNRAS2018} reported photodesorption yield up to \SI{2e-3} molecule per photon at a temperature around \SI{60}{K}, using a microwave discharged hydrogen flow lamp with a strong Ly-$\alpha$ emission component. Assuming this value increases the water line intensities by a factor of the order of two.

Considering the \chem{H_2} internal energy, taking upper limit for the cosmic ray ionization rate and the water photodesorption yield, and using input parameters derived by \cite{JoblinAA2018} otherwise, the resulting intensities for adsorption and desorption model are in good agreement with the observed intensities (see Fig.~\ref{fig:PDRIntensities} and Table~\ref{tab:PDRIntensities}). Input parameters of the adsorption and desorption optimized model are listed in Table~\ref{tab:InputPDR}. We note that the scaling factor used by~\cite{JoblinAA2018} to correct bias for the assumed geometry of the Bar was not applied in this study. 

With this model, the depletion of water by adsorption on grains is compensated in the maximum gas-phase water density region as seen in Fig.~\ref{fig:DensityAds}. Figure~\ref{fig:Tdust} shows that the water ice abundance reaches a plateau at a fractional abundance ($f_{\mathrm{H_2O}}=n_\mathrm{H_2O}/n_\mathrm{H}$) of \num{2e-4} for $A_{\mathrm{V}}>10$ leading to a decrease in the gas-phase water fractional abundance which peaks at \num{2e-7} between $A_{\mathrm{V}}=7$ and $A_{\mathrm{V}}=10$. This gas-grains balance in relation to the depth agrees with what was predicted by \cite{HollenbachApJ2009} with oxygen surface chemistry. However, the fractional abundance derived from our best model is one order of magnitude higher than those reported in this former work for a UV field of $G_0=\num{1e3}$ and a density of \SI{1e4}{cm^{-3}}. Water ice abundance increases from $A_{\mathrm{V}}=3$ where the gas temperature is below \SI{100}{K} and the grain temperature below \SI{40}{K}. The grain temperature evolution with depth is consistent with dust observations~\citep{ArabAA2012}.

Figure~\ref{fig:OutputPDR} shows the populations of gas-phase water rotational levels, which peaks between $A_{\mathrm{V}}=5$ to $A_{\mathrm{V}}=10$ without stratification. The averaged ortho-to-para ratio in this region is equal to \num{2.8(2)} corresponding to a spin temperature of \SI{36(2)}{K} (see Fig.~\ref{fig:OPRPDR}) for a UV field of $G_0=\num{3.1e4}$ in Mathis units and a thermal pressure $P_{th}=\SI{2.8e8}{K.cm^{-3}}$. The spin temperature derived from this model agrees with the gas-phase kinetic temperature expected by \cite{HollenbachApJ2009} for $G_0=\num{1e4}$ and $n_\mathrm{H}=\SI{1e5}{cm^{-3}}$.

\section{Discussion}

\subsection{Self-absorption}

\begin{figure}
\includegraphics[scale=1]{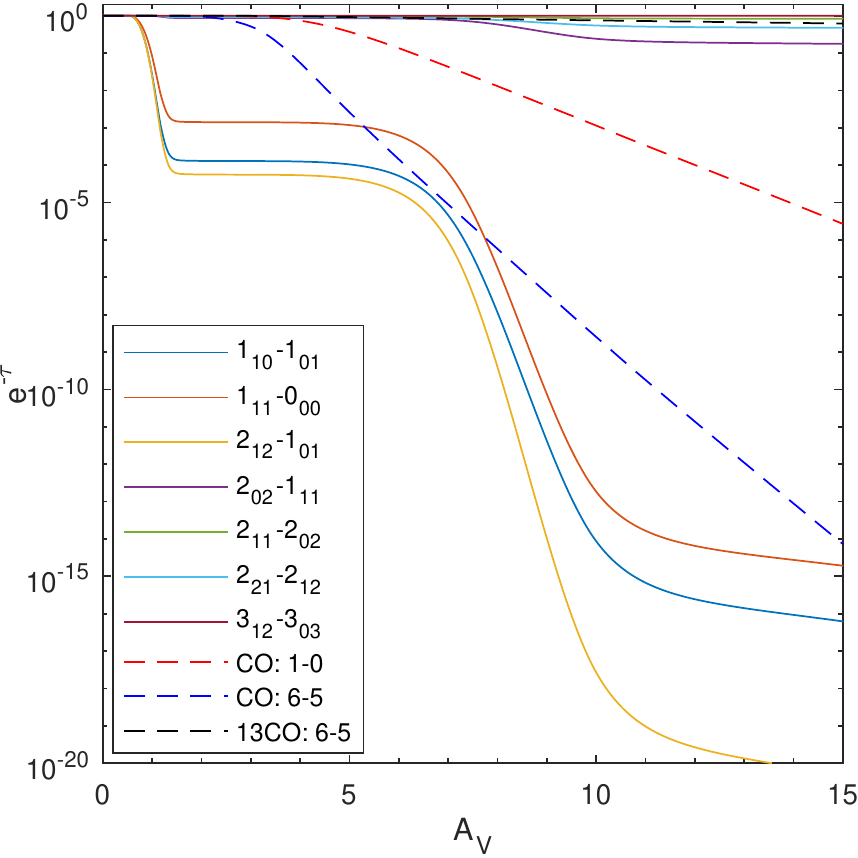}
\caption{\label{fig:OpacityAdsDes} Evolution of the line center opacity of \chem{H_2^{16}O} and \chem{CO} lines as a function of the visual extinction inside the molecular cloud for the adsorption and desorption optimized model with \chem{H_2} internal energy (Table~\ref{tab:InputPDR}).}
\end{figure}

The analysis of water line intensities with the Meudon PDR code brings a justification for the assumption made to derive the observed intensities. The presence of a central dip in line profiles is explained by the self-absorption by foreground gas~\citep{Choi2015}. In this scenario, the flux coming from layers deep inside the cloud is absorbed by the foreground layers when escaping the Bar.

The Meudon PDR code prediction of the line center opacities, defined as $e^{-\tau}$ with $\tau$ the line center optical depth, are given in Fig.~\ref{fig:OpacityAdsDes}. This figure shows that the \trans{1_{10}}{1_{01}}, \trans{1_{11}}{0_{00}}, and \trans{2_{12}}{1_{01}} transitions, which exhibit the deepest self-absorption, are by far optically thicker than the other lines. We have verified that the excitation temperature of the ground state water lines increases rapidly with $A_{\mathrm{V}}$ near the location of the $\tau = 1$ surface ($A_{\mathrm{V}} \sim 1$). The opacity is the highest at the line center and decreases going into the wings. An absorption dip at the central velocity is therefore expected. Thus line profiles predicted by the Meudon PDR code readily reproduce such a dip for the three most intense and broadest lines, while predicting single peak profiles for the other lines.

\subsection{\chem{H_2^{18}O} intensities}

This work has led to the first prediction of \chem{H_2^{18}O} line intensities by the Meudon PDR code. The comparison with observed intensities in Fig.~\ref{fig:PDRIntensities} indicates that the computed intensities are overestimated. Furthermore, our estimated gas-phase column density for \chem{H_2^{18}O} ortho ground state line is two times higher than the one reported by \cite{ChoiAA2014}.

It appears that the gas-phase water \chem{^{16}O}/\chem{^{18}O} isotopic ratio is lower than the typical value of 560 reported by \cite{WilsonARAA1994} for the local ISM and based on \chem{H_2CO} surveys~\citep{GardnerMNRAS1981}. Moreover, the decreasing of the ratio is accentuated in the water emission region down to \num{350}. This could be explained by the fact that only half of the \num{52} oxygen bearing molecules considered by the Meudon PDR code have their oxygen isomers taken into account. Thus a larger \chem{^{18}O} reactants reservoir, in relation to \chem{^{16}O} chemistry, is available to form \chem{H_2^{18}O}. Considering the \chem{^{16}O}/\chem{^{18}O} isotopic ratio of 560, and assuming that the intensity is impacted by the difference between the isotopic ratios in the main water emission region, it appears that the computed \chem{H_2^{18}O} line intensities are overestimated by a factor of \num{1.6}. Applying this correction, the prediction of the \chem{H_2^{18}O} ortho \trans{1_{10}}{1_{01}} intensity goes down to \SI{6.3e-8}{erg.cm^{-2}.s^{-1}.sr^{-1}} in good agreement with the observed intensity. The para \trans{1_{11}}{1_{10}} intensity predicted by the model would thus be around \SI{2.3e-7}{erg.cm^{-2}.s^{-1}.sr^{-1}}. Removing this spurious effect requires experimental data for the rare isotope molecules.

\subsection{OPR}

Water line intensities are well reproduced using the Meudon PDR code, considering either pure gas-phase chemistry model, or by adding adsorption and desorption of water from grains. In both cases, an ortho-to-para ratio close to \num{2.8} is derived, whereas a previous estimate of the \chem{H_2^{18}O} OPR in the Orion Bar leads to a very low value between \num{0.1} and \num{0.5}~\citep{ChoiAA2014}. This low ratio was estimated from the column densities obtained with the RADEX low velocity gradient (LVG) code~\citep{vanderTakAA2007}, assuming a single-slab geometry with a homogeneous density and temperature. As already discussed, the gas-phase \chem{H_2^{18}O} column densities, derived from the Meudon PDR code using our adsorption and desorption optimized model, are probably overestimated due to the departure of the isotopic ratio from the typical local ISM value. Considering the ISM value of \cite{WilsonARAA1994}, the gas-phase \chem{H_2^{18}O} column densities up to $A_{\mathrm{V}}=15$ are $\sim$~\SI{8.2e9}{cm^{-2}} for the para $1_{11}$ level and $\sim$~\SI{2.5e10}{cm^{-2}} for the ortho $1_{10}$ level. The para value is one order of magnitude lower than the one previously reported whereas the ortho value agrees with it~\citep{ChoiAA2014}. Using these two values, the \chem{H_2^{18}O} OPR is estimated to be equal to \num{3.0}. 

Part of the discrepancy between the low OPR reported and our high thermal value derived could be assigned to the improvement of the data reduction pipeline. Indeed, with the latest HIPE version (14.1.0), the \chem{H_2^{18}O} para ground state line, which has a really low signal-to-noise ratio, is barely detected and our upper limit for its intensity is roughly two times lower than the \cite{ChoiAA2014} value\footnote{HIPE version 10.0 was used}. It is worth noting that \cite{NagyAA2017} have not reported the detection of this line, unlike the \chem{H_2^{18}O} ortho ground state transition.

However a rough estimate of the OPR in LTE, using the \chem{H_2^{18}O} ortho line intensity and the para upper limit still gives a low OPR of \num{0.7}. Moreover, considering the two \chem{H_2^{18}O} intensities predicted by the Meudon PDR code, the OPR in LTE would be equal to 1.3. Thus the model used for the analysis has a major impact on the results. RADEX LVG code~\citep{vanderTakAA2007}, used to derive the \chem{H_2^{18}O} OPR, assumes an isothermal homogeneous medium and solves the radiative transfer in non-LTE using radiative and collisional transition rates. Conversely, the Meudon PDR code derives the population considering chemistry, thermal balance and infrared pumping. With this more complete model and assuming an isobaric model, the kinetic temperature, and thus the density, is predicted to be strongly dependent on the depth into the cloud as shown in Figs.~\ref{fig:OPRPDR} and \ref{fig:DensityAds}. Alongside with an expanded dataset, the analysis of water lines intensity in the Orion Bar is consistent with the thermal equilibrium at \SI{36}{K}.

To derive the water OPR, we computed an average over the main gas-phase water reservoir between $A_{\mathrm{V}}=5$ and $A_{\mathrm{V}}=10$. This assumption is justified by the spatial distribution of the water line emission in the Orion Bar. Indeed \chem{OH} and \chem{H_2O} emission seems to be decorrelated whereas \chem{OH} emission matches high-J \chem{CO} ones~\citep{GoicoecheaAA2011,ParikkaAA2017}. As high-J \chem{CO} lines arise from a layer between $A_{\mathrm{V}}=1$ and $A_{\mathrm{V}}=3$~\citep{JoblinAA2018}, the local water density maximum at $A_{\mathrm{V}}=1-2$ should not significantly contribute to the observed water intensities.

This OPR agrees with OPRs at high thermal values from other tracers reported for the Orion Bar. Indeed an OPR of \num{2.8(6)} was derived for \chem{c-C_3H_2}~\citep{CuadradoAA2015} and OPRs of the order of three were inferred for \chem{H_2CO}, \chem{H_2CS} and \chem{H_2CCO}~\citep{HogerheijdeAA1995,CuadradoAA2017}.

This work follows the trend of water OPR reanalysis initiated by better reduction pipeline for archival observations along with the development of more sophisticated models. Thus fewer very low OPRs are confirmed, as for example in the TW Hydrae protoplanetary disk, where the very low OPR reported by \cite{HogerheijdeScience2011} has been shown to be model-dependent and is consistent within the uncertainties with the high-temperature limit~\citep{SalinasAA2016}.

\subsection{Grain surface chemistry}

The introduction of simple adsorption and desorption processes of water has emphasized the major role that grains could have for tracers of the deepest part of the cloud. To overcome the depletion produced by the adsorption of water molecules on grains, several parameters of the code had to be tuned up to their upper acceptable limits. However a careful study of other parameters could mitigate the extreme values adopted for the cosmic ray ionization rate and the water photodesorption yield. 

First, considering the presence of X-ray photons in the Orion Nebula~\citep{GetmanApJS2005,PreibischApJS2005} could lead to a reduction of these parameters. The chemistry initiated by cosmic ray ionization, such as gas-phase water production via \chem{H_3O^{+}}, could  also be activated by X-ray ionization~\citep{GuptaAA2010,CuadradoAA2015}. Furthermore, water X-ray photodesorption appears to be efficient and comparable to far-UV photodesorption~\citep{DupuyNatAstr2018}.

On the other hand, the \ang{60;;} limitation of the Meudon PDR code could lower the predicted intensity in relation to a more edge-on geometry. Indeed, previous studies have estimated the ionization front to be tilted from the line-of-sight by an angle between \ang{3;;} and \ang{20;;}~\citep{WenAPJ1995,HogerheijdeAA1995,JansenAA1995,WalmsleyAA2000} implying a nearly edge-on PDR. \cite{JoblinAA2018} have estimated that a scaling factor of \num{1.3} has to be applied to correct for the assumed geometries of the Bar from a model at a viewing angle of \ang{60;;} and to fit high-J \chem{CO} lines. A more edge-on PDR would be characterized by a scaling factor higher than 1. 

Moreover, the grain size distribution adopted by \cite{JoblinAA2018} is optimized to reproduce the extinction curve toward $\Theta^1$ Ori C~\citep{FitzpatrickApJS1990,MarconiAA1998}. This distribution is kept constant through the cloud, whereas deeper in grains could be larger. Thus in the water-emitting region, the total surface of the grains for water molecules to stick onto would be reduced preventing gas-phase water density from dropping down dramatically.

Furthermore, adding only water adsorption and desorption processes on grains could introduce a bias on the chemistry balance and the radiative transfer computation. Indeed, if a full adsorption and desorption description is adopted, modifying the precursors reservoir by trapping them onto grains could change the chemical state of the cloud. This would also have an impact on the radiative transfer by limiting the UV shielding of the molecules that appear deeper into the cloud, and on the desorption processes that should be enhanced by the increase in the number of UV photons inside the cloud. Moreover, increasing the number of molecules on grains will emphasize the need of considering a full surface chemistry balance. Some PDR models have introduced surface chemistry calculations~\citep{HollenbachApJ2009,EspluguesAA2016} and raised the importance of surface chemistry to explain observed gas-phase abundances. In these models, the main ice water formation pathway begins with oxygen adsorption on grains followed by two reactions with hydrogen atoms. \cite{HollenbachApJ2009} have estimated that in the highest gas-phase water abundance region, \SI{97}{\%} of gas or solid-phase water production is achieved by surface chemistry. However, owing to the high FUV field, the grain temperature is high enough to prevent \chem{O} or \chem{OH} to stick efficiently~\citep{HollenbachApJ2009,MelnickApJ2012}, as seen in Fig.~\ref{fig:Tdust}. Thus the formation of water ice through this pathway should be reduced.

Implementing these processes in the Meudon PDR code is in progress. As already mentioned for isotopic computations, it is only achievable by gathering qualitative and quantitative processes from experimental data. For example adsorption and desorption considerations are mainly limited by the lack of photodesorption yield for astrophysical relevant molecules. In the past few years, several laboratory experiments have been conducted to describe photodesorption processes~\citep{ObergApJ2009,MunozCaroAA2016,DupuyAA2017}. Simulations of water surface chemistry have shown the importance of surface processes for the interstellar gas-phase~\citep{CazauxAA2010}. The development of experimental set-up to address surface chemistry (\citealt{HamaChemRev2013} and references therein) unveiled the various processes that could occur on interstellar grains and affect the physical and chemical state of interstellar objects.

Getting experimental data to deal with nuclear spin effects is crucial to simulate and understand nuclear spin population distributions. The nuclear spin conversion mechanism has been investigated to estimate the lifetime conservation of a potential thermal disequilibrium~\citep{FillionEAS2012,CaccianiPRA2012,TurgeonJPCA2017}, as well as the effect of adsorption and desorption on the OPR~\citep{HamaScience2016}. Furthermore selective reaction or formation of water nuclear spin isomers could also be taken into account by the models~\citep{KilajNatureCom2018}. Considering these processes are the steps to examine the assumptions of water produced at the thermal equilibrium in the Meudon PDR code, used to fit the water observations of the Orion Bar.

\section{Conclusions}

An analysis of a set of seven \chem{H_2^{16}O} lines and one \chem{H_2^{18}O} line measured toward the Orion Bar with \textit{Herschel}/HIFI is performed using the Meudon PDR code. The para \chem{H_2^{18}O} ground state line is not unambiguously detected. Considering an isobaric model and assuming that gas-phase water is formed with an ortho-to-para ratio in agreement with the local thermal equilibrium, an OPR of \num{2.7(1)} is derived.

Investigation of grain effects is made by adding adsorption and desorption processes of water on grains. After tuning the cosmic ray ionization rate and water photodesorption yield up to their upper acceptable limits, a good agreement with the observed intensity is obtained for an OPR of \num{2.8(2)}. This ortho-to-para ratio corresponds to a temperature of \SI{36(2)}{K}, which is much higher than the spin temperature previously reported for \chem{H_2^{18}O} in the Orion Bar and close to the high-temperature limit.

\begin{acknowledgements}
This work was supported by the Programme National de Physique et
Chimie du Milieu Interstellaire (PCMI) of the CNRS/INSU , the INC/INP co-funded by the CEA and the CNES. Financial support from the LabEx MiChem, par of the French state funds managed by the ANR within the investissements d'avenir program under reference ANR-11-10EX-0004-02 is acknowledged. Fundings from the Ile-de-France region DIM ACAV + is acknowledged. Part of this research was carried out at the Jet Propulsion Laboratory, California Institute of Technology, under a contract with the National Aeronautics and Space Administration. 
\end{acknowledgements}

\bibliography{PutaudAAWaterOrionBar}

\begin{thebibliography}{94}
\expandafter\ifx\csname natexlab\endcsname\relax\def\natexlab#1{#1}\fi

\bibitem[{{Ag{\'u}ndez} {et~al.}(2010){Ag{\'u}ndez}, {Goicoechea},
  {Cernicharo}, {Faure}, \& {Roueff}}]{AgundezApJ2010}
{Ag{\'u}ndez}, M., {Goicoechea}, J.~R., {Cernicharo}, J., {Faure}, A., \&
  {Roueff}, E. 2010, \apj, 713, 662

\bibitem[{{Allers} {et~al.}(2005){Allers}, {Jaffe}, {Lacy}, {Draine}, \&
  {Richter}}]{AllersApJ2005}
{Allers}, K.~N., {Jaffe}, D.~T., {Lacy}, J.~H., {Draine}, B.~T., \& {Richter},
  M.~J. 2005, \apj, 630, 368

\bibitem[{{Andree-Labsch} {et~al.}(2017){Andree-Labsch}, {Ossenkopf-Okada}, \&
  {R{\"o}llig}}]{AndreeLabschAA2017}
{Andree-Labsch}, S., {Ossenkopf-Okada}, V., \& {R{\"o}llig}, M. 2017, \aap,
  598, A2

\bibitem[{{Arab} {et~al.}(2012){Arab}, {Abergel}, {Habart}, {Bernard-Salas},
  {Ayasso}, {Dassas}, {Martin}, \& {White}}]{ArabAA2012}
{Arab}, H., {Abergel}, A., {Habart}, E., {et~al.} 2012, \aap, 541, A19

\bibitem[{{Bergin} {et~al.}(2010){Bergin}, {Phillips}, {Comito}, {Crockett},
  {Lis}, {Schilke}, {Wang}, {Bell}, {Blake}, {Bumble}, {Caux}, {Cabrit},
  {Ceccarelli}, {Cernicharo}, {Daniel}, {de Graauw}, {Dubernet},
  {Emprechtinger}, {Encrenaz}, {Falgarone}, {Gerin}, {Giesen}, {Goicoechea},
  {Goldsmith}, {Gupta}, {Hartogh}, {Helmich}, {Herbst}, {Joblin}, {Johnstone},
  {Kawamura}, {Langer}, {Latter}, {Lord}, {Maret}, {Martin}, {Melnick},
  {Menten}, {Morris}, {M{\"u}ller}, {Murphy}, {Neufeld}, {Ossenkopf}, {Pagani},
  {Pearson}, {P{\'e}rault}, {Plume}, {Roelfsema}, {Qin}, {Salez}, {Schlemmer},
  {Stutzki}, {Tielens}, {Trappe}, {van der Tak}, {Vastel}, {Yorke}, {Yu}, \&
  {Zmuidzinas}}]{BerginAA2010}
{Bergin}, E.~A., {Phillips}, T.~G., {Comito}, C., {et~al.} 2010, \aap, 521, L20

\bibitem[{{Bron} {et~al.}(2018){Bron}, {Ag{\'u}ndez}, {Goicoechea}, \&
  {Cernicharo}}]{BronAA2018}
{Bron}, E., {Ag{\'u}ndez}, M., {Goicoechea}, J.~R., \& {Cernicharo}, J. 2018,
  arXiv e-prints, arXiv:1801.01547

\bibitem[{{Cacciani} {et~al.}(2012){Cacciani}, {Cosl\'eou}, \&
  {Khelkhal}}]{CaccianiPRA2012}
{Cacciani}, P., {Cosl\'eou}, J., \& {Khelkhal}, M. 2012, \pra, 85, 012521

\bibitem[{{Cazaux} {et~al.}(2010){Cazaux}, {Cobut}, {Marseille}, {Spaans}, \&
  {Caselli}}]{CazauxAA2010}
{Cazaux}, S., {Cobut}, V., {Marseille}, M., {Spaans}, M., \& {Caselli}, P.
  2010, \aap, 522, A74

\bibitem[{{Choi}(2015)}]{Choi2015}
{Choi}, Y. 2015, PhD thesis, Groningen

\bibitem[{{Choi} {et~al.}(2014){Choi}, {van der Tak}, {Bergin}, \&
  {Plume}}]{ChoiAA2014}
{Choi}, Y., {van der Tak}, F.~F.~S., {Bergin}, E.~A., \& {Plume}, R. 2014,
  \aap, 572, L10

\bibitem[{{Cruz-Diaz} {et~al.}(2018){Cruz-Diaz}, {Mart{\'\i}n-Dom{\'e}nech},
  {Moreno}, {Mu{\~n}oz Caro}, \& {Chen}}]{CruzDiazMNRAS2018}
{Cruz-Diaz}, G.~A., {Mart{\'\i}n-Dom{\'e}nech}, R., {Moreno}, E., {Mu{\~n}oz
  Caro}, G.~M., \& {Chen}, Y.-J. 2018, \mnras, 474, 3080

\bibitem[{{Cuadrado} {et~al.}(2017){Cuadrado}, {Goicoechea}, {Cernicharo},
  {Fuente}, {Pety}, \& {Tercero}}]{CuadradoAA2017}
{Cuadrado}, S., {Goicoechea}, J.~R., {Cernicharo}, J., {et~al.} 2017, \aap,
  603, A124

\bibitem[{{Cuadrado} {et~al.}(2015){Cuadrado}, {Goicoechea}, {Pilleri},
  {Cernicharo}, {Fuente}, \& {Joblin}}]{CuadradoAA2015}
{Cuadrado}, S., {Goicoechea}, J.~R., {Pilleri}, P., {et~al.} 2015, \aap, 575,
  A82

\bibitem[{{de Graauw} {et~al.}(2010){de Graauw}, {Helmich}, {Phillips},
  {Stutzki}, {Caux}, {Whyborn}, {Dieleman}, {Roelfsema}, {Aarts}, {Assendorp},
  {Bachiller}, {Baechtold}, {Barcia}, {Beintema}, {Belitsky}, {Benz}, {Bieber},
  {Boogert}, {Borys}, {Bumble}, {Ca{\"\i}s}, {Caris}, {Cerulli-Irelli},
  {Chattopadhyay}, {Cherednichenko}, {Ciechanowicz}, {Coeur-Joly}, {Comito},
  {Cros}, {de Jonge}, {de Lange}, {Delforges}, {Delorme}, {den Boggende},
  {Desbat}, {Diez-Gonz{\'a}lez}, {di Giorgio}, {Dubbeldam}, {Edwards},
  {Eggens}, {Erickson}, {Evers}, {Fich}, {Finn}, {Franke}, {Gaier}, {Gal},
  {Gao}, {Gallego}, {Gauffre}, {Gill}, {Glenz}, {Golstein}, {Goulooze},
  {Gunsing}, {G{\"u}sten}, {Hartogh}, {Hatch}, {Higgins}, {Honingh}, {Huisman},
  {Jackson}, {Jacobs}, {Jacobs}, {Jarchow}, {Javadi}, {Jellema}, {Justen},
  {Karpov}, {Kasemann}, {Kawamura}, {Keizer}, {Kester}, {Klapwijk}, {Klein},
  {Kollberg}, {Kooi}, {Kooiman}, {Kopf}, {Krause}, {Krieg}, {Kramer},
  {Kruizenga}, {Kuhn}, {Laauwen}, {Lai}, {Larsson}, {Leduc}, {Leinz}, {Lin},
  {Liseau}, {Liu}, {Loose}, {L{\'o}pez-Fernandez}, {Lord}, {Luinge}, {Marston},
  {Mart{\'\i}n-Pintado}, {Maestrini}, {Maiwald}, {McCoey}, {Mehdi}, {Megej},
  {Melchior}, {Meinsma}, {Merkel}, {Michalska}, {Monstein}, {Moratschke},
  {Morris}, {Muller}, {Murphy}, {Naber}, {Natale}, {Nowosielski}, {Nuzzolo},
  {Olberg}, {Olbrich}, {Orfei}, {Orleanski}, {Ossenkopf}, {Peacock}, {Pearson},
  {Peron}, {Phillip-May}, {Piazzo}, {Planesas}, {Rataj}, {Ravera}, {Risacher},
  {Salez}, {Samoska}, {Saraceno}, {Schieder}, {Schlecht}, {Schl{\"o}der},
  {Schm{\"u}lling}, {Schultz}, {Schuster}, {Siebertz}, {Smit}, {Szczerba},
  {Shipman}, {Steinmetz}, {Stern}, {Stokroos}, {Teipen}, {Teyssier}, {Tils},
  {Trappe}, {van Baaren}, {van Leeuwen}, {van de Stadt}, {Visser}, {Wildeman},
  {Wafelbakker}, {Ward}, {Wesselius}, {Wild}, {Wulff}, {Wunsch}, {Tielens},
  {Zaal}, {Zirath}, {Zmuidzinas}, \& {Zwart}}]{deGraauwAA2010}
{de Graauw}, T., {Helmich}, F.~P., {Phillips}, T.~G., {et~al.} 2010, \aap, 518,
  L6

\bibitem[{{Dupuy} {et~al.}(2018){Dupuy}, {Bertin}, {F{\'e}raud}, {Hassenfratz},
  {Michaut}, {Putaud}, {Philippe}, {Jeseck}, {Angelucci}, {Cimino}, {Baglin},
  {Romanzin}, \& {Fillion}}]{DupuyNatAstr2018}
{Dupuy}, R., {Bertin}, M., {F{\'e}raud}, G., {et~al.} 2018, Nat.~Astron., 2,
  796

\bibitem[{{Dupuy} {et~al.}(2017){Dupuy}, {Bertin}, {F\'eraud}, {Michaut},
  {Jeseck}, {Doronin}, {Philippe}, {Romanzin}, \& {Fillion}}]{DupuyAA2017}
{Dupuy}, R., {Bertin}, M., {F\'eraud}, G., {et~al.} 2017, \aap, 603, A61

\bibitem[{{Esplugues} {et~al.}(2016){Esplugues}, {Cazaux}, {Meijerink},
  {Spaans}, \& {Caselli}}]{EspluguesAA2016}
{Esplugues}, G.~B., {Cazaux}, S., {Meijerink}, R., {Spaans}, M., \& {Caselli},
  P. 2016, \aap, 591, A52

\bibitem[{{Fillion} {et~al.}(2012){Fillion}, {Bertin}, {Lekic}, {Moudens},
  {Philippe}, \& {Michaut}}]{FillionEAS2012}
{Fillion}, J.-H., {Bertin}, M., {Lekic}, A., {et~al.} 2012, EAS Pub.~Ser., 58,
  307

\bibitem[{{Fitzpatrick} \& {Massa}(1990)}]{FitzpatrickApJS1990}
{Fitzpatrick}, E.~L. \& {Massa}, D. 1990, \apjs, 72, 163

\bibitem[{{Fuente} {et~al.}(1996){Fuente}, {Rodriguez-Franco}, \&
  {Martin-Pintado}}]{FuenteAA1996}
{Fuente}, A., {Rodriguez-Franco}, A., \& {Martin-Pintado}, J. 1996, \aap, 312,
  599

\bibitem[{{Gardner} \& {Whiteoak}(1981)}]{GardnerMNRAS1981}
{Gardner}, F.~F. \& {Whiteoak}, J.~B. 1981, \mnras, 194, 37P

\bibitem[{{Garrod} {et~al.}(2009){Garrod}, {Vasyunin}, {Semenov}, {Wiebe}, \&
  {Henning}}]{GarrodApJ2009}
{Garrod}, R.~T., {Vasyunin}, A.~I., {Semenov}, D.~A., {Wiebe}, D.~S., \&
  {Henning}, T. 2009, \apjl, 700, L43

\bibitem[{{Gerin} {et~al.}(2010){Gerin}, {De Luca}, {Black}, {Goicoechea},
  {Herbst}, {Neufeld}, {Falgarone}, {Godard}, {Pearson}, {Lis}, {Phillips},
  {Bell}, {Sonnentrucker}, {Boulanger}, {Cernicharo}, {Coutens}, {Dartois},
  {Encrenaz}, {Giesen}, {Goldsmith}, {Gupta}, {Gry}, {Hennebelle},
  {Hily-Blant}, {Joblin}, {Kazmierczak}, {Kolos}, {Krelowski},
  {Martin-Pintado}, {Monje}, {Mookerjea}, {Perault}, {Persson}, {Plume},
  {Rimmer}, {Salez}, {Schmidt}, {Stutzki}, {Teyssier}, {Vastel}, {Yu},
  {Contursi}, {Menten}, {Geballe}, {Schlemmer}, {Shipman}, {Tielens},
  {Philipp-May}, {Cros}, {Zmuidzinas}, {Samoska}, {Klein}, \&
  {Lorenzani}}]{GerinAA2010}
{Gerin}, M., {De Luca}, M., {Black}, J., {et~al.} 2010, \aap, 518, L110

\bibitem[{{Getman} {et~al.}(2005){Getman}, {Feigelson}, {Grosso},
  {McCaughrean}, {Micela}, {Broos}, {Garmire}, \& {Townsley}}]{GetmanApJS2005}
{Getman}, K.~V., {Feigelson}, E.~D., {Grosso}, N., {et~al.} 2005, \apjs, 160,
  353

\bibitem[{{Goicoechea} {et~al.}(2017){Goicoechea}, {Cuadrado}, {Pety}, {Bron},
  {Black}, {Cernicharo}, {Chapillon}, {Fuente}, \& {Gerin}}]{GoicoecheaAA2017}
{Goicoechea}, J.~R., {Cuadrado}, S., {Pety}, J., {et~al.} 2017, \aap, 601, L9

\bibitem[{{Goicoechea} {et~al.}(2011){Goicoechea}, {Joblin}, {Contursi},
  {Bern{\'e}}, {Cernicharo}, {Gerin}, {Le Bourlot}, {Bergin}, {Bell}, \&
  {R{\"o}llig}}]{GoicoecheaAA2011}
{Goicoechea}, J.~R., {Joblin}, C., {Contursi}, A., {et~al.} 2011, \aap, 530,
  L16

\bibitem[{{Goicoechea} {et~al.}(2016){Goicoechea}, {Pety}, {Cuadrado},
  {Cernicharo}, {Chapillon}, {Fuente}, {Gerin}, {Joblin}, {Marcelino}, \&
  {Pilleri}}]{GoicoecheaNature2016}
{Goicoechea}, J.~R., {Pety}, J., {Cuadrado}, S., {et~al.} 2016, \nat, 537, 207

\bibitem[{{Goicoechea} {et~al.}(2015){Goicoechea}, {Teyssier}, {Etxaluze},
  {Goldsmith}, {Ossenkopf}, {Gerin}, {Bergin}, {Black}, {Cernicharo},
  {Cuadrado}, {Encrenaz}, {Falgarone}, {Fuente}, {Hacar}, {Lis}, {Marcelino},
  {Melnick}, {M{\"u}ller}, {Persson}, {Pety}, {R{\"o}llig}, {Schilke}, {Simon},
  {Snell}, \& {Stutzki}}]{GoicoecheaAPJ2015}
{Goicoechea}, J.~R., {Teyssier}, D., {Etxaluze}, M., {et~al.} 2015, \apj, 812,
  75

\bibitem[{{Gonzalez Garcia} {et~al.}(2008){Gonzalez Garcia}, {Le Bourlot}, {Le
  Petit}, \& {Roueff}}]{GonzalezGarciaAA2008}
{Gonzalez Garcia}, M., {Le Bourlot}, J., {Le Petit}, F., \& {Roueff}, E. 2008,
  \aap, 485, 127

\bibitem[{{Gordon} {et~al.}(2017){Gordon}, {Rothman}, {Hill}, Kochanov, Tan,
  Bernath, Birk, Boudon, Campargue, Chance, Drouin, Flaud, Gamache, Hodges,
  Jacquemart, Perevalov, Perrin, Shine, Smith, Tennyson, Toon, Tran, Tyuterev,
  Barbe, Császár, Devi, Furtenbacher, Harrison, Hartmann, Jolly, Johnson,
  Karman, Kleiner, Kyuberis, Loos, Lyulin, Massie, Mikhailenko, Moazzen-Ahmadi,
  Müller, Naumenko, Nikitin, Polyansky, Rey, Rotger, Sharpe, Sung, Starikova,
  Tashkun, Auwera, Wagner, Wilzewski, Wcisło, Yu, \& Zak}]{Hitran2016}
{Gordon}, I.~E., {Rothman}, L.~S., {Hill}, C., {et~al.} 2017, \jqsrt, 203, 3

\bibitem[{{Green} {et~al.}(1993){Green}, {Maluendes}, \&
  {McLean}}]{GreenApJS1993}
{Green}, S., {Maluendes}, S., \& {McLean}, A.~D. 1993, \apjs, 85, 181

\bibitem[{{Gupta} {et~al.}(2010){Gupta}, {Rimmer}, {Pearson}, {Yu}, {Herbst},
  {Harada}, {Bergin}, {Neufeld}, {Melnick}, {Bachiller}, {Baechtold}, {Bell},
  {Blake}, {Caux}, {Ceccarelli}, {Cernicharo}, {Chattopadhyay}, {Comito},
  {Cabrit}, {Crockett}, {Daniel}, {Falgarone}, {Diez-Gonzalez}, {Dubernet},
  {Erickson}, {Emprechtinger}, {Encrenaz}, {Gerin}, {Gill}, {Giesen},
  {Goicoechea}, {Goldsmith}, {Joblin}, {Johnstone}, {Langer}, {Larsson},
  {Latter}, {Lin}, {Lis}, {Liseau}, {Lord}, {Maiwald}, {Maret}, {Martin},
  {Martin-Pintado}, {Menten}, {Morris}, {M\"uller}, {Murphy}, {Nordh},
  {Olberg}, {Ossenkopf}, {Pagani}, {P\'erault}, {Phillips}, {Plume}, {Qin},
  {Salez}, {Samoska}, {Schilke}, {Schlecht}, {Schlemmer}, {Szczerba},
  {Stutzki}, {Trappe}, {van der Tak}, {Vastel}, {Wang}, {Yorke}, {Zmuidzinas},
  {Boogert}, {G\"usten}, {Hartogh}, {Honingh}, {Karpov}, {Kooi}, {Krieg},
  {Schieder}, \& {Zaal}}]{GuptaAA2010}
{Gupta}, H., {Rimmer}, P., {Pearson}, J.~C., {et~al.} 2010, \aap, 521, L47

\bibitem[{{Habart}(2011)}]{HabartProp2011}
{Habart}, E. 2011, {OT2\_ehabart\_4: Unveiling the origin and excitation
  mechanisms of the warm CO, OH and CH+}, Herschel Space Observatory Proposal,
  id.2039

\bibitem[{{Habart} {et~al.}(2010){Habart}, {Dartois}, {Abergel}, {Baluteau},
  {Naylor}, {Polehampton}, {Joblin}, {Ade}, {Anderson}, {Andr\'e}, {Arab},
  {Bernard}, {Blagrave}, {Bontemps}, {Boulanger}, {Cohen}, {Compiegne}, {Cox},
  {Davis}, {Emery}, {Fulton}, {Gry}, {Huang}, {Jones}, {Kirk}, {Lagache},
  {Lim}, {Madden}, {Makiwa}, {Martin}, {Miville-Desch\^enes}, {Molinari},
  {Moseley}, {Motte}, {Okumura}, {Pinheiro Gon\c{c}alves}, {Rodon}, {Russeil},
  {Saraceno}, {Sidher}, {Spencer}, {Swinyard}, {Ward-Thompson}, {White}, \&
  {Zavagno}}]{HabartAA2010}
{Habart}, E., {Dartois}, E., {Abergel}, A., {et~al.} 2010, \aap, 518, L116

\bibitem[{{Habing}(1968)}]{HabingBAIN1968}
{Habing}, H.~J. 1968, \bain, 19, 421

\bibitem[{{Hama} {et~al.}(2016){Hama}, {Kouchi}, \&
  {Watanabe}}]{HamaScience2016}
{Hama}, T., {Kouchi}, A., \& {Watanabe}, N. 2016, Science, 351, 65

\bibitem[{{Hama} \& {Watanabe}(2013)}]{HamaChemRev2013}
{Hama}, T. \& {Watanabe}, N. 2013, Chem.~Rev., 113, 8783

\bibitem[{{Herr{\'a}ez-Aguilar} {et~al.}(2014){Herr{\'a}ez-Aguilar},
  {Jambrina}, {Men{\'e}ndez}, {Aldegunde}, {Warmbier}, \&
  {Aoiz}}]{HerraezAguilarPCCP2014}
{Herr{\'a}ez-Aguilar}, D., {Jambrina}, P.~G., {Men{\'e}ndez}, M., {et~al.}
  2014, Phys.~Chem.~Chem.~Phys., 16, 24800

\bibitem[{{Hogerheijde} {et~al.}(2011){Hogerheijde}, {Bergin}, {Brinch},
  {Cleeves}, {Fogel}, {Blake}, {Dominik}, {Lis}, {Melnick}, {Neufeld},
  {Pani{\'c}}, {Pearson}, {Kristensen}, {Y{\i}ld{\i}z}, \& {van
  Dishoeck}}]{HogerheijdeScience2011}
{Hogerheijde}, M.~R., {Bergin}, E.~A., {Brinch}, C., {et~al.} 2011, Science,
  334, 338

\bibitem[{{Hogerheijde} {et~al.}(1995){Hogerheijde}, {Jansen}, \& {van
  Dishoeck}}]{HogerheijdeAA1995}
{Hogerheijde}, M.~R., {Jansen}, D.~J., \& {van Dishoeck}, E.~F. 1995, \aap,
  294, 792

\bibitem[{{Hollenbach} {et~al.}(2009){Hollenbach}, {Kaufman}, {Bergin}, \&
  {Melnick}}]{HollenbachApJ2009}
{Hollenbach}, D.~J., {Kaufman}, M.~J., {Bergin}, E.~A., \& {Melnick}, G.~J.
  2009, \apj, 690, 1497

\bibitem[{{Hollenbach} \& {Tielens}(1997)}]{HollenbachARAA1997}
{Hollenbach}, D.~J. \& {Tielens}, A.~G.~G.~M. 1997, \araa, 35, 179

\bibitem[{{Jansen} {et~al.}(1995){Jansen}, {Spaans}, {Hogerheijde}, \& {van
  Dishoeck}}]{JansenAA1995}
{Jansen}, D.~J., {Spaans}, M., {Hogerheijde}, M.~R., \& {van Dishoeck}, E.~F.
  1995, \aap, 303, 541

\bibitem[{{Joblin} {et~al.}(2018){Joblin}, {Bron}, {Pinto}, {Pilleri}, {Le
  Petit}, {Gerin}, {Le Bourlot}, {Fuente}, {Berne}, {Goicoechea}, {Habart},
  {K{\"o}hler}, {Teyssier}, {Nagy}, {Montillaud}, {Vastel}, {Cernicharo},
  {R{\"o}llig}, {Ossenkopf-Okada}, \& {Bergin}}]{JoblinAA2018}
{Joblin}, C., {Bron}, E., {Pinto}, C., {et~al.} 2018, \aap, 615, A129

\bibitem[{{Kilaj} {et~al.}(2018){Kilaj}, {Gao}, {Rösch}, {Rivero}, {Küpper},
  \& {Willitsch}}]{KilajNatureCom2018}
{Kilaj}, A., {Gao}, H., {Rösch}, D., {et~al.} 2018, Nat.~Commun., 9

\bibitem[{{Kong} {et~al.}(2018){Kong}, {Arce}, {Feddersen}, {Carpenter},
  {Nakamura}, {Shimajiri}, {Isella}, {Ossenkopf-Okada}, {Sargent},
  {S{\'a}nchez-Monge}, {Suri}, {Kauffmann}, {Pillai}, {Pineda}, {Koda},
  {Bally}, {Lis}, {Padoan}, {Klessen}, {Mairs}, {Goodman}, {Goldsmith},
  {McGehee}, {Schilke}, {Teuben}, {Jos{\'e} Maureira}, {Hara}, {Ginsburg},
  {Burkhart}, {Smith}, {Schmiedeke}, {Pineda}, {Ishii}, {Sasaki}, {Kawabe},
  {Urasawa}, {Oyamada}, \& {Tanabe}}]{KongAPJS2018}
{Kong}, S., {Arce}, H.~G., {Feddersen}, J.~R., {et~al.} 2018, \apjs, 236, 25

\bibitem[{{Larsson} {et~al.}(2003){Larsson}, {Liseau}, {Bergman}, {Bernath},
  {Black}, {Booth}, {Buat}, {Curry}, {Encrenaz}, {Falgarone}, {Feldman},
  {Fich}, {Flor\'en}, {Frisk}, {Gerin}, {Gregersen}, {Harju}, {Hasegawa},
  {Johansson}, {Kwok}, {Lecacheux}, {Liljestr\"om}, {Mattila}, {Mitchell},
  {Nordh}, {Olberg}, {Olofsson}, {Pagani}, {Plume}, {Ristorcelli}, {Sandqvist},
  {F. v. Sch\'eele}, {Tothill}, {Volk}, \& {Wilson}}]{LarssonAA2003}
{Larsson}, B., {Liseau}, R., {Bergman}, P., {et~al.} 2003, \aap, 402, L69

\bibitem[{{Le Bourlot} {et~al.}(2012){Le Bourlot}, {Le Petit}, {Pinto},
  {Roueff}, \& {Roy}}]{LeBourlotAA2012}
{Le Bourlot}, J., {Le Petit}, F., {Pinto}, C., {Roueff}, E., \& {Roy}, F. 2012,
  \aap, 541, A76

\bibitem[{{Le Petit} {et~al.}(2009){Le Petit}, {Barzel}, {Biham}, {Roueff}, \&
  {Le Bourlot}}]{LePetitAA2009}
{Le Petit}, F., {Barzel}, B., {Biham}, O., {Roueff}, E., \& {Le Bourlot}, J.
  2009, \aap, 505, 1153

\bibitem[{{Le Petit} {et~al.}(2006){Le Petit}, {Nehm{\'e}}, {Le Bourlot}, \&
  {Roueff}}]{LePetitApJSS2006}
{Le Petit}, F., {Nehm{\'e}}, C., {Le Bourlot}, J., \& {Roueff}, E. 2006, \apjs,
  164, 506

\bibitem[{{Leurini} {et~al.}(2006){Leurini}, {Rolffs}, {Thorwirth}, {Parise},
  {Schilke}, {Comito}, {Wyrowski}, {G\"usten}, {Bergman}, {Menten}, \&
  {Nyman}}]{LeuriniAA2006}
{Leurini}, S., {Rolffs}, R., {Thorwirth}, S., {et~al.} 2006, \aap, 454, L47

\bibitem[{{Lis} {et~al.}(2013){Lis}, {Bergin}, {Schilke}, \& {van
  Dishoeck}}]{LisJPCA2013}
{Lis}, D.~C., {Bergin}, E.~A., {Schilke}, P., \& {van Dishoeck}, E.~F. 2013,
  J.~Phys.~Chem.~A, 117, 9661

\bibitem[{{Lis} \& {Schilke}(2003)}]{LisApJl2003}
{Lis}, D.~C. \& {Schilke}, P. 2003, \apjl, 597, L145

\bibitem[{{Loison} {et~al.}(2019){Loison}, {Wakelam}, {Gratier}, {Hickson},
  {Bacmann}, {Ag{\`u}ndez}, {Marcelino}, {Cernicharo}, {Guzman}, {Gerin},
  {Goicoechea}, {Roueff}, {Petit}, {Pety}, {Fuente}, \&
  {Riviere-Marichalar}}]{LoisonMNRAS2019}
{Loison}, J.-C., {Wakelam}, V., {Gratier}, P., {et~al.} 2019, \mnras, 485, 5777

\bibitem[{{Marconi} {et~al.}(1998){Marconi}, {Testi}, {Natta}, \&
  {Walmsley}}]{MarconiAA1998}
{Marconi}, A., {Testi}, L., {Natta}, A., \& {Walmsley}, C.~M. 1998, \aap, 330,
  696

\bibitem[{{Mathis} {et~al.}(1983){Mathis}, {Mezger}, \&
  {Panagia}}]{MathisAA1983}
{Mathis}, J.~S., {Mezger}, P.~G., \& {Panagia}, N. 1983, \aap, 128, 212

\bibitem[{{Melnick} {et~al.}(2012){Melnick}, {Tolls}, {Goldsmith}, {Kaufman},
  {Hollenbach}, {Black}, {Encrenaz}, {Falgarone}, {Gerin}, {Hjalmarson}, {Li},
  {Lis}, {Liseau}, {Neufeld}, {Pagani}, {Snell}, {van der Tak}, \& {van
  Dishoeck}}]{MelnickApJ2012}
{Melnick}, G.~J., {Tolls}, V., {Goldsmith}, P.~F., {et~al.} 2012, \apj, 752, 26

\bibitem[{{Menten} {et~al.}(2007){Menten}, {Reid}, {Forbrich}, \&
  {Brunthaler}}]{MentenAA2007}
{Menten}, K.~M., {Reid}, M.~J., {Forbrich}, J., \& {Brunthaler}, A. 2007, \aap,
  474, 515

\bibitem[{{Mu\~noz Caro} {et~al.}(2016){Mu\~noz Caro}, {Chen}, {Aparicio},
  {Jim\'enez-Escobar}, {Rosu-Finsen}, {Lasne}, \&
  {McCoustra}}]{MunozCaroAA2016}
{Mu\~noz Caro}, G.~M., {Chen}, Y.-J., {Aparicio}, S., {et~al.} 2016, \aap, 589,
  A19

\bibitem[{{Mueller} {et~al.}(2014){Mueller}, {Jellema}, {Olberg}, {Moreno}, \&
  {Teyssier}}]{Mueller2012}
{Mueller}, M., {Jellema}, W., {Olberg}, M., {Moreno}, R., \& {Teyssier}, D.
  2014, {The HIFI Beam: Release 1 - Release Note for Astronomers},
  Tech.~Rep.~HIFI-ICC-RP-2014-001, SRON Gronigen

\bibitem[{{Nagy} {et~al.}(2017){Nagy}, {Choi}, {Ossenkopf-Okada}, {van der
  Tak}, {Bergin}, {Gerin}, {Joblin}, {R{\"o}llig}, {Simon}, \&
  {Stutzki}}]{NagyAA2017}
{Nagy}, Z., {Choi}, Y., {Ossenkopf-Okada}, V., {et~al.} 2017, \aap, 599, A22

\bibitem[{{{\"O}berg} {et~al.}(2009){{\"O}berg}, {Linnartz}, {Visser}, \& {van
  Dishoeck}}]{ObergApJ2009}
{{\"O}berg}, K.~I., {Linnartz}, H., {Visser}, R., \& {van Dishoeck}, E.~F.
  2009, \apj, 693, 1209

\bibitem[{{Ossenkopf} {et~al.}(2013){Ossenkopf}, {R\"ollig}, {Neufeld},
  {Pilleri}, {Lis}, {Fuente}, {van der Tak}, \& {Bergin}}]{OssenkopfAA2013}
{Ossenkopf}, V., {R\"ollig}, M., {Neufeld}, D.~A., {et~al.} 2013, \aap, 550,
  A57

\bibitem[{{Ott}(2010)}]{Ott2010}
{Ott}, S. 2010, in Astronomical Data Analysis Software and Systems XIX, Vol.
  434, ASP Conf.~Ser., ed. Y.~{Mizumoto}, K.-I. {Morita}, \& M.~{Ohishi}, 139

\bibitem[{{Padovani} {et~al.}(2018){Padovani}, {Galli}, {Ivlev}, {Caselli}, \&
  {Ferrara}}]{PadovaniAA2018}
{Padovani}, M., {Galli}, D., {Ivlev}, A.~V., {Caselli}, P., \& {Ferrara}, A.
  2018, \aap, 619, A144

\bibitem[{{Parikka} {et~al.}(2017){Parikka}, {Habart}, {Bernard-Salas},
  {Goicoechea}, {Abergel}, {Pilleri}, {Dartois}, {Joblin}, {Gerin}, \&
  {Godard}}]{ParikkaAA2017}
{Parikka}, A., {Habart}, E., {Bernard-Salas}, J., {et~al.} 2017, \aap, 599, A20

\bibitem[{{Parmar} {et~al.}(1991){Parmar}, {Lacy}, \&
  {Achtermann}}]{ParmarApJl1991}
{Parmar}, P.~S., {Lacy}, J.~H., \& {Achtermann}, J.~M. 1991, \apjl, 372, L25

\bibitem[{{Pilbratt} {et~al.}(2010){Pilbratt}, {Riedinger}, {Passvogel},
  {Crone}, {Doyle}, {Gageur}, {Heras}, {Jewell}, {Metcalfe}, {Ott}, \&
  {Schmidt}}]{PilbrattAA2010}
{Pilbratt}, G.~L., {Riedinger}, J.~R., {Passvogel}, T., {et~al.} 2010, \aap,
  518, L1

\bibitem[{{Prasad} \& {Tarafdar}(1983)}]{PrasadApJ1983}
{Prasad}, S.~S. \& {Tarafdar}, S.~P. 1983, \apj, 267, 603

\bibitem[{{Preibisch} {et~al.}(2005){Preibisch}, {McCaughrean}, {Grosso},
  {Feigelson}, {Flaccomio}, {Getman}, {Hillenbrand}, {Meeus}, {Micela},
  {Sciortino}, \& {Stelzer}}]{PreibischApJS2005}
{Preibisch}, T., {McCaughrean}, M.~J., {Grosso}, N., {et~al.} 2005, \apjs, 160,
  582

\bibitem[{{Rodriguez-Franco} {et~al.}(1998){Rodriguez-Franco},
  {Martin-Pintado}, \& {Fuente}}]{RodriguezFrancoAA1998}
{Rodriguez-Franco}, A., {Martin-Pintado}, J., \& {Fuente}, A. 1998, \aap, 329,
  1097

\bibitem[{{Roelfsema} {et~al.}(2012){Roelfsema}, {Helmich}, {Teyssier},
  {Ossenkopf}, {Morris}, {Olberg}, {Shipman}, {Risacher}, {Akyilmaz},
  {Assendorp}, {Avruch}, {Beintema}, {Biver}, {Boogert}, {Borys}, {Braine},
  {Caris}, {Caux}, {Cernicharo}, {Coeur-Joly}, {Comito}, {de Lange},
  {Delforge}, {Dieleman}, {Dubbeldam}, {de Graauw}, {Edwards}, {Fich},
  {Flederus}, {Gal}, {di Giorgio}, {Herpin}, {Higgins}, {Hoac}, {Huisman},
  {Jarchow}, {Jellema}, {de Jonge}, {Kester}, {Klein}, {Kooi}, {Kramer},
  {Laauwen}, {Larsson}, {Leinz}, {Lord}, {Lorenzani}, {Luinge}, {Marston},
  {Mart{\'\i}n-Pintado}, {McCoey}, {Melchior}, {Michalska}, {Moreno},
  {M{\"u}ller}, {Nowosielski}, {Okada}, {Orlea{\'n}ski}, {Phillips}, {Pearson},
  {Rabois}, {Ravera}, {Rector}, {Rengel}, {Sagawa}, {Salomons}, {S{\'a}nchez-
  Su{\'a}rez}, {Schieder}, {Schl{\"o}der}, {Schm{\"u}lling}, {Soldati},
  {Stutzki}, {Thomas}, {Tielens}, {Vastel}, {Wildeman}, {Xie}, {Xilouris},
  {Wafelbakker}, {Whyborn}, {Zaal}, {Bell}, {Bjerkeli}, {De Beck},
  {Cavali{\'e}}, {Crockett}, {Hily-Blant}, {Kama}, {Kaminski}, {Lefl{\'o}ch},
  {Lombaert}, {de Luca}, {Makai}, {Marseille}, {Nagy}, {Pacheco}, {van der
  Wiel}, {Wang}, \& {Y{\i}ld{\i}z}}]{RoelfsemaAA2012}
{Roelfsema}, P.~R., {Helmich}, F.~P., {Teyssier}, D., {et~al.} 2012, \aap, 537,
  A17

\bibitem[{{Salinas} {et~al.}(2016){Salinas}, {Hogerheijde}, {Bergin},
  {Cleeves}, {Brinch}, {Blake}, {Lis}, {Melnick}, {Pani{\'c}}, {Pearson},
  {Kristensen}, {Y{\i}ld{\i}z}, \& {van Dishoeck}}]{SalinasAA2016}
{Salinas}, V.~N., {Hogerheijde}, M.~R., {Bergin}, E.~A., {et~al.} 2016, \aap,
  591, A122

\bibitem[{{Shipman} {et~al.}(2017){Shipman}, {Beaulieu}, {Teyssier}, {Morris},
  {Rengel}, {McCoey}, {Edwards}, {Kester}, {Lorenzani}, {Coeur-Joly},
  {Melchior}, {Xie}, {Sanchez}, {Zaal}, {Avruch}, {Borys}, {Braine}, {Comito},
  {Delforge}, {Herpin}, {Hoac}, {Kwon}, {Lord}, {Marston}, {Mueller}, {Olberg},
  {Ossenkopf}, {Puga}, \& {Akyilmaz-Yabaci}}]{ShipmanAA2017}
{Shipman}, R.~F., {Beaulieu}, S.~F., {Teyssier}, D., {et~al.} 2017, A\&A, 608,
  A49

\bibitem[{{Sternberg} \& {Dalgarno}(1989)}]{SternbergApJ1989}
{Sternberg}, A. \& {Dalgarno}, A. 1989, \apj, 338, 197

\bibitem[{{Stoerzer} {et~al.}(1995){Stoerzer}, {Stutzki}, \&
  {Sternberg}}]{StoerzerAA1995}
{Stoerzer}, H., {Stutzki}, J., \& {Sternberg}, A. 1995, \aap, 296, L9

\bibitem[{{Tauber} {et~al.}(1994){Tauber}, {Tielens}, {Meixner}, \&
  {Goldsmith}}]{TauberApJ1994}
{Tauber}, J.~A., {Tielens}, A.~G.~G.~M., {Meixner}, M., \& {Goldsmith}, P.~F.
  1994, \apj, 422, 136

\bibitem[{{Teyssier} {et~al.}(2017){Teyssier}, {Avruch}, {Beaulieu}, {Braine},
  {Marston}, {Morris}, {Olberg}, {Rengel}, \& {Shipman}}]{TeyssierHB2017}
{Teyssier}, D., {Avruch}, I., {Beaulieu}, S., {et~al.} 2017, HIFI Handbook,
  HERSCHEL-HSC-DOC-2097

\bibitem[{{Tielens} \& {Hollenbach}(1985)}]{TielensApJ1985}
{Tielens}, A.~G.~G.~M. \& {Hollenbach}, D. 1985, \apj, 291, 747

\bibitem[{{Turgeon} {et~al.}(2017){Turgeon}, {Vermette}, {Alexandrowicz},
  {Peperstraete}, {Philippe}, {Bertin}, {Fillion}, {Michaut}, \&
  {Ayotte}}]{TurgeonJPCA2017}
{Turgeon}, P.-A., {Vermette}, J., {Alexandrowicz}, G., {et~al.} 2017,
  J.~Phys.~Chem.~A, 121, 1571, pMID: 28157310

\bibitem[{{van der Tak} {et~al.}(2007){van der Tak}, {Black}, {Sch\"oier},
  {Jansen}, \& {van Dishoeck}}]{vanderTakAA2007}
{van der Tak}, F.~F.~S., {Black}, J.~H., {Sch\"oier}, F.~L., {Jansen}, D.~J.,
  \& {van Dishoeck}, E.~F. 2007, \aap, 468, 627

\bibitem[{{van der Tak} {et~al.}(2013){van der Tak}, {Nagy}, {Ossenkopf},
  {Makai}, {Black}, {Faure}, {Gerin}, \& {Bergin}}]{vanderTakAA2013}
{van der Tak}, F.~F.~S., {Nagy}, Z., {Ossenkopf}, V., {et~al.} 2013, \aap, 560,
  A95

\bibitem[{{van der Tak} {et~al.}(2012){van der Tak}, {Ossenkopf}, {Nagy},
  {Faure}, {R{\"o}llig}, \& {Bergin}}]{vanderTakAA2012}
{van der Tak}, F.~F.~S., {Ossenkopf}, V., {Nagy}, Z., {et~al.} 2012, \aap, 537,
  L10

\bibitem[{{van der Werf} {et~al.}(1996){van der Werf}, {Stutzki}, {Sternberg},
  \& {Krabbe}}]{vanderWerfAA1996}
{van der Werf}, P.~P., {Stutzki}, J., {Sternberg}, A., \& {Krabbe}, A. 1996,
  \aap, 313, 633

\bibitem[{{van Dishoeck} {et~al.}(2013){van Dishoeck}, {Herbst}, \&
  {Neufeld}}]{vanDishoeckChemRev2013}
{van Dishoeck}, E.~F., {Herbst}, E., \& {Neufeld}, D.~A. 2013, Chem.~Rev., 113,
  9043

\bibitem[{{Wakelam} {et~al.}(2017){Wakelam}, {Loison}, {Mereau}, \&
  {Ruaud}}]{WakelamMolAstr2017}
{Wakelam}, V., {Loison}, J.-C., {Mereau}, R., \& {Ruaud}, M. 2017,
  Mol.~Astrophys., 6, 22

\bibitem[{{Walmsley} {et~al.}(2000){Walmsley}, {Natta}, {Oliva}, \&
  {Testi}}]{WalmsleyAA2000}
{Walmsley}, C.~M., {Natta}, A., {Oliva}, E., \& {Testi}, L. 2000, \aap, 364,
  301

\bibitem[{{Wen} \& {O'Dell}(1995)}]{WenAPJ1995}
{Wen}, Z. \& {O'Dell}, C.~R. 1995, \apj, 438, 784

\bibitem[{{Wilson} \& {Rood}(1994)}]{WilsonARAA1994}
{Wilson}, T.~L. \& {Rood}, R. 1994, \araa, 32, 191

\bibitem[{{Wu} {et~al.}(2018){Wu}, {Bron}, {Onaka}, {Le Petit}, {Galliano},
  {Languignon}, {Nakamura}, \& {Okada}}]{WuAA2018}
{Wu}, R., {Bron}, E., {Onaka}, T., {et~al.} 2018, \aap, 618, A53

\bibitem[{{Wyrowski} {et~al.}(1997){Wyrowski}, {Schilke}, {Hofner}, \&
  {Walmsley}}]{WyrowskiAA1997}
{Wyrowski}, F., {Schilke}, P., {Hofner}, P., \& {Walmsley}, C.~M. 1997, \apj,
  487, L171

\bibitem[{{Young Owl} {et~al.}(2000){Young Owl}, {Meixner}, {Wolfire},
  {Tielens}, \& {Tauber}}]{YoungOwlApJ2000}
{Young Owl}, R.~C., {Meixner}, M.~M., {Wolfire}, M., {Tielens}, A. G. G.~M., \&
  {Tauber}, J. 2000, \apj, 540, 886

\bibitem[{{Zanchet} {et~al.}(2013{\natexlab{a}}){Zanchet}, {Ag{\'u}ndez},
  {Herrero}, {Aguado}, \& {Roncero}}]{ZanchetAJ2013}
{Zanchet}, A., {Ag{\'u}ndez}, M., {Herrero}, V.~J., {Aguado}, A., \& {Roncero},
  O. 2013{\natexlab{a}}, \aj, 146, 125

\bibitem[{{Zanchet} {et~al.}(2013{\natexlab{b}}){Zanchet}, {Godard}, {Bulut},
  {Roncero}, {Halvick}, \& {Cernicharo}}]{ZanchetApJ2013}
{Zanchet}, A., {Godard}, B., {Bulut}, N., {et~al.} 2013{\natexlab{b}}, \apj,
  766, 80

\end{thebibliography}

\begin{appendix}
\onecolumn{
\section{Observed lines and calibrations \label{app:LineUsed}}

\begin{table*}[!h]
\centering
\caption{Line intensities integrated with a linear baseline fit outside the line, except for \chem{H_2^{18}O} \trans{1_{10}}{1_{01}} where the frequency switch observing mode leads to optical standing waves that had to be corrected by a fouth order polynom. See Table~\ref{tab:ObsUsed} for the definition of a and b labeled transitions. \label{tab:IntensityCal}}
\begin{tabular}{clcccccccccc}
\hline \hline
\multicolumn{2}{c}{Transition}                                                 &Pol.&HPBW\tablefootmark{a}&$\eta_{mb}$\tablefootmark{a}&\multicolumn{2}{c}{~RA-DEC\tablefootmark{b}}&$\Delta V_{int}$\tablefootmark{c}&$\int T_{mb} \diff V$\tablefootmark{} &$\Delta V_{Line}$\tablefootmark{d}&$\Omega$\tablefootmark{e}& Cal.\tablefootmark{f}\\
                                  &                                            &    &(\si{''})            &                            &\multicolumn{2}{c}{(\si{''})}            &(\si{km.s^{-1}})                 &(\si{K.km.s^{-1}})                    &(\si{km.s^{-1}})                  &                         &(\si{\%})               \\\hline
\multirow{16}{*}{\chem{H_2^{16}O}}&\multirow{2}{*}{\trans{1_{10}}{1_{01}}-a}   &H   &\num{37.15}          & 0.622\pmt0.07              &1.6                &4.8                 &[6.7-14.5]                       &21.02\pmt0.22                         &4.3\pmt0.3                        &0.74\pmt0.11             &5.8                   \\ 
                                  &                                            &V   &\num{37.49}          & 0.616\pmt0.07              &-0.2               &-1.6                &[6.7-14.5]                       &22.58\pmt0.23                         &4.6\pmt0.3                        &0.79\pmt0.10             &5.8                   \\
                                  &\multirow{2}{*}{\trans{1_{10}}{1_{01}}-b}   &H   &\num{37.15}          & 0.622\pmt0.07              &0.9                &3.4                 &[6.7-14.5]                       &20.37\pmt0.24                         &4.6\pmt0.5                        &0.75\pmt0.11             &5.3                   \\ 
                                  &                                            &V   &\num{37.49}          & 0.616\pmt0.07              &-2.7               &-2.2                &[6.7-14.5]                       &20.56\pmt0.25                         &4.3\pmt0.8                        &0.80\pmt0.09             &5.3                   \\
                                  &\multirow{2}{*}{\trans{1_{11}}{0_{00}}}     &H   &\num{18.88}          & 0.632\pmt0.08              &-1.0               &1.5                 &[6.7-14.5]                       &9.63\pmt0.44                          &4.4\pmt0.8                        &0.78\pmt0.09             &5.4                   \\ 
                                  &                                            &V   &\num{18.71}          & 0.636\pmt0.08              &2.2                &0.0                 &[6.7-14.5]                       &10.93\pmt0.45                         &4.3\pmt0.5                        &0.85\pmt0.05             &5.8                   \\ 
                                  &\multirow{2}{*}{\trans{2_{12}}{1_{01}}-a}   &H   &\num{12.58}          & 0.593\pmt0.07              &0.2                &0.8                 &[7.2-14.0]                       &10.41\pmt0.46                         &5.2\pmt0.4                        &0.78\pmt0.07             &7.1                   \\ 
                                  &                                            &V   &\num{12.41}          & 0.572\pmt0.07              &0.4                &0.1                 &[7.2-14.0]                       &10.10\pmt0.47                         &4.9\pmt1.1                        &0.82\pmt0.07             &6.9                   \\ 
                                  &\multirow{2}{*}{\trans{2_{12}}{1_{01}}-b}   &H   &\num{12.58}          & 0.593\pmt0.07              &-1.9               &-0.3                &[7.2-14.0]                       &6.76\pmt1.33                          &4.9\pmt1.4                        &0.80\pmt0.06             &5.1                   \\ 
                                  &                                            &V   &\num{12.41}          & 0.572\pmt0.07              &-2.0               &0.5                 &[7.2-14.0]                       &8.38\pmt1.77                          &4.0\pmt0.3                        &0.76\pmt0.08             &4.8                   \\ 
                                  &\multirow{2}{*}{\trans{2_{02}}{1_{11}}}     &H   &\num{21.28}          & 0.634\pmt0.08              &-1.2               &2.0                 &[7.7-13.5]                       &6.47\pmt0.28                          &2.9\pmt0.8                        &0.78\pmt0.09             &5.4                   \\ 
                                  &                                            &V   &\num{21.09}          & 0.637\pmt0.08              &1.8                &0.1                 &[7.7-13.5]                       &9.64\pmt0.38                          &3.0\pmt0.8                        &0.84\pmt0.06             &5.7                   \\ 
                                  &\multirow{2}{*}{\trans{2_{11}}{2_{02}}}     &H   &\num{28.00}          & 0.640\pmt0.09              &-0.4               &3.0                 &[8.2-13.0]                       &6.51\pmt0.26                          &2.6\pmt0.6                        &0.77\pmt0.11             &5.5                   \\  
                                  &                                            &V   &\num{27.91}          & 0.659\pmt0.09              &0.9                &-1.3                &[8.2-13.0]                       &9.13\pmt0.22                          &2.4\pmt0.4                        &0.82\pmt0.08             &5.7                   \\ 
                                  &\multirow{2}{*}{\trans{2_{21}}{2_{12}}}     &H   &\num{12.65}          & 0.579\pmt0.07              &0.2                &10.8                &[8.2-13.0]                       &2.04\pmt0.45                          &1.4\pmt1.0                        &0.78\pmt0.07             &7.1                   \\  
                                  &                                            &V   &\num{12.48}          & 0.572\pmt0.07              &0.4                &0.1                 &[8.2-13.0]                       &3.20\pmt0.55                          &2.4\pmt0.5                        &0.82\pmt0.07             &6.9                   \\ 
                                  &\multirow{2}{*}{\trans{3_{12}}{3_{03}}}     &H   &\num{19.16}          & 0.632\pmt0.08              &-1.0               &1.5                 &[9.2-12.0]                       &1.43\pmt0.24                          &2.1\pmt0.6                        &0.78\pmt0.09             &5.4                   \\ 
                                  &                                            &V   &\num{18.98}          & 0.636\pmt0.08              &2.2                &0.0                 &[9.2-12.0]                       &2.27\pmt0.23                          &1.9\pmt0.6                        &0.85\pmt0.05             &5.8                   \\ \hline
\multirow{4}{*}{\chem{H_2^{18}O}} &\multirow{2}{*}{\trans{1_{10}}{1_{01}}}     &H   &\num{37.77}          & 0.622\pmt0.07              &-0.7               &-1.5                &[8.2-13.0]                       &0.31\pmt0.02                          &1.9\pmt0.5                        &0.79\pmt0.10             &5.2                   \\ 
                                  &                                            &V   &\num{38.12}          & 0.616\pmt0.07              &1.9                &4.6                 &[8.2-13.0]                       &0.26\pmt0.01                          &1.9\pmt0.3                        &0.74\pmt0.11             &5.3                   \\
                                  &\multirow{2}{*}{\trans{1_{11}}{0_{00}}}     &H   &\num{19.08}          & 0.632\pmt0.08              &-1.0               &1.5                 &[8.2-13.0]                       &0.43\pmt0.30                          &                                  &0.78\pmt0.09             &5.4                   \\ 
                                  &                                            &V   &\num{18.91}          & 0.636\pmt0.08              &2.2                &0.0                 &[8.2-13.0]                       &0.25\pmt0.30                          &                                  &0.85\pmt0.05             &5.8                   \\
\hline
\end{tabular}
\tablefoot{
\tablefoottext{a}{Half-power beam width and main-beam efficiency of the \textit{Herschel} telescope at the water line frequency~\citep{Mueller2012}.}
\tablefoottext{b}{Right ascension and declination offsets from the \chem{CO^{+}} peak $(\alpha_{2000}=05^{\si{h}}35^{\si{m}}20.61^{\si{s}},~\delta_{2000}=-05\si{\degree}25\si{'}14.0\si{''})$.}
\tablefoottext{c}{Integration velocity range.}
\tablefoottext{d}{Observed full width at half maximum of the lines.}
\tablefoottext{e}{Beam coupling factor.}
\tablefoottext{f}{Absolute calibration uncertainty.}
}
\end{table*}
}

\begin{figure*}
\centering
\subfloat{\includegraphics[scale=0.84]{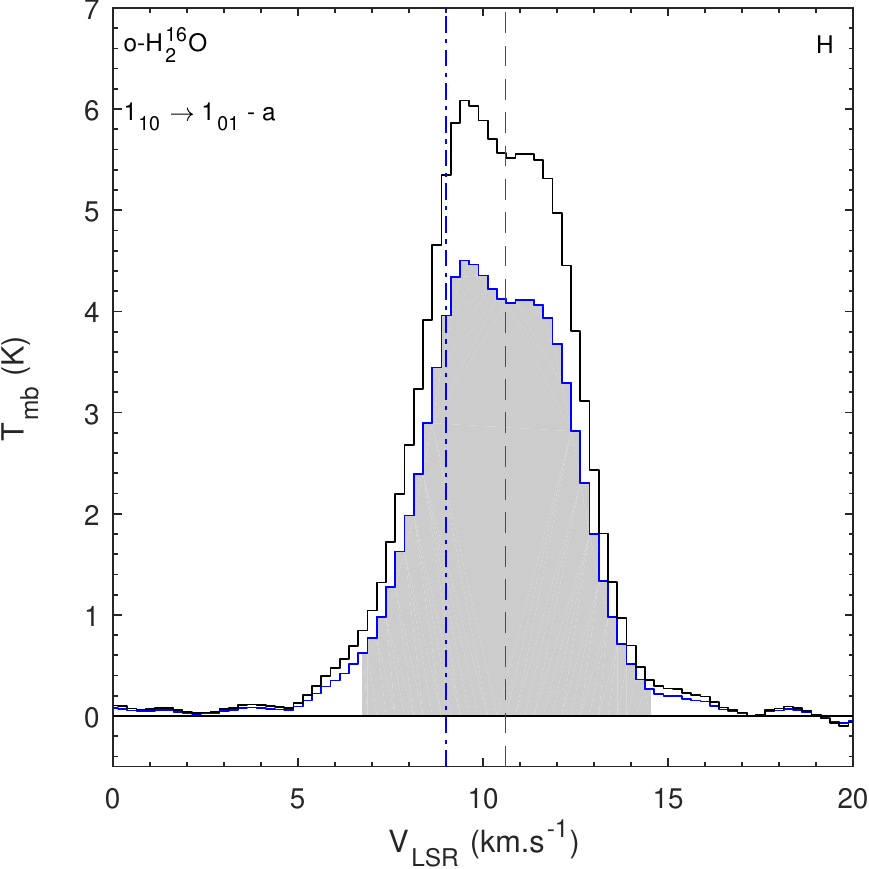}} \qquad
\subfloat{\includegraphics[scale=0.84]{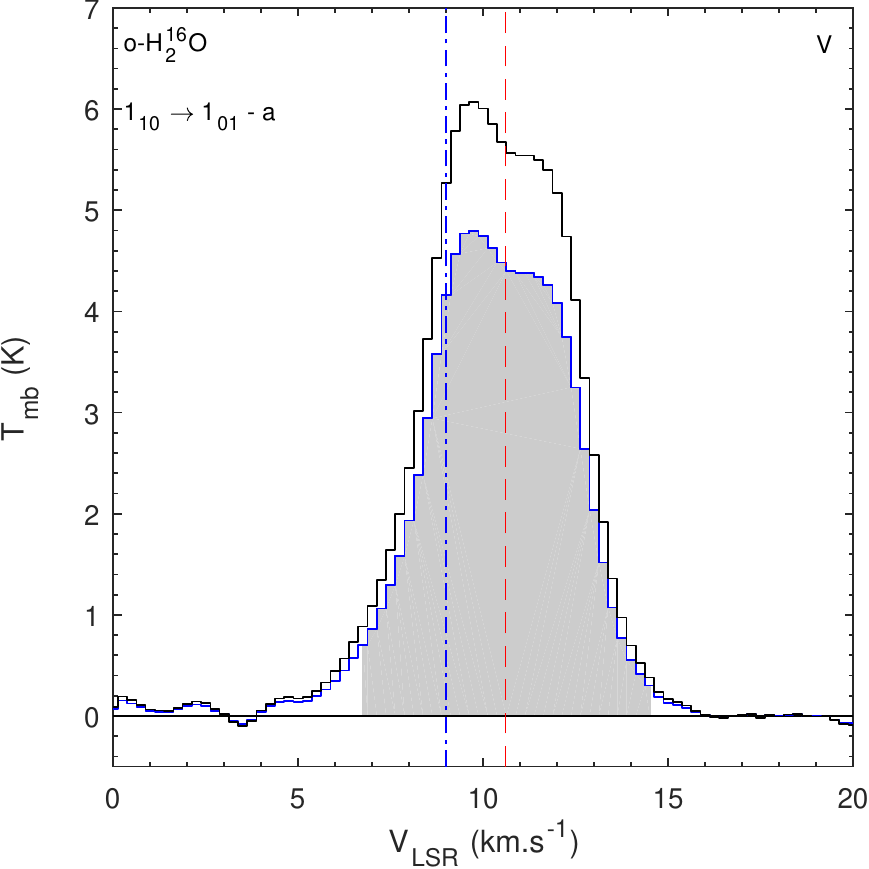}}\\
\subfloat{\includegraphics[scale=0.84]{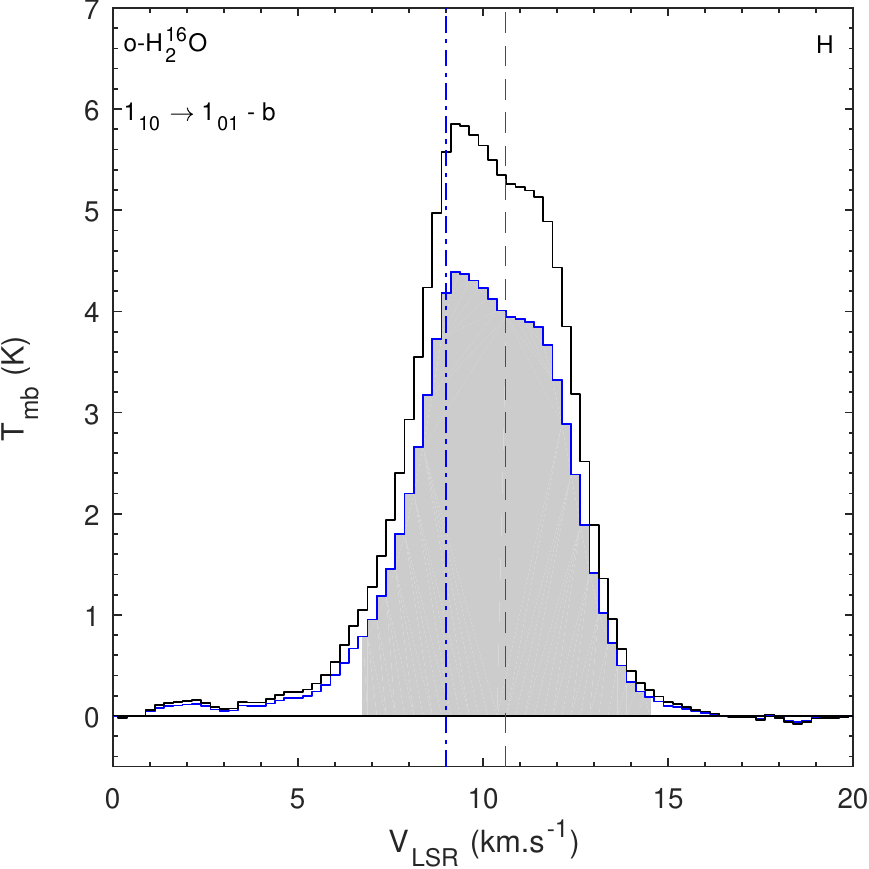}} \qquad
\subfloat{\includegraphics[scale=0.84]{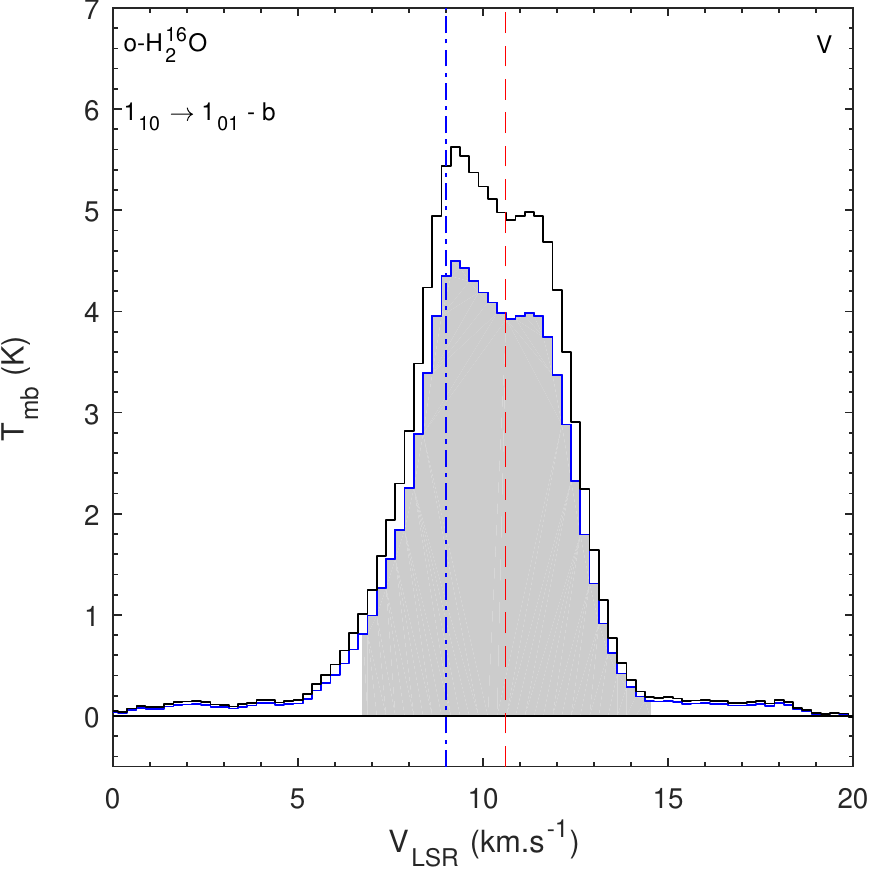}}\\
\subfloat{\includegraphics[scale=0.84]{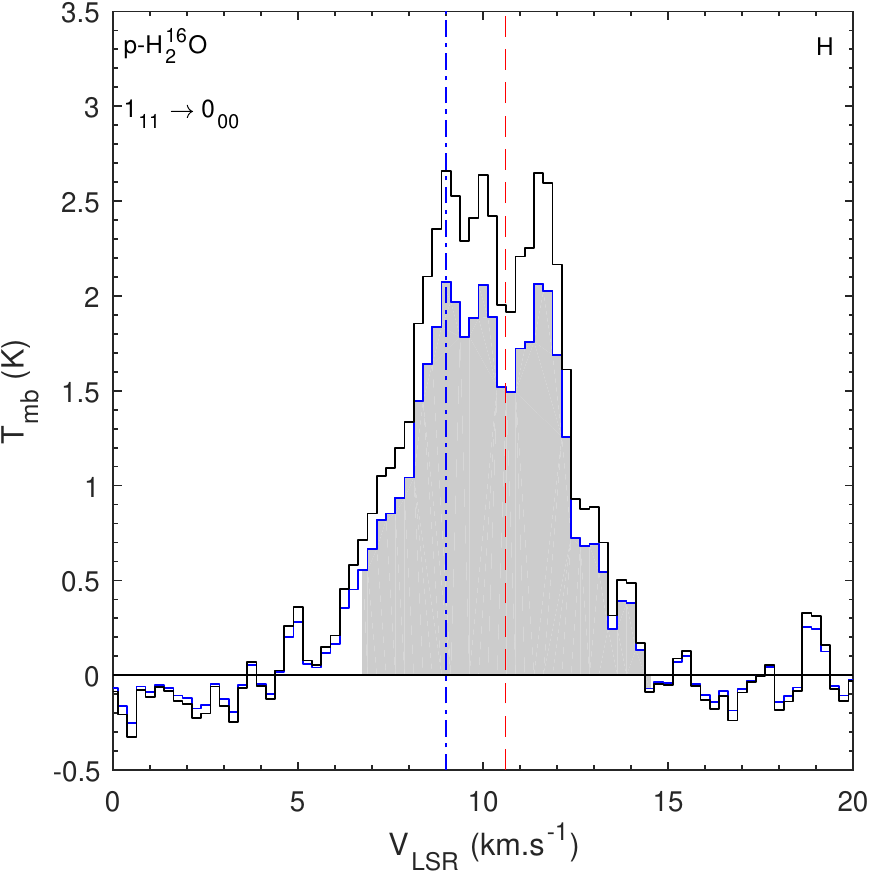}} \qquad
\subfloat{\includegraphics[scale=0.84]{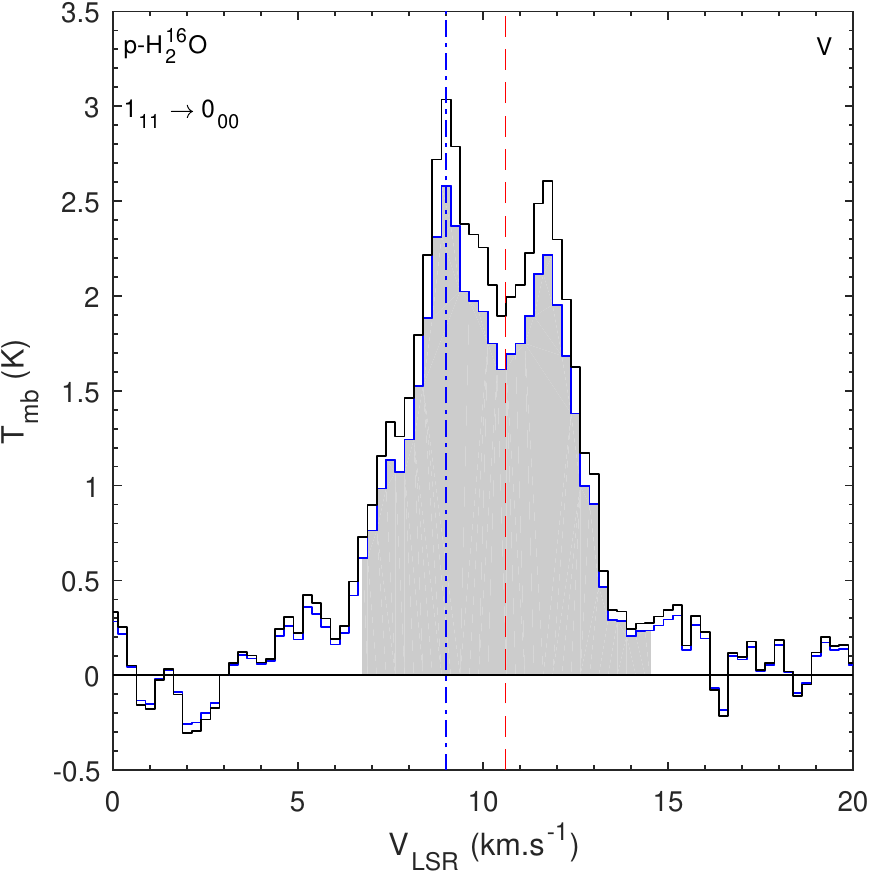}} \\\caption{\label{fig:LineUsed1} Line profile of \chem{H_2^{16}O} for H (left panel) and V (right panel) polarization. The blue solid line is the spectrum corrected for HIFI calibration, the black solid line is the spectrum corrected for the beam coupling factor, the blue dash-dotted line and red dashed line represent Orion Bar and Orion Ridge velocity features and the gray area is the velocity integration range. See Tables~\ref{tab:ObsUsed} and \ref{tab:IntensityCal} for the definition of a and b labeled transitions.}
\end{figure*}

\begin{figure*}
\centering
\subfloat{\includegraphics[scale=0.84]{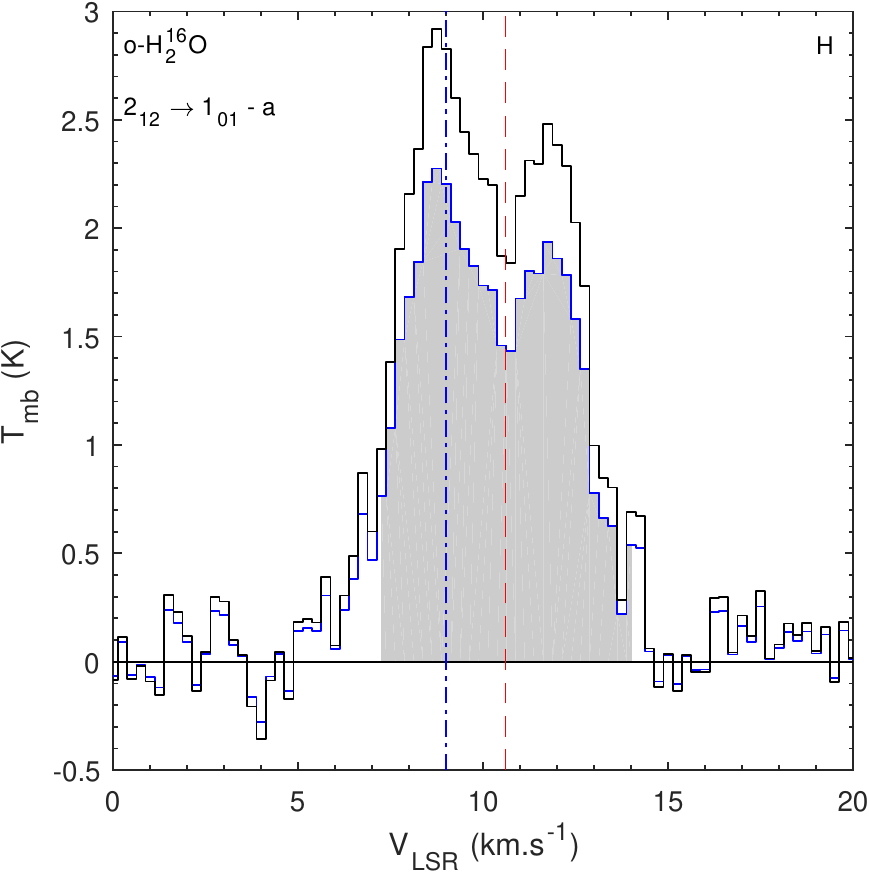}} \qquad
\subfloat{\includegraphics[scale=0.84]{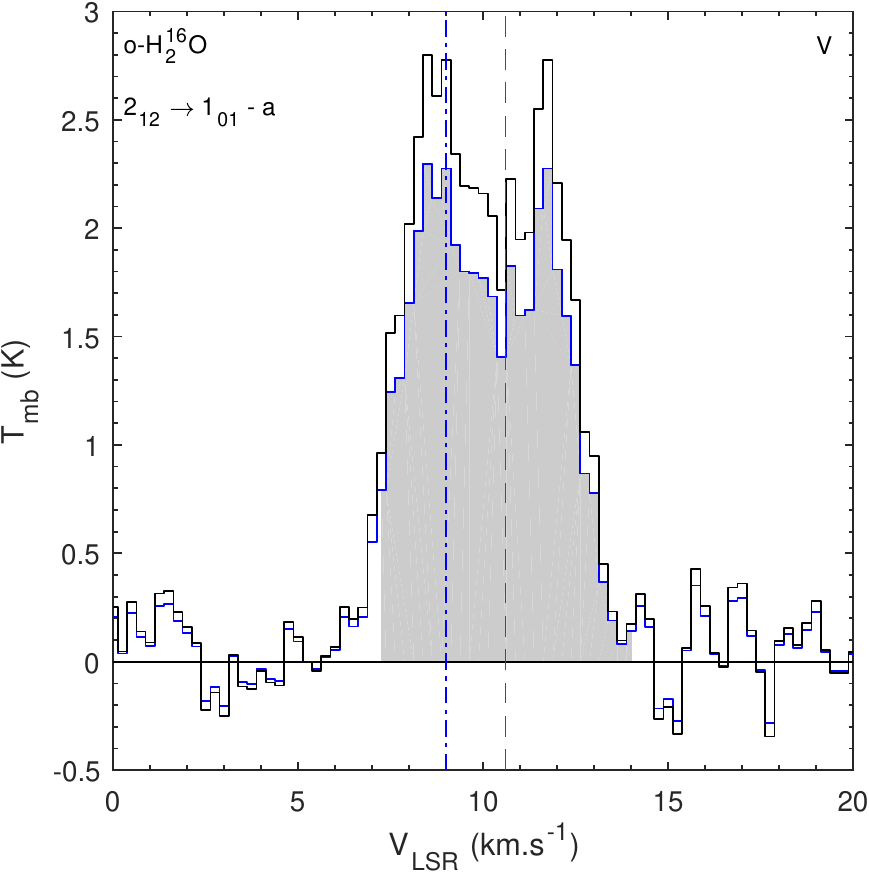}}\\
\subfloat{\includegraphics[scale=0.84]{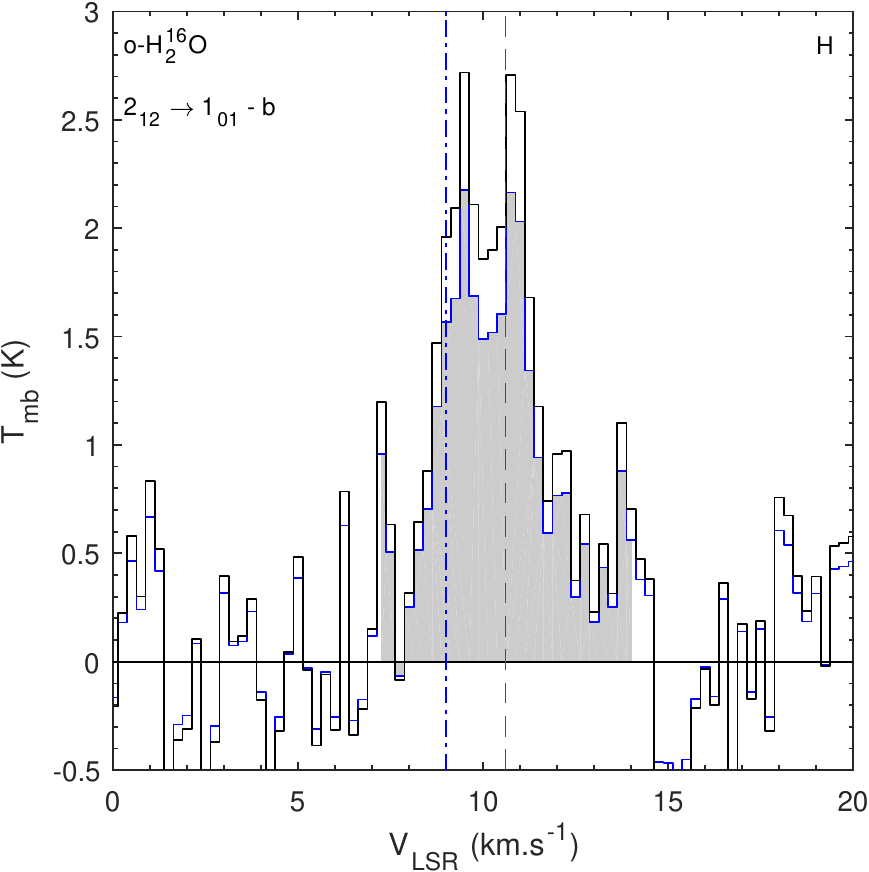}} \qquad
\subfloat{\includegraphics[scale=0.84]{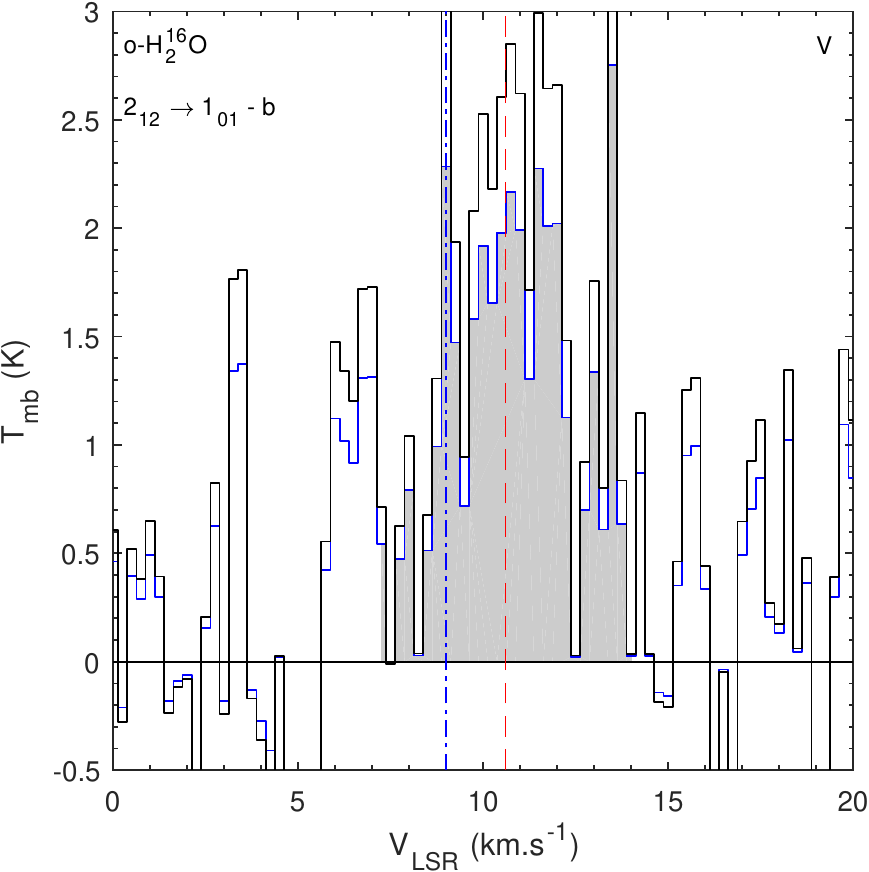}}\\
\subfloat{\includegraphics[scale=0.84]{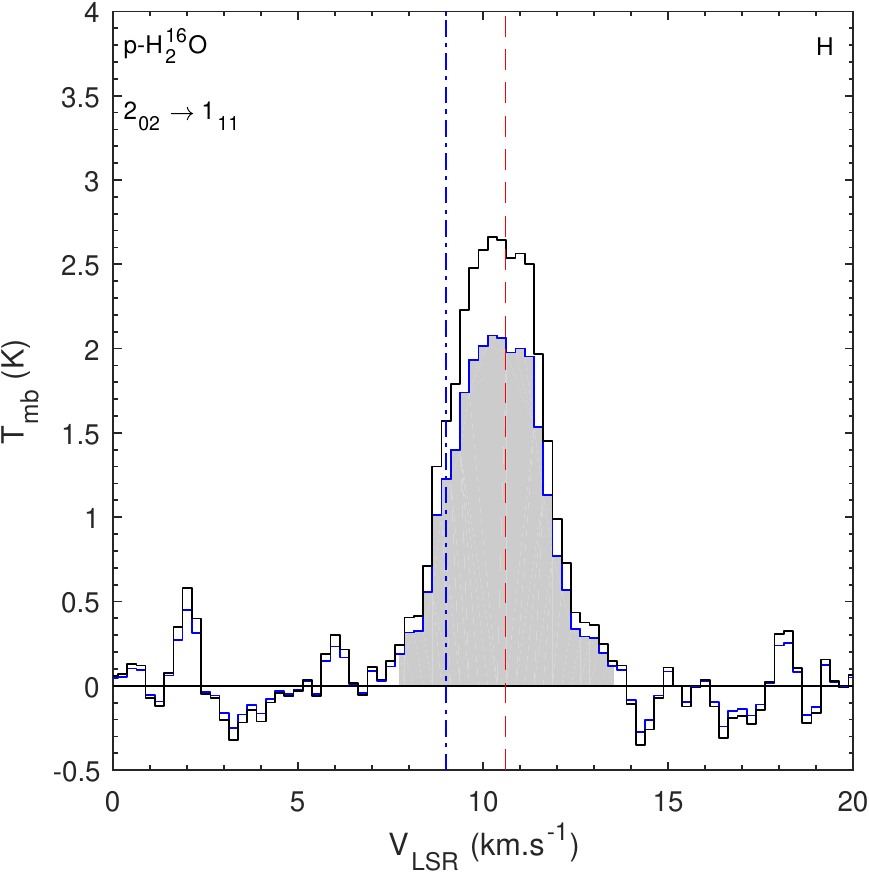}} \qquad
\subfloat{\includegraphics[scale=0.84]{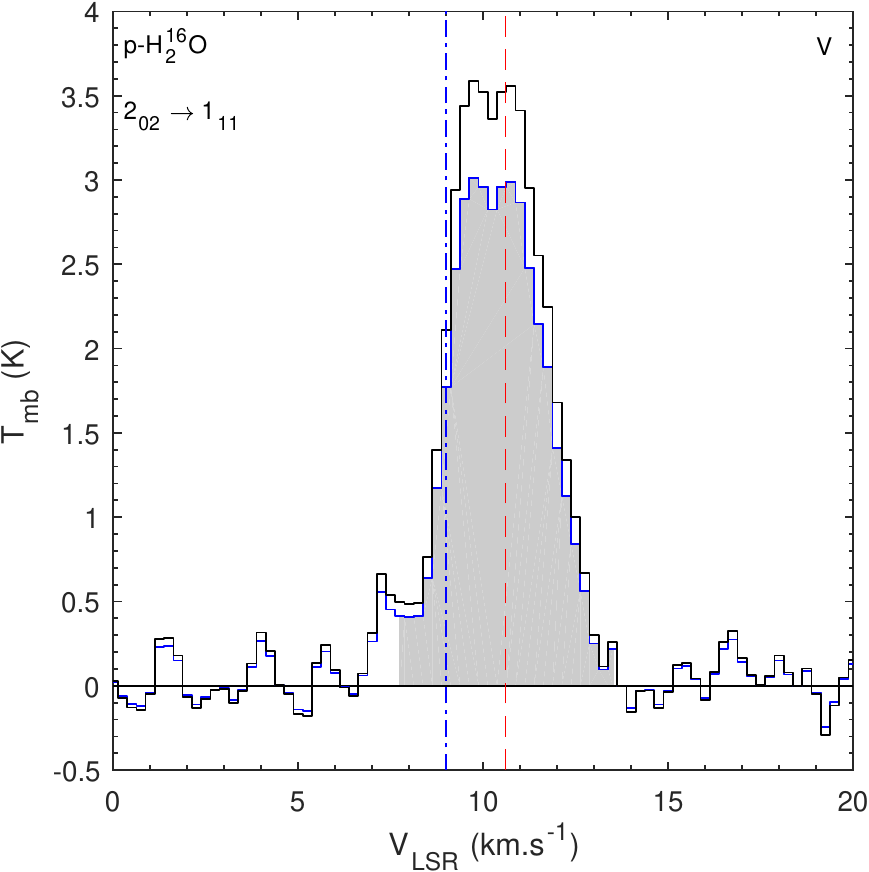}}\\\caption{\label{fig:LineUsed1} Line profile of \chem{H_2^{16}O} for H (left panel) and V (right panel) polarization. The blue solid line is the spectrum corrected for HIFI calibration, the black solid line is the spectrum corrected for the beam coupling factor, the blue dash-dotted line and red dashed line lines represent Orion Bar and Orion Ridge velocity features and the gray area is the velocity integration range. See Tables~\ref{tab:ObsUsed} and \ref{tab:IntensityCal} for the definition of a and b labeled transitions.}
\end{figure*}

\begin{figure*}
\centering
\subfloat{\includegraphics[scale=0.84]{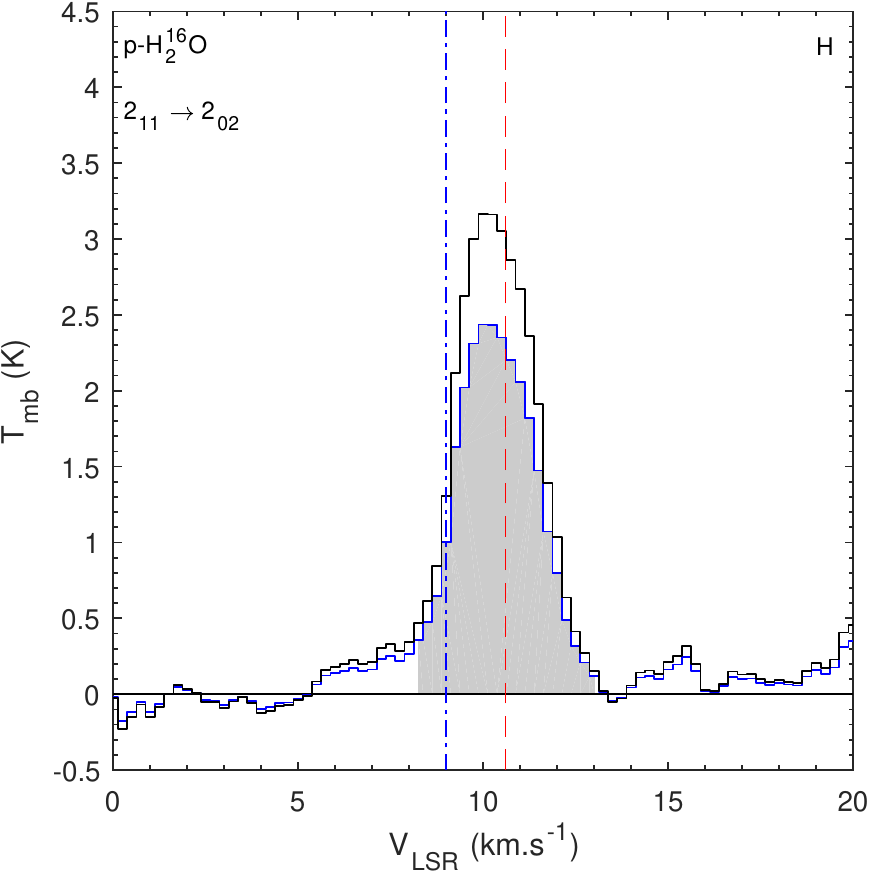}} \qquad
\subfloat{\includegraphics[scale=0.84]{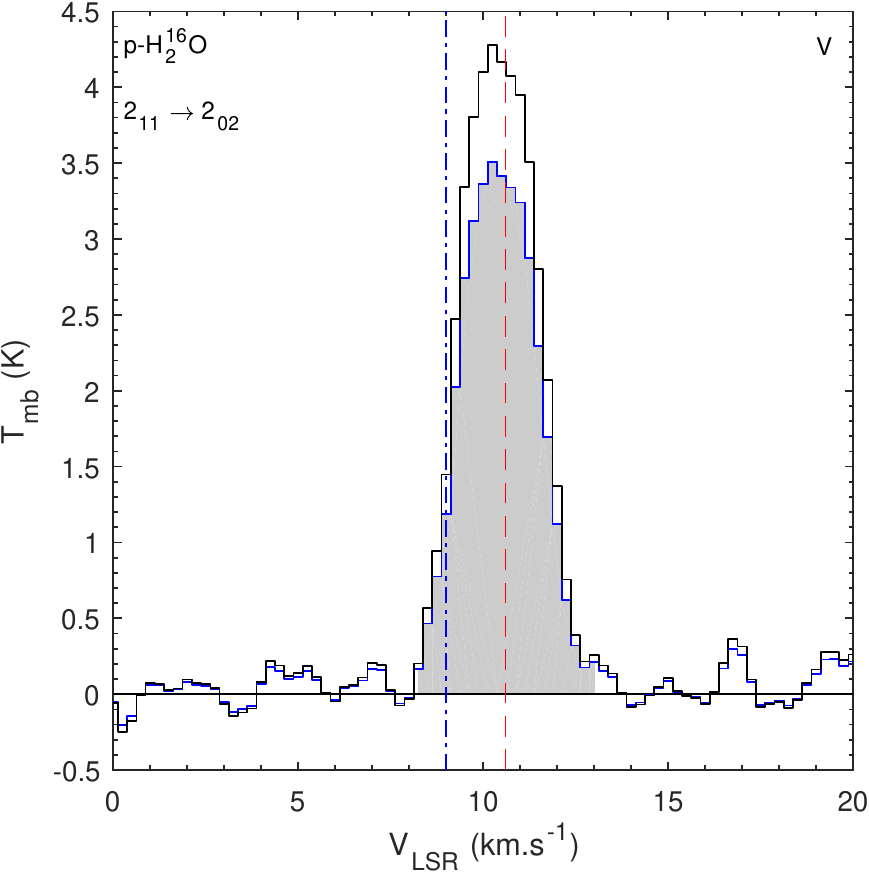}}\\
\subfloat{\includegraphics[scale=0.84]{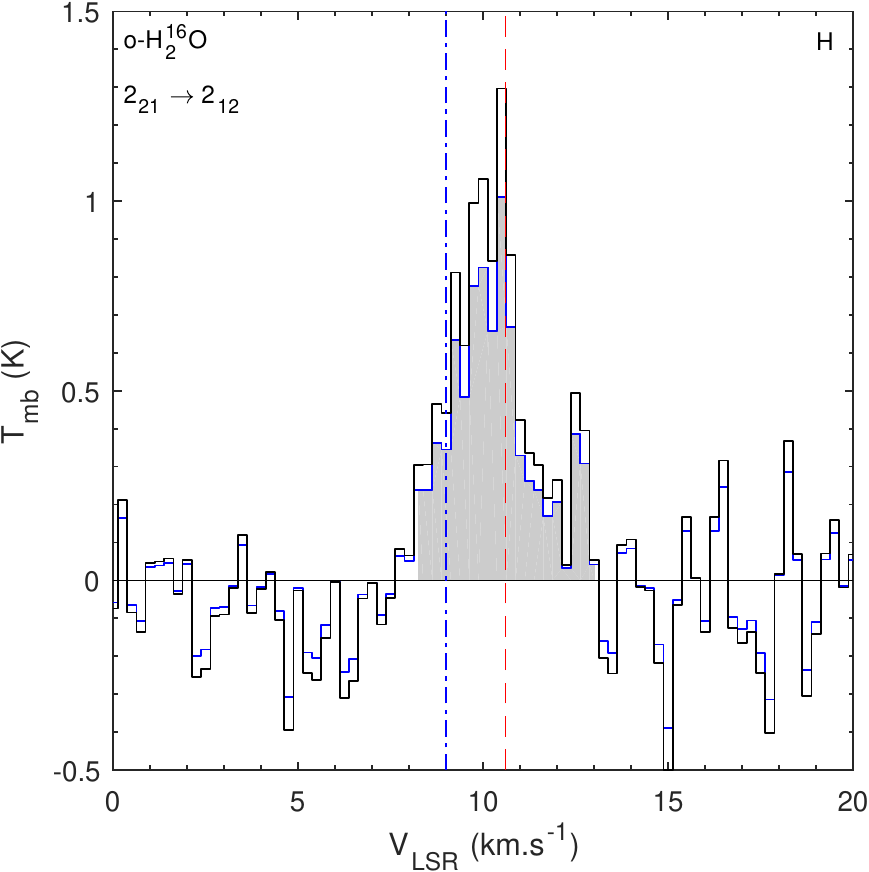}} \qquad
\subfloat{\includegraphics[scale=0.84]{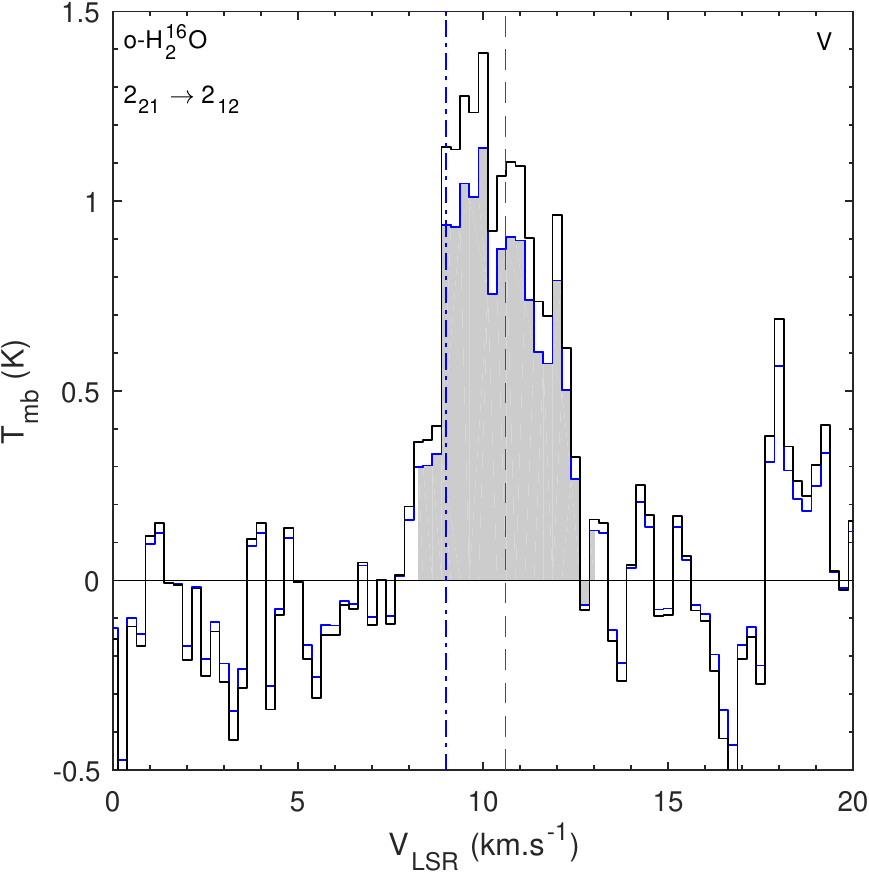}}\\
\subfloat{\includegraphics[scale=0.84]{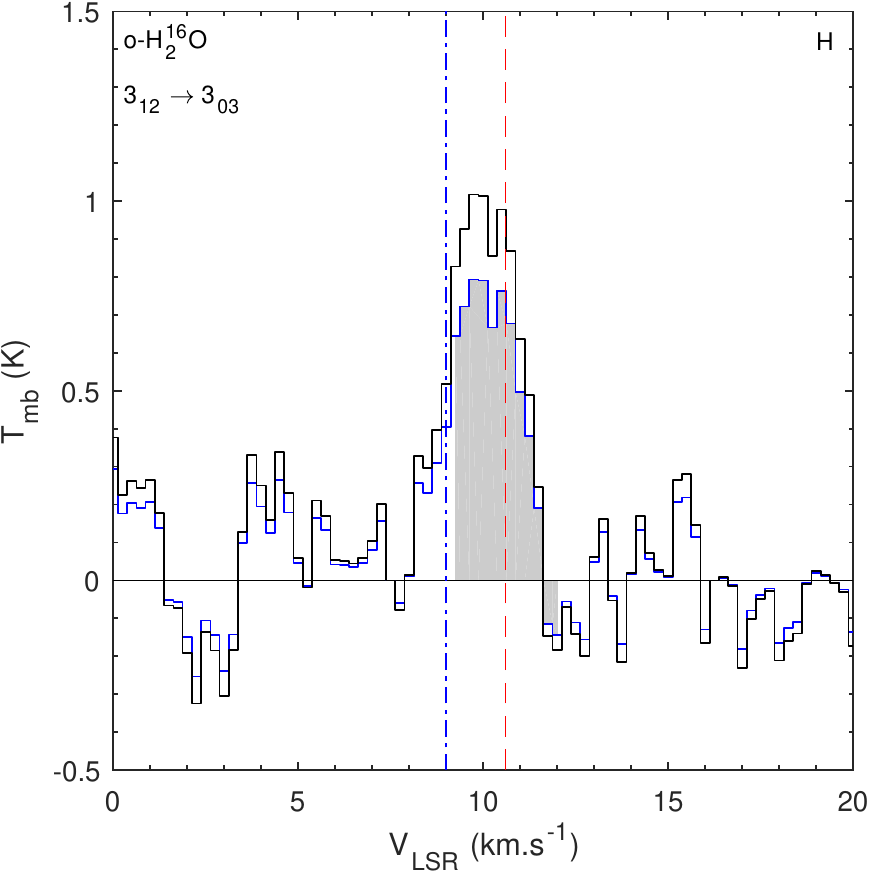}} \qquad
\subfloat{\includegraphics[scale=0.84]{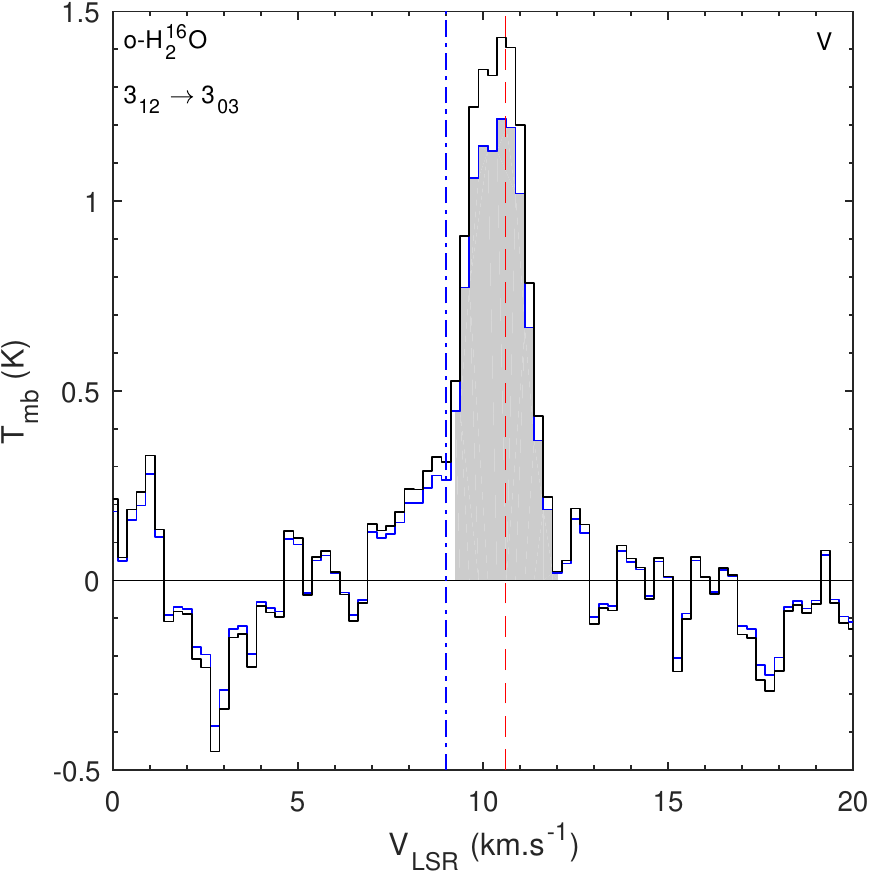}}\\

\caption{\label{fig:LineUsed1} Line profile of \chem{H_2^{16}O} for H (left panel) and V (right panel) polarization. The blue solid line is the spectrum corrected for HIFI calibration, the black solid line is the spectrum corrected for the beam coupling factor, the blue dash-dotted line0gray area is the velocity integration range.}
\end{figure*}

\begin{figure*}
\centering
\subfloat{\includegraphics[scale=0.84]{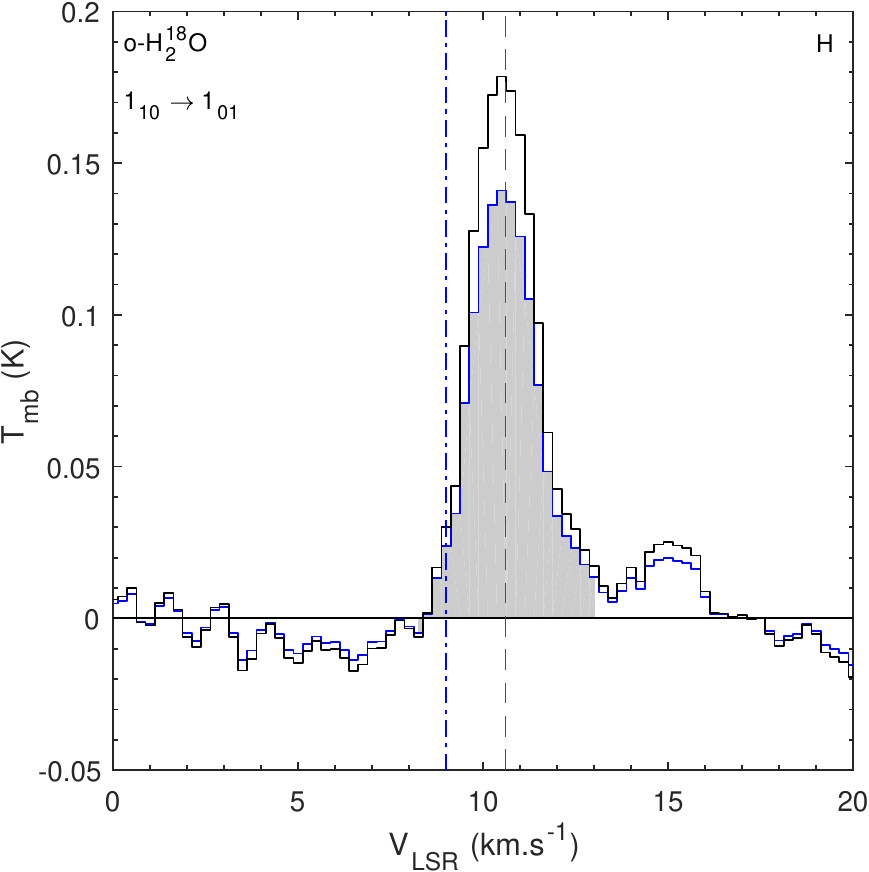}} \qquad
\subfloat{\includegraphics[scale=0.84]{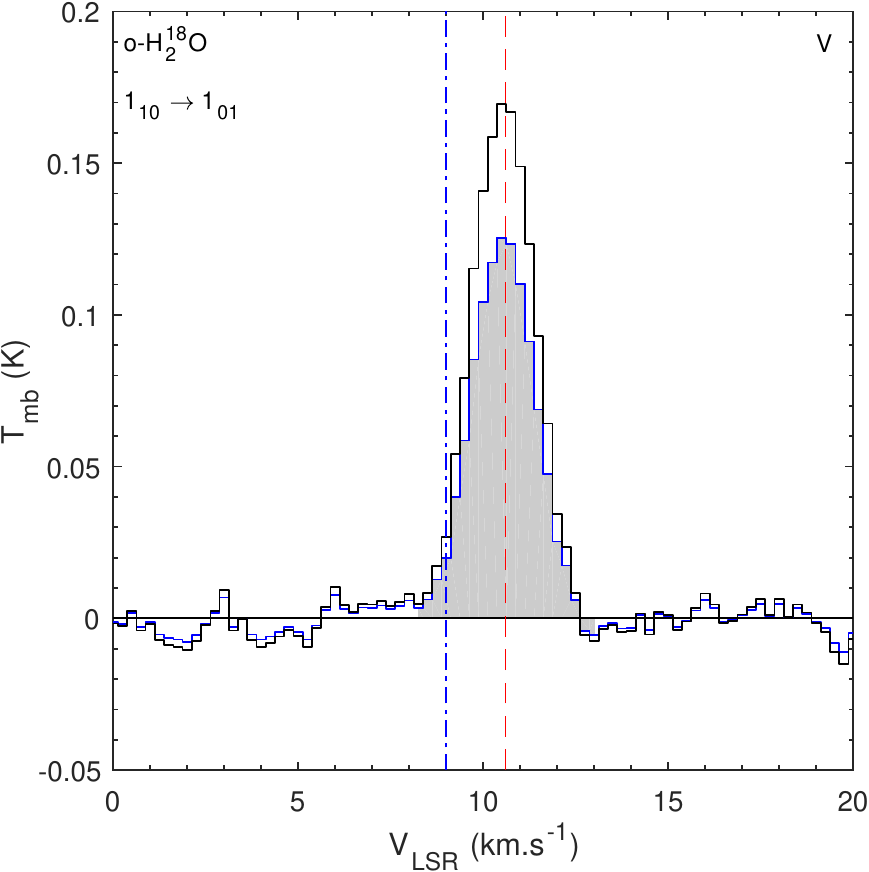}}\\
\subfloat{\includegraphics[scale=0.84]{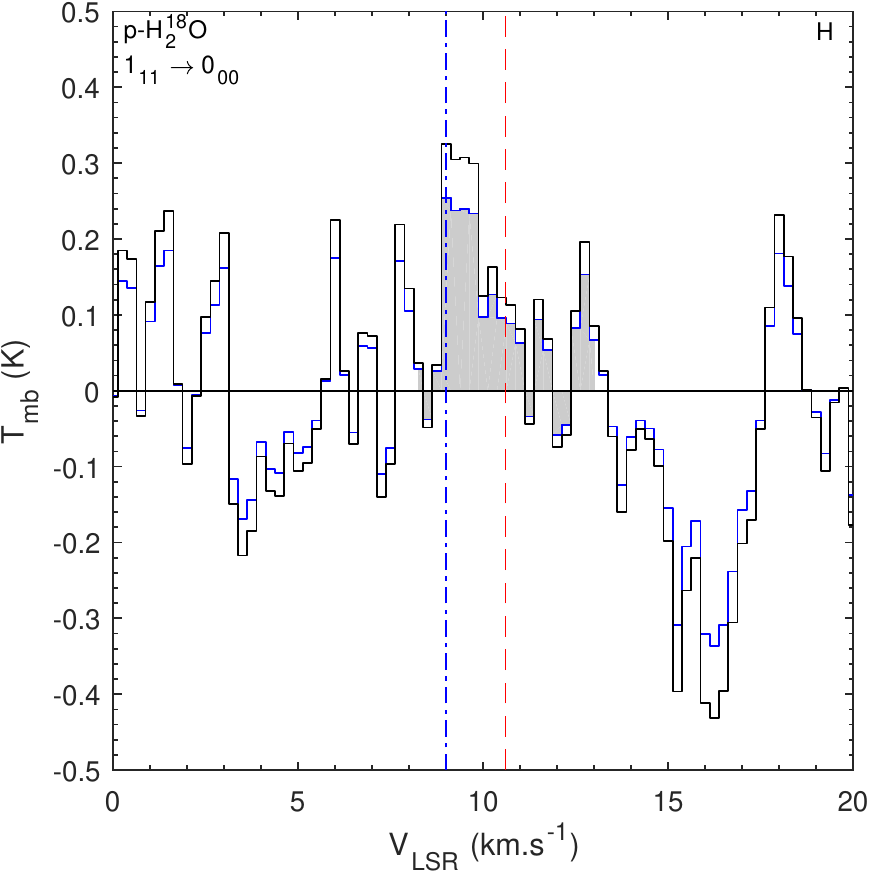}} \qquad
\subfloat{\includegraphics[scale=0.84]{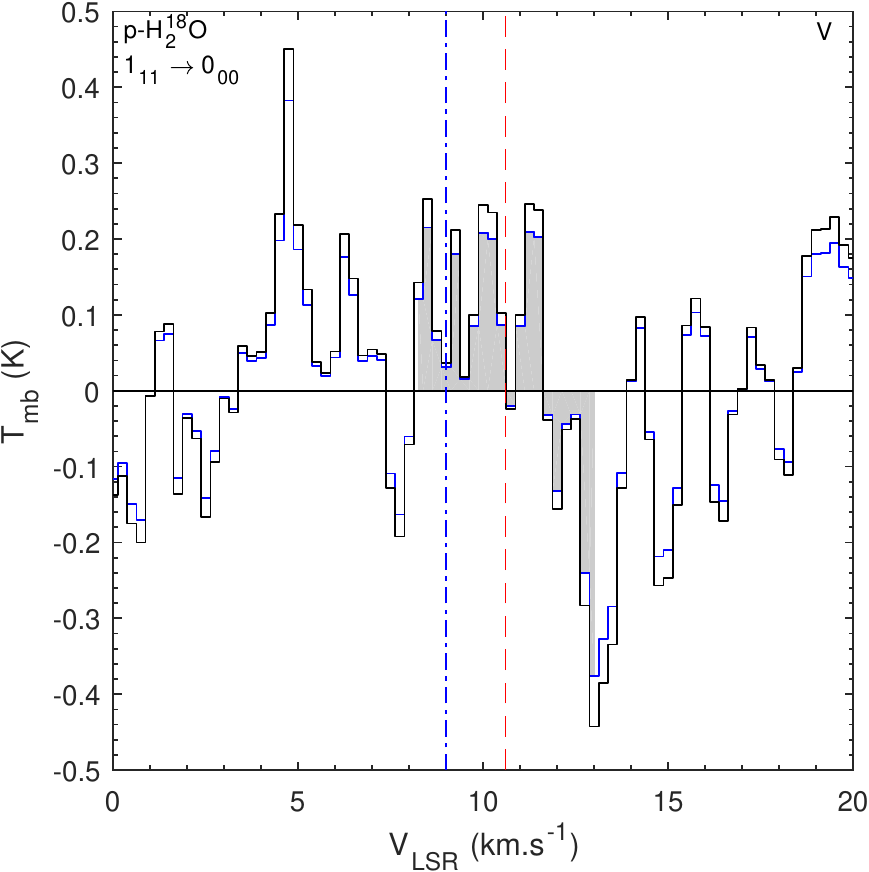}}\\
\caption{\label{fig:LineUsed1} Line profile of \chem{H_2^{18}O} for H (left panel) and V (right panel) polarization. The blue solid line is the spectrum corrected for HIFI calibration, the black solid line is the spectrum corrected for the beam coupling factor, the blue dash-dotted line and red dashed line represent Orion Bar and Orion Ridge velocity features and the gray area is the velocity integration range.}
\end{figure*}

\begin{figure*}
\centering
\subfloat{\includegraphics[scale=0.39]{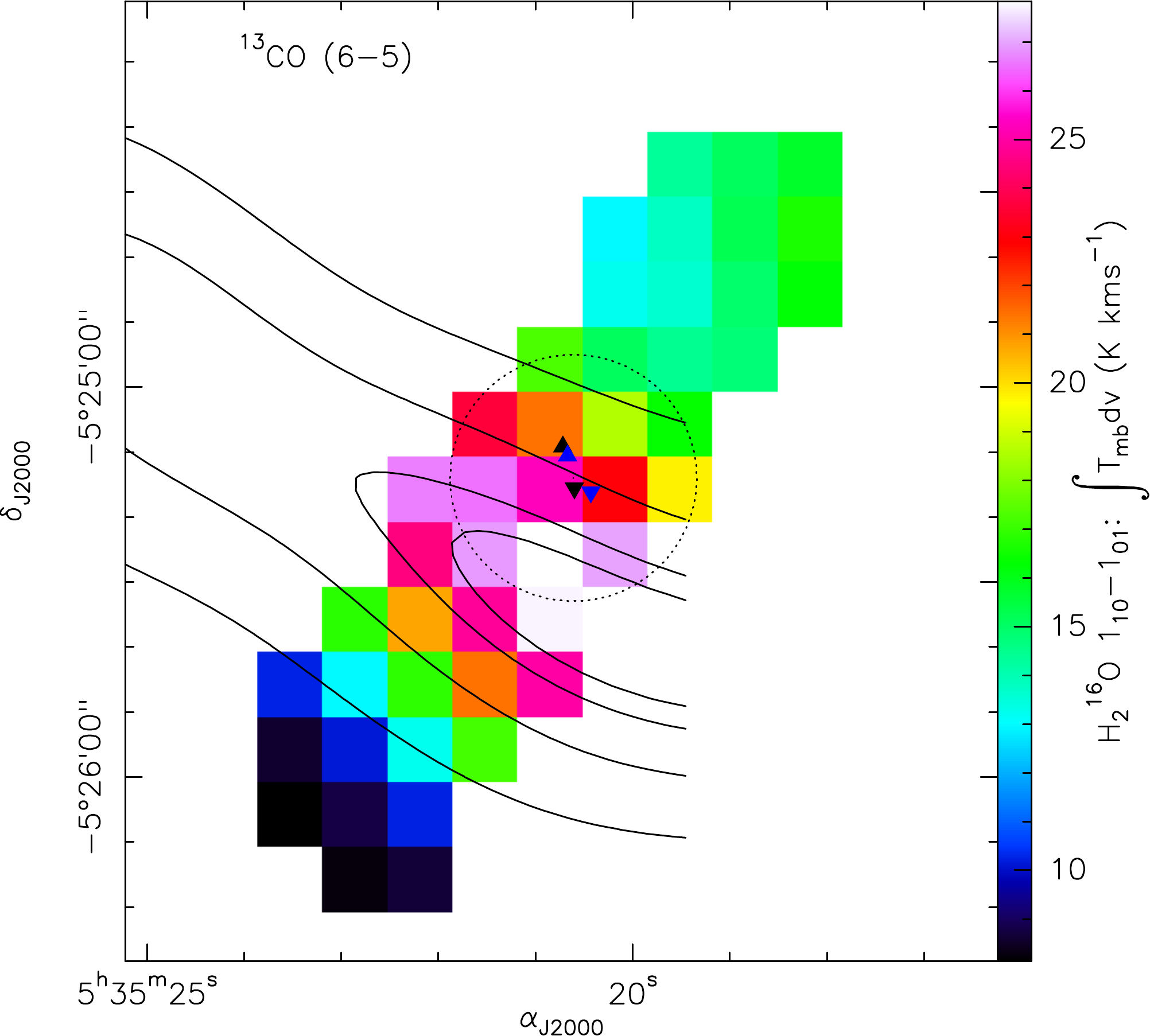}}\qquad
\subfloat{\includegraphics[scale=0.39]{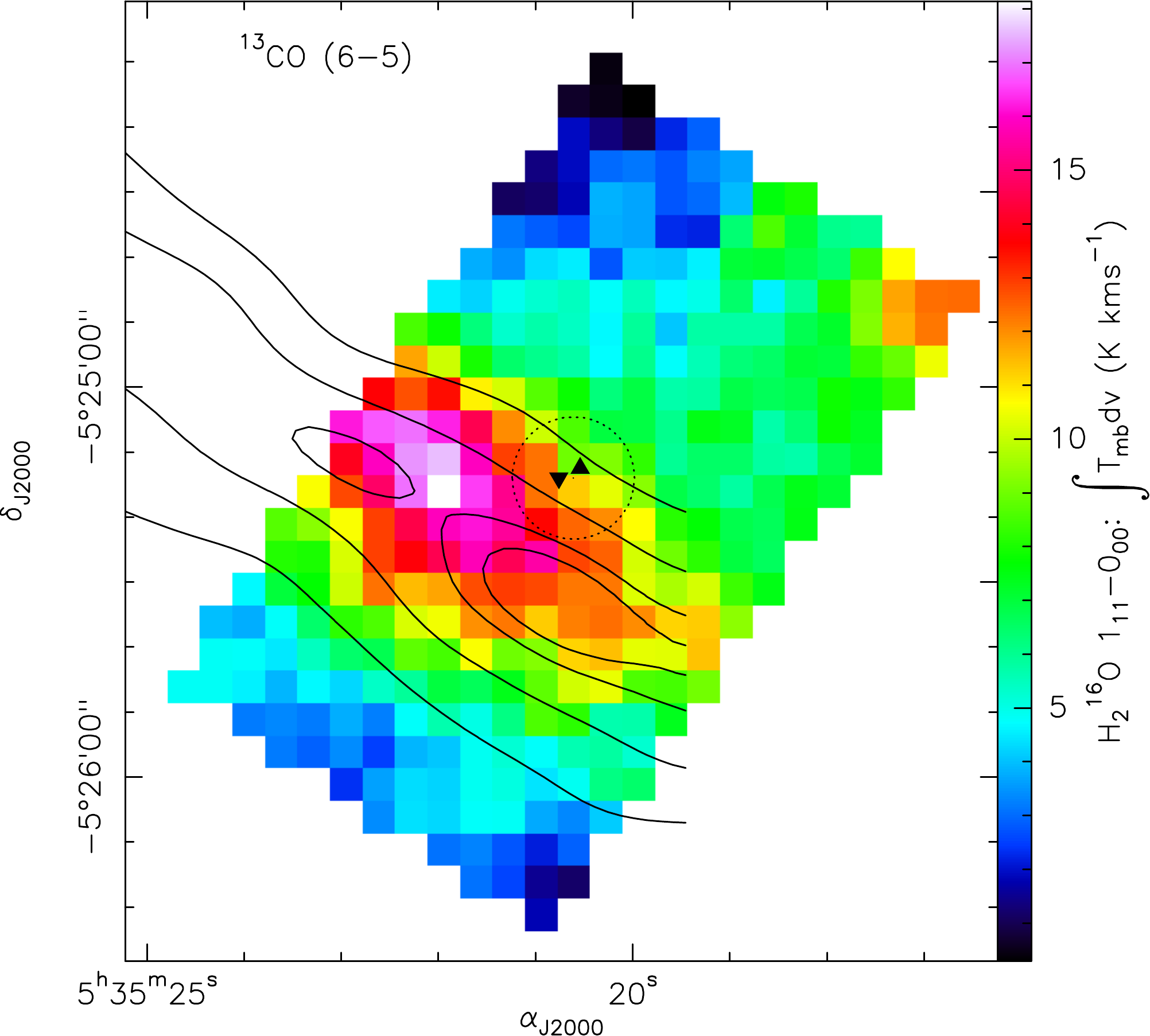}}\\
\subfloat{\includegraphics[scale=0.39]{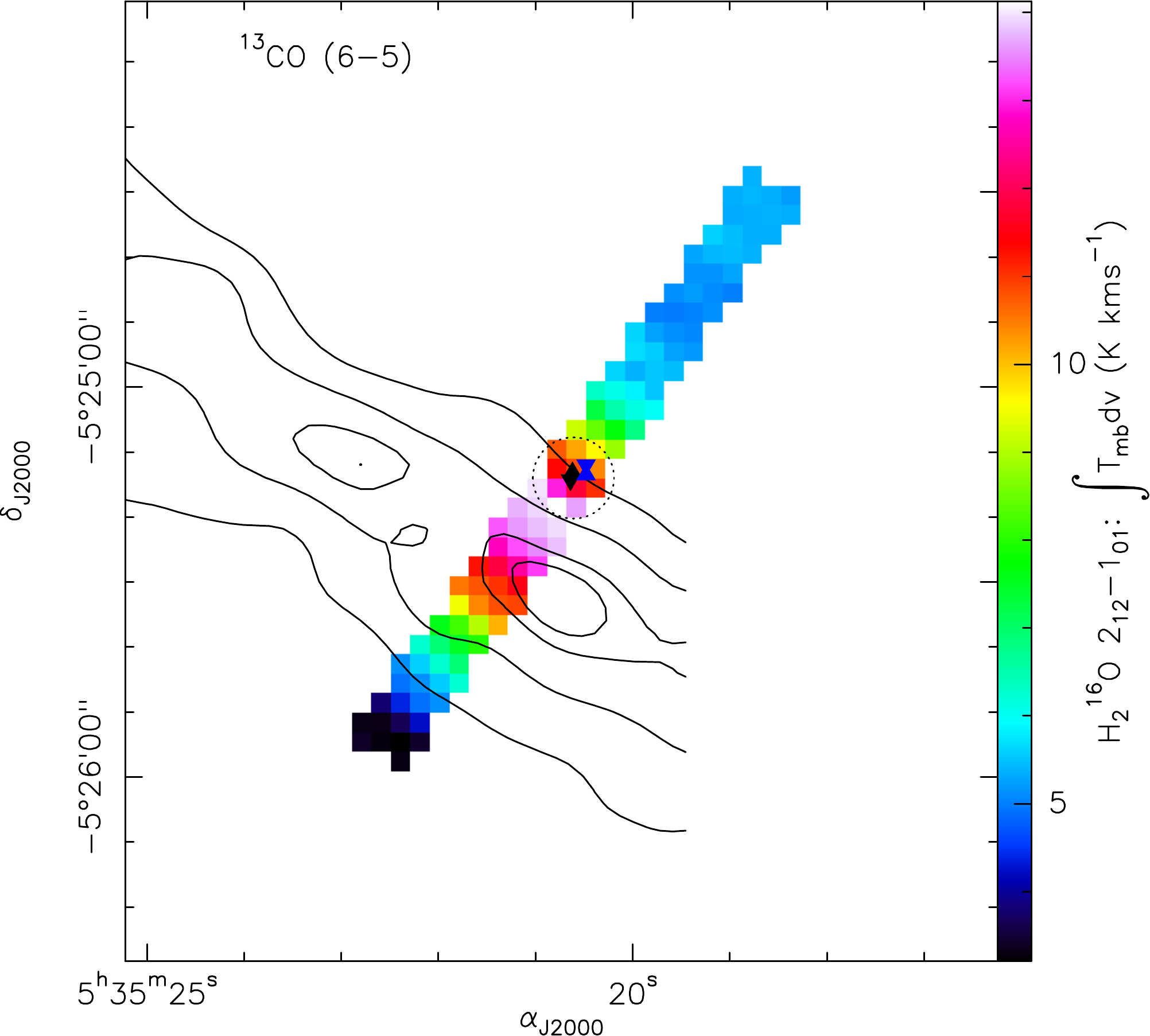}}\qquad
\subfloat{\includegraphics[scale=0.39]{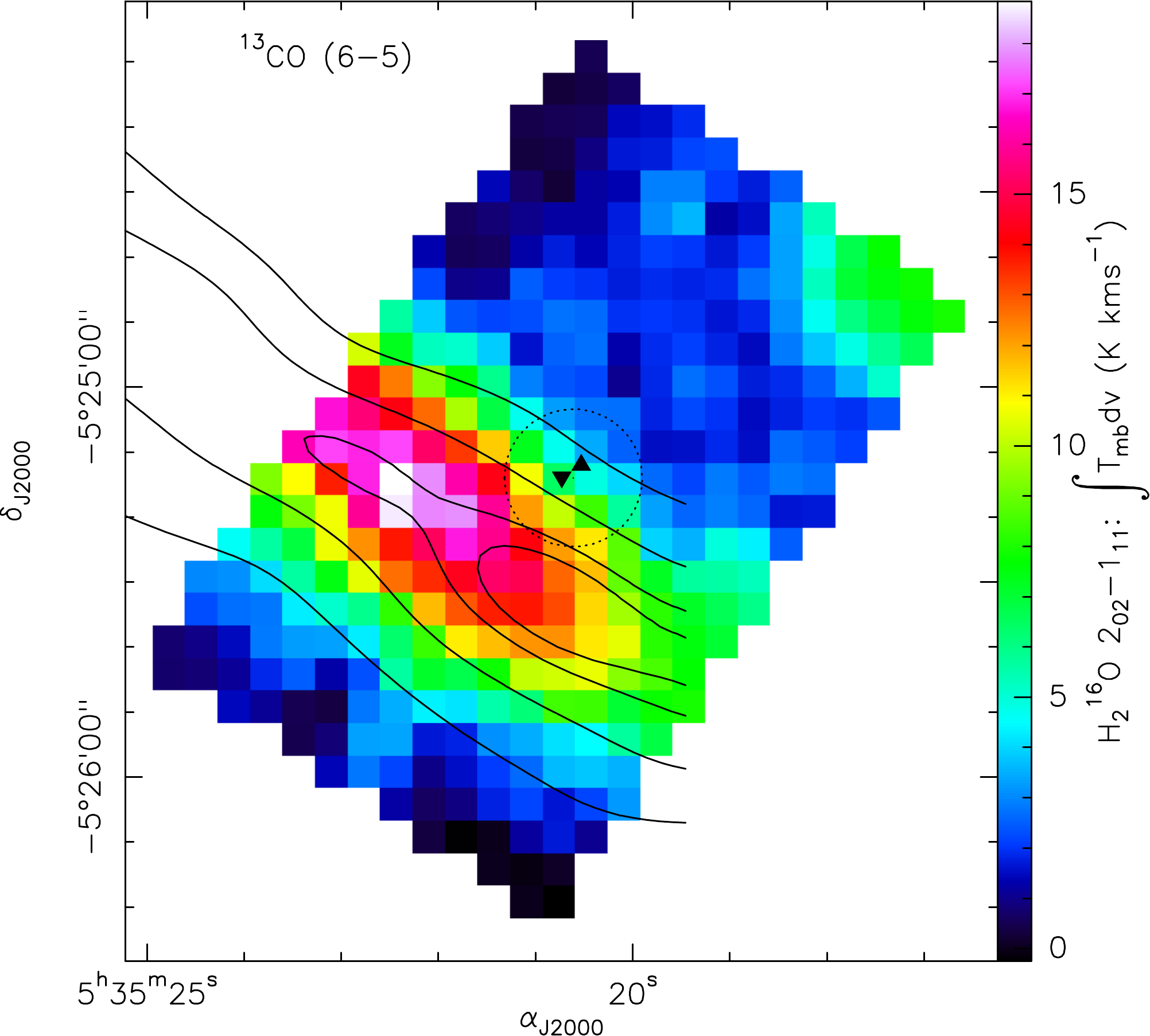}} \\
\subfloat{\includegraphics[scale=0.39]{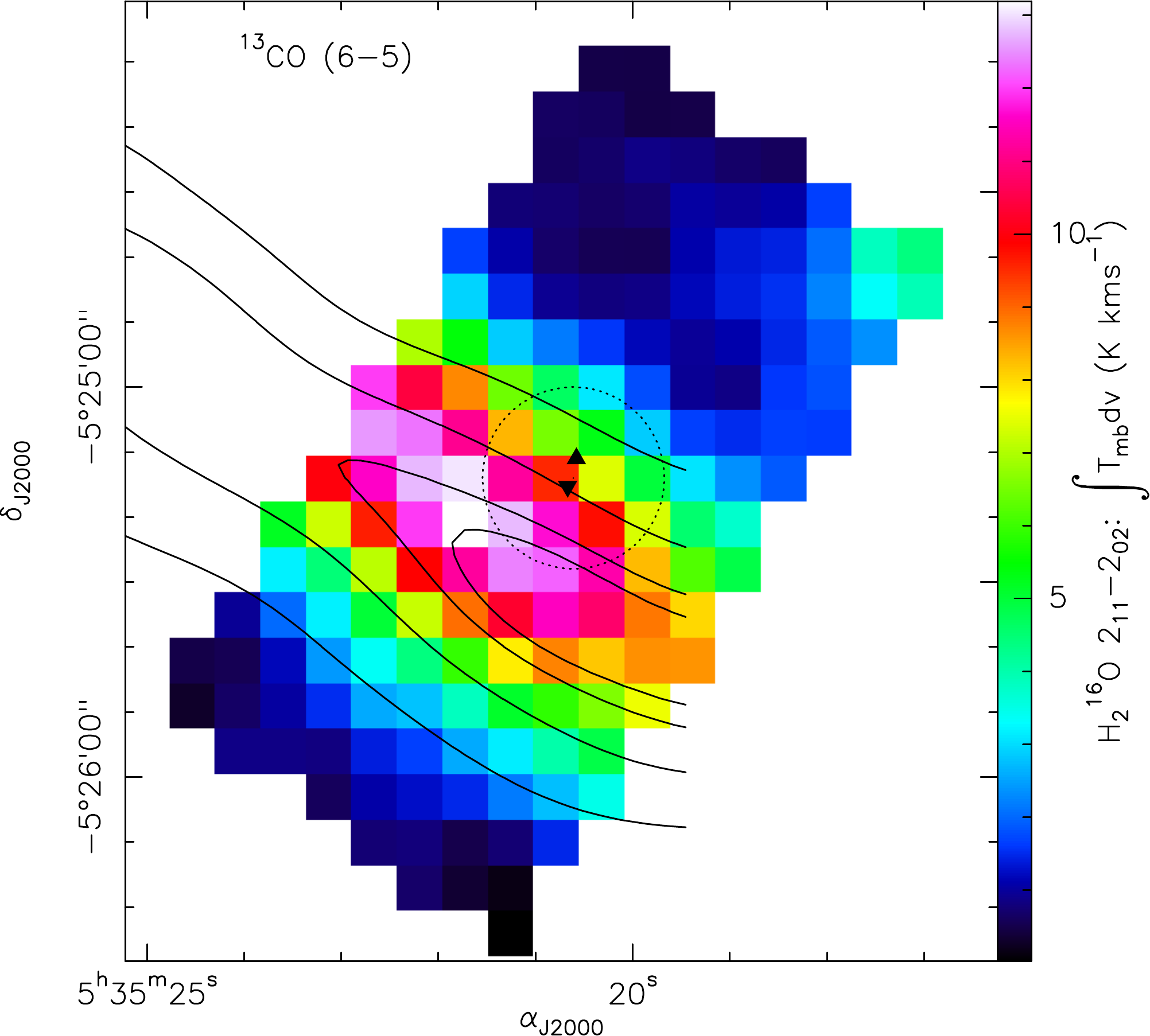}}\qquad
\subfloat{\includegraphics[scale=0.39]{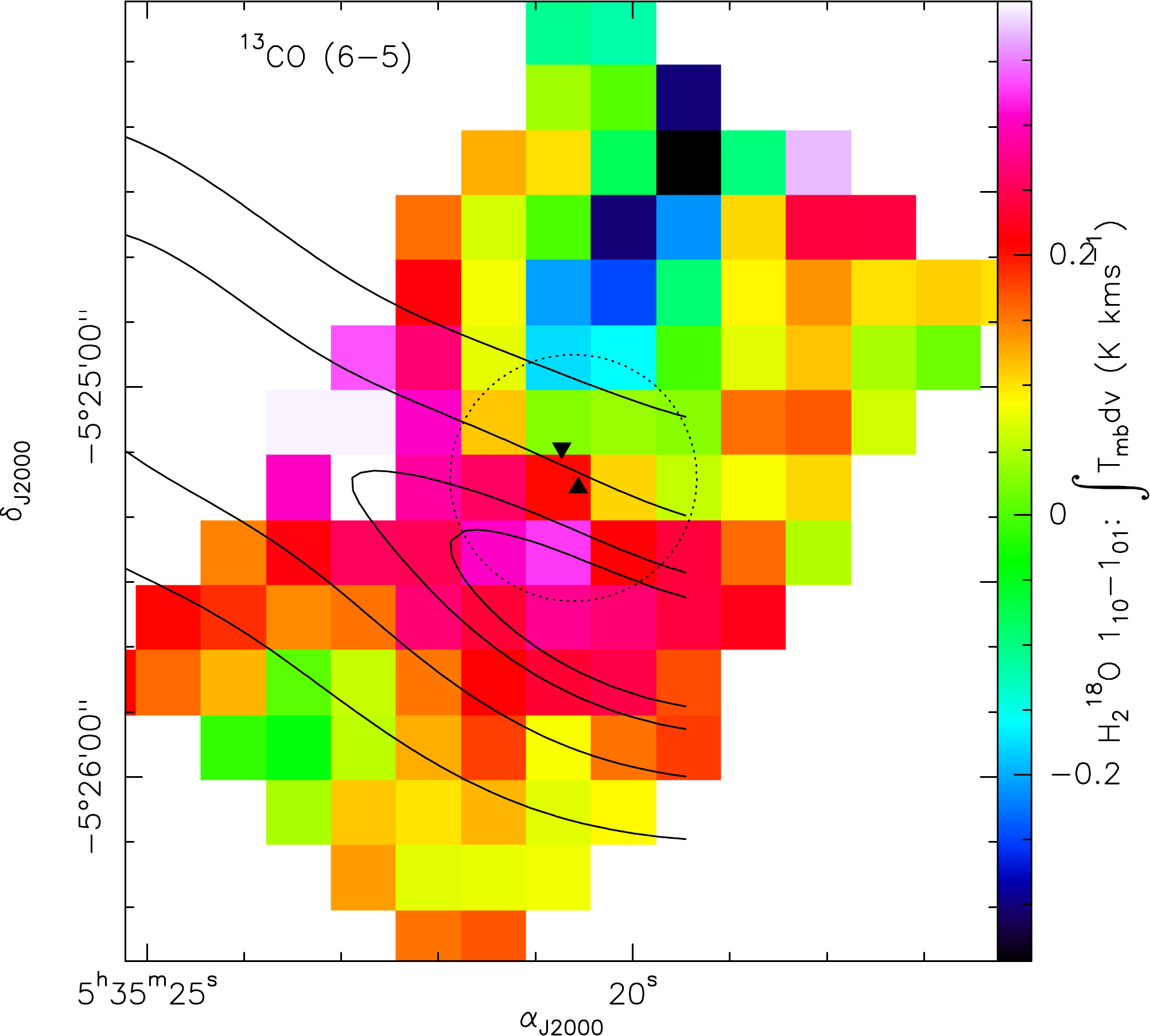}}\\
\caption{\label{fig:WaterOffset} Water emission maps averaged for H and V polarizations. Integration ranges are given in Table~\ref{tab:IntensityCal}. Black solid contours represent the emission of \chem{^{13}CO} \trans{6}{5} at \si{95}{\%}, \si{90}{\%}, \si{75}{\%} and \si{50}{\%} of the maximum intensity integrated from \SI{2}{km.s^{-1}} to \SI{17}{km.s^{-1}} observed with the CSO telescope and convolved at the averaged water line H and V beam width (Table~\ref{tab:IntensityCal}). Upward and downward triangles mark the targeted observations for H and V polarizations (blue ones are for the b labeled transitions, see Tables~\ref{tab:ObsUsed} and \ref{tab:IntensityCal}). Black dashed circle is the HIFI HPBW at the water frequency centered on the \chem{CO^{+}} peak.}
\end{figure*}

\end{appendix}
\end{document}